\documentclass[a4paper,10pt]{article}
\usepackage[position=t,singlelinecheck=off]{subfig}
\usepackage{amssymb,amsmath,amsfonts}
\usepackage{graphics,graphicx,float,braket}
\usepackage{epstopdf}
\usepackage[labelfont=bf]{caption}
\usepackage{array}
\newcolumntype{L}[1]{>{\raggedright\let\newline\\\arraybackslash\hspace{0pt}}m{#1}}
\newcolumntype{C}[1]{>{\centering\let\newline\\\arraybackslash\hspace{0pt}}m{#1}}
\newcolumntype{R}[1]{>{\raggedleft\let\newline\\\arraybackslash\hspace{0pt}}m{#1}}

\def\beq{\begin{equation}}
\def\eeq{\end{equation}}
\def\bea{\begin{eqnarray}}
\def\eea{\end{eqnarray}}
\def\nn{\nonumber}

\def\lo{\left(}
\def\rc{\right)}

\def\lg{\left\lgroup}
\def\rg{\right\rgroup}
\makeatletter
 
  \def\@cite#1#2{${\mbox{#1\if@tempswa , #2\fi}}$}
\makeatother

\makeatletter
\newcommand{\lambdabar}{{\mathchoice
  {\smash@bar\textfont\displaystyle{0.3}{1.2}\lambda}
  {\smash@bar\textfont\textstyle{0.25}{1.2}\lambda}
  {\smash@bar\scriptfont\scriptstyle{0.25}{1.2}\lambda}
  {\smash@bar\scriptscriptfont\scriptscriptstyle{0.25}{1.2}\lambda}
}}
\newcommand{\smash@bar}[4]{%
  \smash{\rlap{\raisebox{-#3\fontdimen5#10}{$\m@th#2\mkern#4mu\mathchar'26$}}}%
}
\makeatother
 \date{}
\begin{document}
 \begin{center}
 {\LARGE \sf{Study of the spin kitten states in a  strongly coupled spin-oscillator system}} \\
 
 \bigskip \bigskip\bigskip M. Balamurugan$^1$, R. Chakrabarti$^1$, B. Virgin Jenisha$^{1,2}$, V. Yogesh$^1$\\

 \begin{small}
 \bigskip
 \textit{$^1$
Department of Theoretical Physics, 
 University of Madras, \\
  Guindy Campus, 
 Chennai 600 025, India \\}
 \bigskip
 \textit{$^2$
Department of Physics, 
 Government College of Engineering, Srirangam, \\
Tamil Nadu- 620 012, India \\}
\end{small}
 \end{center}
 \vfill
 \begin{abstract}
\noindent Utilizing an adiabatic approximation method a bipartite qudit-oscillator Hamiltonian is explicitly studied for  low spin values in  both  strong and ultrastrong coupling regimes. The quasiprobability densities on the hybrid factorized phase space are introduced. Integrating over a sector of the composite phase space, the quasiprobability distributions of the complementary subsystem are recovered. In the strong coupling regime the qudit entropy displays a pattern of quasiperiodic  collapses and  revivals, where  the  locally minimum
 nonzero configurations appearing at rational fractions of the revival time  correspond to the spin kitten states. Starting with a bipartite \textit{factorizable} initial  state the evolution to the nonclassical transitory spin kitten states are displayed via the diagonal spin 
 $\mathrm{P}_{\mathcal{Q}}$-representation. The formation of transient spin kitten states is further substantiated by constructing 
 the spin tomogram  that employs  the positive definite probability distributions  embodying the diagonal elements of the corresponding density matrix in an arbitrarily rotated frame. As another manifestation of nonclassicality the emergence of the spin squeezed states during the bipartite evolution is observed. In the ultrastrong coupling domain a large number of interaction dependent modes and their 
 harmonics are generated. The consequent randomization of the phases  eliminates the quasiperiodicity of the system which is now 
 driven towards a stabilization of the entropy  accompanied with stochastic fluctuations around its stabilized value. Both in the strong and ultrastrong coupling realms antibunching of the photoemission events are realized particularly for the small spin values.
 \end{abstract}
 \newpage
 
\section{Introduction}
\label{Intro}
\noindent
Recently much interest has developed towards   experimental and theoretical studies on  hybrid interacting  spin-oscillator systems 
going beyond the rotating wave approximation [\cite{JC1963}] that complies with the preservation of the total excitation number. While the said approximation remains valid in a weak coupling regime endowed
with a small detuning between the spin and the oscillator frequencies, recent experimental realizations  use a varied set of tools to explore systems with strong and ultrastrong coupling between the degrees of freedom. For instance, a nanoelectromechanical resonator capacitively coupled to a Cooper pair box driven by microwave currents [\cite{ABS2002}, \cite{LaHaye2009}], a flux biased quantum circuit that utilizes the large inductance of a Josephson junction to generate an ultrastrong coupling with a coplanar waveguide resonator [\cite{Niemczyk2010}, \cite{Forn-Diaz2010}], and a quantum semiconductor microcavity embedding doped quantum wells [\cite{Anappara2009}, \cite{Todorov2010}] lie in this category. Specifically, the superconducting two level (qubit) as well as multilevel  (qudit) systems and circuits acting as artificial atoms are adaptable for a wide range of parameters. This flexibility  makes them the preferred building  blocks for quantum simulators [\cite{YN2005}-\cite{LMK2016}]. Multilevel superconducting circuit has been recently
considered [\cite{KFMM2015}] for implementing quantum gates. In principle, the entangled multilevel  quantum  systems store significantly more information, and have less networking problems compared to their 
two level  counterparts. An experimental demonstration of the nonclassical properties of a photonic qudit state has been attained [\cite{Lapkiewicz2011}]. Moreover, hybrid quantum circuits integrating multilevel atoms, spins,
cavity photons, and superconducting qudits coupled with nanoelectromechanical resonators hold much promise for  realization of the
quantum information network [\cite{ZAYN2014}].

\par

On the other hand the atomic coherent state [\cite{R1971}, \cite{ACGT1972}] provides a description for the collective atomic quantum processes such as superradiance [\cite{AT1976}-\cite{CNL2011}] and resonance fluorescence [\cite{ABNV1977}, \cite{D1980}] that require  quantum correlations in an atomic ensemble. One crucial instance of nonclassical properties is evident in the formation of the  
Schr\"{o}dinger cat and kitten states [\cite{HR2006}] that embody a coherent superposition of two or more distinguishable states of a mesoscopic system.  These states have been studied [\cite{APS1997}] in an ensemble of two level atoms interacting with a dispersive cavity mode in the context of the rotating wave approximation. Atomic cat-type states are known [\cite{GG1997}] to display 
interference-induced properties such as enhancement or reduction in the rates of spontaneous and stimulated emission. Employing two
hyperfine ground states of a beryllium ion the authors of Ref. [\cite{Leibfried2005}] considered a collection of ions confined in an electromagnetic trap and controlled with a classical laser beam. The cat states representing equal superposition of  two maximally distinct states have been observed [\cite{Leibfried2005}] up to six ions.  A probabilistic scheme has been proposed [\cite{McConnell2013}] for obtaining pure entangled spin states in large atomic ensembles  where the transmitted photons undergo a weak random Faraday rotation caused by the quantum noise of the atomic spin. Detection of two or more photons emerging from the ensemble with polarization vector orthogonal to the corresponding incoming polarization signals formation [\cite{McConnell2013}] of atomic 
Schr\"{o}dinger cat states. Experimental realization of atomic cat states characterized by coherent superposition of electronic spin states of opposite orientation has been observed [\cite{Chalopin2018}] in samples of dysprosium atoms undergoing AC Stark shift effected by detuned spin-light interaction. More recently, using superconducting transmon qubits coupled via a coplanar waveguide bus resonator 
the authors of Ref. [\cite{Song2019}] constructed Schr\"{o}dinger kitten states consisting up to $20$ qubits. 

\par

Another feature of nonclassicality is expressed by the spin squeezed states [\cite{KU1991}-\cite{KU1993}], which owe their origin to the nonlinear spin-spin effective interaction in the theory. These states  have been extensively utilized in the study of quantum phase transitions [\cite{Miranowicz2002}, \cite{WWYJ2002}], quantum chaos [\cite{SWYZ2006}], Bose-Einstein condensate [\cite{CLMZ1998}, \cite{JFIB2018}], and arrays of superconducting qubits [\cite{TAN2008}, \cite{ZYL2014}]. Recently one photon-two atom excitation process has been considered [\cite{YNSZ2019}] towards engendering optimal squeezing  in an ensemble of $N$ spins coupled to a single cavity mode. The spin squeezed state improves the precision measurement of magnetometry beyond the standard quantum limit [\cite{PMM2005}].  Interestingly, employing the quantum state transfer from the nonclassical light to the cold atoms the generation of macroscopic spin squeezed ensemble of atoms has been experimentally observed [\cite{HSSP1999}]. An extensive recent review of the spin squeezing is given in Ref. [\cite{MWSN2011}]. 

\par

In the setting described above, here we  study the evolution of a hybrid bipartite state which is a linear combination of the qudit spin coherent states tensored with the squeezed coherent states of the field mode. A suitable adiabatic approximation pioneered in Refs. [\cite{IGMS2005}, \cite{AN2010}] allows us to investigate the combined structure  for a strong spin-photon coupling as well as a large detuning of the associated frequencies. To analyze the system we introduce the bipartite quasiprobability distributions in the
composite phase space of the qudit and  field variable. Tracing over one  degree of freedom reproduces the  phase space quasiprobability distributions [\cite{A1981}, \cite{Gerry2005}] of the coupled complementary subsystem. Starting with a factorized state of the bipartite system nonclassical states such as the transitory qudit kitten states dynamically emerge corresponding to the local minima  of  
time evolution of the entropy in the strong interaction regime. We also employ the tomographic  procedure towards  reproducing the states of the qudit by utilizing the close correspondence between the quasiprobability distributions and the probability density related to the diagonal elements of the spin density matrix in an arbitrarily rotated frame [\cite{MM1997}, \cite{DM1997}]. The transient spin kitten states are also evident in the tomographic depictions considered here for the components of the spin variable. Moreover, quantum 
fluctuations triggered by the nonlinear terms in the effective Hamiltonian of the spin degree of freedom turn spin coherent states to   short-lived squeezed spin states. Specifically towards illustrating the above construction we explicitly discuss the low spin 
$s=1, \tfrac{3}{2}$ cases. For a higher spin variable \textit{more complex kitten states} emerge as more spin configurations participate in the interference pattern. In the ultrastrong spin-oscillator coupling regime the realization of a large number of interaction modes spread over a wide range of time scales abolishes the phase correlations necessary for the manifestation of the spin kitten states, which, consequently, disappear. In addition, it causes materialization of a steady state value of the entropy, which is, however,
subjected to rapid stochastic fluctuations.  
\section{Hamiltonian and its approximate diagonalization}
\label{Diagonalize}
\setcounter{equation}{0}
The bipartite qudit-oscillator Hamiltonian  reads
\beq
 H=  - \Delta S_{\mathsf x} + \omega a^{\dagger} a + \lambda S_{\mathsf z} (a+a^{\dagger}).
 \label{H-Sp}
\eeq
The spin variables 
$\{S_{\mathcal X}| {\mathcal X} = \mathsf{x, y, z}\}$ obey the  $su(2)$ algebra $\{S_{\pm}\equiv 
S_{\mathsf {x}} \pm i S_{\mathsf {y}}; [S_z, S_{\pm}] = \pm S_{\pm}, [S_{+}, S_{-}] = 2 S_{\mathsf{z}}\}$ and
maintain the standard irreducible representations
\beq
S_{\mathsf{z}} |s, m \rangle = m |s, m\rangle, \;
S_{\pm}|s, m\rangle = \sqrt{(s \mp m) (s \pm m + 1)}\, |s, m \pm 1\rangle;\; s = 0, \tfrac{1}{2}, 1,  \ldots;
 \; m = -s, -s+1, \ldots, s, 
\label{S-alg}
\eeq
whereas the oscillator degree of freedom is characterized as follows:
$\{a, a^{\dagger}, \hat {n} \equiv  a^{\dagger} a; [a,  a^{\dagger}] = \mathbb{I}:  \hat {n} |n\rangle = n |n\rangle, a |n\rangle = \sqrt{n}\, |n - 1\rangle,  a^{\dagger} |n\rangle = \sqrt{n + 1}\, |n + 1\rangle\}$. 
Employing a variational method the Hamiltonian (\ref{H-Sp}) has been previously investigated [\cite{LL2013}], and in the vicinity of the resonance configuration its approximate ground state has been determined [\cite{LL2013}]. On the other hand
the adiabatic approximation [\cite{IGMS2005}, \cite{AN2010}] considered here employs a separation of the time scales between the fast  moving oscillator with frequency $\omega$, and the slow moving qudit possessing an energy gap $\Delta \ll \omega$. The qudit-oscillator coupling is parametrized by $\lambda$. The (ultra)strong interaction  regime $\lambda \lesssim \omega$  necessitates incorporating terms in the Hamiltonian (\ref{H-Sp}) that do not preserve the excitation number.  Under the said approximation the oscillator and the qudit parts of the Hamiltonian, respectively, assume the form 
\beq
 H_{\mathcal O} = \omega a^{\dagger} a + \lambda\, \bra{s, m} S_{\mathsf{z}} \ket{s, m} \,(a+a^{\dagger}), \;\; 
\bra{s, m} S_{\mathsf{z}} \ket{s, m} =m;\qquad  
 H_{\mathcal Q} = - \Delta S_{\mathsf{x}}.
 \label{H-qudit-osc} 
\eeq
The eigenenergies and eigenstates of the oscillator Hamiltonian $ H_{\mathcal O}$ read
\beq
 \varepsilon_{n,m}=\omega(n-(m\widetilde{\lambda})^2), \quad \widetilde{\lambda}= \tfrac{\lambda}{\omega}, \quad
 |n_m \rangle = \mathrm{D} (m \widetilde{\lambda})^{\dagger} |n\rangle,
 \label{H-osc} 
\eeq
where the displacement operator is denoted by $ \mathrm{D}(\alpha) = \exp (\alpha a^{\dagger} - \alpha^{*} a), \alpha = \mathrm {Re}(\alpha) + i 
\,\mathrm {Im} (\alpha)$. Under the adiabatic approximation the  Hamiltonian $H$ now assumes the block diagonal form where the $n$-th photonic manifold is expressed via the basis set  $|s,m \rangle |n_m \rangle \equiv |s, m; n_m \rangle $. The approximate energy eigenstates for the said $n$-th photonic manifold in the examples of spin variables $s= 1, \frac{3}{2}$ are described below.

\par

  For the $s=1$ case the $n$-th photonic block of the Hamiltonian is given by
\beq
 H_n^{(s=1)} = \omega
 \begin{pmatrix}
 n- \widetilde{\lambda}^2 &  \Delta_n & 0 \\
 \Delta_n & n & \Delta_n \\
 0 & \Delta_n & n- \widetilde{\lambda}^2
 \end{pmatrix},
 \label{H-n-1}
\eeq
where the scaled and renormalized qudit gap parameter is listed as $\Delta_n= - \tfrac{\Delta}{\sqrt{2} \omega}  \exp\lo-\frac{\widetilde{\lambda}^2}{2}\rc L_n^{0}(\widetilde{\lambda}^2)$. The Laguerre polynomial maintains the standard expansion $L_{n}^{k} (\mathsf{x}) =\sum_{\ell}\frac{(- 1)^{\ell}}{\ell !}
\binom{n + k}{n - \ell} \mathsf{x}^{\ell}$.
In the subspace of the $n$-th block the eigenenergies $E_{\jmath,n}^{(1)}\;, \jmath \in \{0,\pm \}$ of the Hamiltonian (\ref{H-n-1}) may be given by 
\beq
E_{0,n}^{(1)} = \omega (n- \widetilde{\lambda}^2), 
\quad 
 E_{\pm,n}^{(1)} = \omega \lo n- \tfrac{1}{2} \lo\widetilde{\lambda}^2 \mp \delta_n \rc\rc, \quad
 \delta_n= \sqrt{8\Delta_n^2+\widetilde{\lambda}^4}
 \label{H-n-1-eigenvalue}
\eeq
and the corresponding normalized eigenvectors read
\bea
 |E_{0,n}^{(1)}\rangle \! \! &=& \! \!\tfrac{1}{\sqrt{2}}(|1,1;n_{1}\rangle-|1,-1;n_{-1}\rangle),\nn \\
 |E_{\pm,n}^{(1)}\rangle \! \!&=&\! \! \tfrac{1}{\sqrt{\mathcal{N}^{(1)}_{\pm, n}}}
  (2 \Delta_n\,|1,1;n_{1}\rangle\,+\,(\widetilde{\lambda}^2 \pm \delta_n)\,|1,0;n\rangle\,
  +\,2 \Delta_n \;|1,-1;n_{-1}\rangle), 
 \;\mathcal{N}^{(1)}_{\pm, n} =2 \,\delta_n \lo\delta_n \pm \widetilde{\lambda}^2\rc.
 \label{H-n-1-eigenvector}
\eea
In a specific $n$-th photonic block the above basis set fulfills the orthocompleteness relations: $\left\langle E_{\jmath,n}^{(1)}\right|\left.E_{\ell,n}^{(1)}\right\rangle = \delta_{\jmath, \ell},\;$ where $\jmath, \ell \in \{0, \pm\},$ and $\displaystyle{\sum_{\jmath}} 
\left|E_{\jmath,n}^{(1)}\right\rangle \left\langle E_{\jmath,n}^{(1)}\right| = \mathbb{I}$.

\par

The example of the $n$-th photonic block Hamiltonian for the spin $s = \frac{3}{2}$ may be discussed similarly:
\beq
H_n^{(s=\frac{3}{2})} = \omega
\begin{pmatrix}
	n-{\left(\tfrac{3 \widetilde{\lambda}}{2}\right)}^2 & \sqrt{\frac{3}{2}} \Delta_n & 0 & 0\\
	\sqrt{\frac{3}{2}}\Delta_n & n-{\left(\tfrac{\widetilde{\lambda}}{2}\right)}^2 & \sqrt{2}\,\Delta_n & 0 \\
	0 & \sqrt{2}\,\Delta_n & n-{\left(\tfrac{\widetilde{\lambda}}{2}\right)}^2& \sqrt{\frac{3}{2}}\Delta_n \\
	0 & 0 & \sqrt{\frac{3}{2}}\Delta_n & n-{\left(\tfrac{3 \widetilde{\lambda}}{2}\right)}^2
\end{pmatrix}.
\label{H-n-3/2}
\eeq
The eigenenergies of the Hamiltonian (\ref{H-n-3/2}) are listed as
\beq
E^{\pm, \frac{3}{2}}_{\ell,n}=\! \omega \left(n-\tfrac{5\widetilde{\lambda}^{2}}{4}+ \ell\;\tfrac{\Delta_n}{\sqrt{2}} \pm \chi^{(\ell)}_{n}\right), \;
\chi^{(\ell)}_{n}=\sqrt{\widetilde{\lambda}^{4}+ \ell \;\sqrt{2}\,\widetilde{\lambda}^{2}\,\Delta_{n}+2\,\Delta_{n}^{2}},
\quad \ell \in \{\pm 1\},
\label{3/2-eigenvalue}
\eeq
while the corresponding eigenvectors read
\bea
\left|E^{\pm, \frac{3}{2}}_{\ell,n}\right\rangle  \!\!\!\! &=& \!\!\!\! \tfrac{1}{\mathcal{N}_{\ell,n}^{\pm,\frac{3}{2}}} \; 
\lg\left|\tfrac{3}{2},\tfrac{3}{2};n_{_{\frac{3}{2}}}\right\rangle 
+ \Gamma_{\ell}^{(\pm)} \left|\tfrac{3}{2},\tfrac{1}{2};n_{_{\frac{1}{2}}}\right\rangle 
+ \ell \; \Gamma_{\ell}^{(\pm)} \left|\tfrac{3}{2}, -\tfrac{1}{2} ;n_{_{-\frac{1}{2}}}\right\rangle 
+ \ell\; \left|\tfrac{3}{2}, -\tfrac{3}{2};n_{_{-\frac{3}{2}}}\right\rangle \rg
\label{3/2-eigenvector}
\eea
and the coefficients are given by
\beq
\Gamma_{\ell}^{(\pm)}  =  \tfrac{1}{\sqrt{3}} \left( \ell \pm \sqrt{2} \; \tfrac{{}^{
		\mbox{\fontsize{10.28}{21.6}\selectfont\( \chi_{n}^{(\ell)} \)}} \! \pm \widetilde{\lambda}^{2}}{\Delta_{n}}\right), \;\;
\mathcal{N}_{\ell,n}^{\pm, \frac{3}{2}} = \sqrt{2} \left( 1 + \left( \Gamma_{\ell}^{(\pm)} \right)^{2} \right)^{\frac{1}{2}}.
\label{N-3/2}
\eeq
The orthocompleteness relations for the states (\ref{3/2-eigenvector}) hold in the $n$-th photonic sector: 
 $\left\langle E_{\jmath^{\prime} ,n}^{\kappa^{\prime}, \frac{3}{2}}\right|\left.E_{\jmath,n}^{\kappa, \frac{3}{2}}\right\rangle = \delta_{\jmath, \jmath^{\prime}}\;\delta_{\kappa, \kappa^{\prime}},\;$ where $\jmath, \jmath^{\prime} \in \{\pm 1\}, \kappa, \kappa^{\prime} \in \{\pm\}$ and $\displaystyle{\sum_{\jmath, \kappa}} 
\left|E_{\jmath,n}^{\kappa, \frac{3}{2}}\right\rangle \left\langle E_{\jmath,n}^{\kappa, \frac{3}{2}}\right| = \mathbb{I}$.
\section{Initial state and its evolution via the adiabatic approximation}
\label{InitialState}
\setcounter{equation}{0}
The generalized quasi-Bell bipartite entangled initial state is chosen as
\beq
 |\psi_{s}(0)\rangle = \mathcal{N}_{s} \left( | \mathfrak{z}\rangle_{(s)}\;\; |\alpha, \xi \rangle + \mathrm{c} \,| \mathfrak{-z}\rangle_{(s)} \;\;|-\alpha, \xi \rangle \right) ,
 \label{state-t0}
\eeq
where $|\mathfrak{z}\rangle_{(s)}$ is the qudit spin-$s$ coherent state [\cite{ACGT1972}]. Its expansion  via the eigenstates of 
the generator $S_{\mathsf{z}}$ reads 
\beq
 |\mathfrak{z}\rangle_{(s)} =\tfrac{1}{\left(1+|\mathfrak{z}|^2\right)^s} \sum_{m=-s}^{s} \tbinom{2s}{s+m}^{\frac{1}{2}} 
\mathfrak{z}^{s+m}\, |s,m \rangle.
\label{s-coherent}
\eeq
The polar coordinate $\lo\mathfrak{z}=\tan\lo\frac{\widetilde{\theta}}{2}\rc \,\exp(-i\widetilde{\phi})\rc$ allows us to recast the  sum (\ref{s-coherent}) in terms of the spherical phase space variables as
\beq
|\mathfrak{z}\rangle_{(s)}  \equiv |\widetilde{\theta},\widetilde{\phi} \rangle_{(s)} 
=\sum_{m=-s}^{s} \tbinom{2s}{s+m}^{\frac{1}{2}} \,
\lo\sin \tfrac{\widetilde{\theta}}{2}\rc^{s+m} \,
\lo \cos \tfrac{\widetilde{\theta}}{2}\rc^{s-m}
 \,\exp(-i(s+m)\widetilde{\phi})\,|s,m \rangle.
 \label{s-polar}
\eeq
The squeezed oscillator coherent state [\cite{Gerry2005}] is structured as $|\alpha,\xi\rangle \equiv \mathrm{D} (\alpha)  \mathrm{S}(\xi) |0\rangle,\;
\mathrm{S}(\xi)  = 
\exp  \lo\frac{\xi^{*} a^{2} - \xi a^{\dagger 2}}{2}\rc$, where $\xi (= r \exp (i \zeta)) \in \mathbb{C}$.
Employing the parameters $\mu = \cosh r, \nu = \sinh r \;\exp (i \zeta)$ its mode expansion is given by
\bea
|\alpha, \xi  \rangle = \sum_{n=0}^{\infty} \mathcal{S}_{n}(\alpha, \xi) |n \rangle,
\quad  \mathcal{S}_{n}(\alpha, \xi)=\tfrac{1}{\sqrt{n!\;\mu}} 
\left(\tfrac{\nu}{2\mu}\right)^{\frac{n}{2}}\; \exp\left(-\tfrac{1}{2}\,|\alpha|^2-\tfrac{\nu}{2\,\mu}\, \alpha^*{}^2 \right)\, 
 \mathrm{H}_n\left(\tfrac{\mu \alpha + \nu \alpha^{*}}{\sqrt{2\,\mu\,\nu}}\right),
\label{squeezed-mode}
\eea
where the Hermite polynomials obey the sum rule: $\exp\lo 2 \mathcal{X} t - t^{2}\rc = \sum_{n =0}^{\infty}
 H_{n}(\mathcal{X})\, \tfrac{t^{n}}{n!}$. For a large value of the parameter $|\alpha|^{2}\gg 1$ the oscillator coherent state may be regarded as macroscopic in nature. The normalization constant for the initial state (\ref{state-t0}) reads $\mathcal{N}_{s}= \left( 1 +|\mathrm{c}|^{2} + 
 2 \left( \tfrac{1-|\mathfrak{z}|^{2}}{1+|\mathfrak{z}|^{2}} \right)^{2s} \exp(-2|\alpha \mu + \alpha^{*} \nu|^{2}) \; \mathrm{Re}(\mathrm{c})) \right)^{-\frac{1}{2}}$. The parameter $\mathrm{c} \in \mathbb{C}$ appearing in the linear combination (\ref{state-t0}) allows us to suitably select the initial state. For instance, the choice $\mathrm{c} = 0$ leads to the factorized bipartite state at 
 $t=0$, and therefore the transient formation of the nonclassial Schr\"{o}dinger kitten states discussed in Sec. \ref{nonclassicality} owes its origin to \textit{dynamical effects}.
 
 \par
 
 For the spin $s=1$ case our approximate diagonalization via the basis states $\{|E_{\jmath,n}^{(1)}\rangle \,|\jmath \in (0, \pm ); n = 0, 1, \ldots\}$ given in (\ref{H-n-1-eigenvector})
permits us to extract the  time evolution of the corresponding   initial state (\ref{state-t0}):
\beq
 |\psi_{_{1}}(t)\rangle =  \sum_{\jmath=-1}^{1} \sum_{n=0}^{\infty}\;\mathcal{A}_{\jmath,n}^{(1)} \;\exp(-i E_{\jmath,n}^{(1)} t)\; |E_{\jmath,n}^{(1)}\rangle, 
   \label{state-t-expn}
\eeq
where the projectors of the initial state (\ref{state-t0}) on the approximate eigenvector basis (\ref{H-n-1-eigenvector}) read
\bea
\mathcal{A}_{0,n}^{(1)} \equiv \langle E_{0,n}^{(1)} | \psi_{_{1}}(0) \rangle \!\!\!\! &=& \!\!\!\! \tfrac{\mathcal{N}_{1}}{\sqrt{2}\,(1+|\mathfrak{z}|^2)} \lg 
(\mathfrak{z}^2 - (-1)^{n} \mathrm{c}) \, \exp\big( -i\widetilde{\lambda}\, \mathrm{Im}(\alpha)\big)\, \mathcal{S}_{n}(\alpha_+, \xi) \right. \nn \\
&& \left. - (1- (-1)^{n}  \mathrm{c}\, \mathfrak{z}^2)
\exp\big( i\widetilde{\lambda}\, \mathrm{Im}(\alpha)\big)\, \mathcal{S}_{n}(\alpha_-, \xi)\rg,\nn \\
\mathcal{A}_{\pm,n}^{(1)} \equiv \langle E_{\pm,n}^{(1)} | \psi_{_{1}}(0) \rangle \!\!\!\! &=& \!\!\!\! \tfrac{\mathcal{N}_{1}}{\sqrt{\mathcal{N}^{(1)}_{\pm,n}}\,(1+|\mathfrak{z}|^2)} \left\lgroup 
2  \Delta_n\, (\mathfrak{z}^2 + (-1)^{n} \mathrm{c}) \,\exp \big(-i \widetilde{\lambda} \mathrm{Im}(\alpha)\big) \,\mathcal{S}_{n}(\alpha_+, \xi) \right.\nn \\
 && \left. + \; \sqrt{2} (\widetilde{\lambda}^2 \pm \delta_n) \, \mathfrak{z}\, (1 - (-1)^{n} \mathrm{c})
\mathcal{S}_{n}(\alpha, \xi) \right. \nn \\
&& \left. + \; 2 \Delta_{n}\, (1 + (-1)^{n} \mathrm{c}\, \mathfrak{z}^2)
\exp\big( i\widetilde{\lambda}\, \mathrm{Im}(\alpha)\big)\, \mathcal{S}_{n}(\alpha_-, \xi)
\rg, \quad \alpha_{\pm} = \alpha \pm \widetilde{\lambda}.  
\label{t-coefficients}
\eea
For the spin $s$ example the evolving bipartite state $|\psi_{_{s}}(t)\rangle$ produces, up to the approximation considered here,
 the  pure state  density matrix as follows:
\beq
\rho^{(s)}(t)=|\psi_{_{s}}(t)\rangle \langle \psi_{_{s}}(t) |. 
\label{DenMaHybrid}
\eeq 
Partial tracing on the oscillator degrees of freedom contained in the above density matrix 
$\rho^{(s=1)}(t)$  yields the corresponding qudit reduced density matrix
$\rho^{(1)}_{\mathcal{Q}}(t)=\mathrm{Tr}_{\mathcal{O}}(\rho^{(1)}(t))$:
\beq
 \rho^{(1)}_{\mathcal{Q}}(t)= \sum_{n,\widetilde{n}=0}^{\infty} 
 \begin{pmatrix}
  \mathcal{B}_{+,n}^{(1)}(t) \mathcal{B}_{+,\widetilde{n}}^{(1)}(t)^{*}\, \delta_{n \widetilde{n}} &
  \mathcal{B}_{+,n}^{(1)}(t) \mathcal{B}_{0,\widetilde{n}}^{(1)}(t)^{*}\, \mathcal{G}_{\widetilde{n} n}(-\widetilde{\lambda})
  & \mathcal{B}_{+,n}^{(1)}(t) \mathcal{B}_{-,\widetilde{n}}^{(1)}(t)^{*}\, \mathcal{G}_{\widetilde{n} n}(-2 \widetilde{\lambda}) \\
    \mathcal{B}_{0,n}^{(1)}(t)\mathcal{B}_{+,\widetilde{n}}^{(1)}(t)^{*}\,  \mathcal{G}_{\widetilde{n} n}(\widetilde{\lambda})  & 
    \mathcal{B}_{0,n}^{(1)}(t)\mathcal{B}_{0,\widetilde{n}}^{(1)}(t)^{*}\, \delta_{n\widetilde{n}}
  & \mathcal{B}_{0,n}^{(1)}(t) \mathcal{B}_{-,\widetilde{n}}^{(1)}(t)^{*}\, \mathcal{G}_{\widetilde{n} n}(- \widetilde{\lambda}) \\
      \mathcal{B}_{-,n}^{(1)}(t) \mathcal{B}_{+,\widetilde{n}}^{(1)}(t)^{*}\,  \mathcal{G}_{\widetilde{n} n}(2 \widetilde{\lambda})  
      & \mathcal{B}_{-,n}^{(1)}(t) \mathcal{B}_{0,\widetilde{n}}^{(1)}(t)^{*}\,
       \mathcal{G}_{\widetilde{n} n}(\widetilde{\lambda}) 
  & \mathcal{B}_{-,n}^{(1)}(t)\mathcal{B}_{-,\widetilde{n}}^{(1)}(t)^{*}\,\delta_{n\widetilde{n}}
 \end{pmatrix},
 \label{density-1}
\eeq
where the elements are expressed via the sum of the factorized time dependent components as
\bea
 \mathcal{B}_{0,n}^{(1)} (t)&=& \tfrac{1}{\sqrt{\mathcal{N}_{n}^{+}}}\,  (\widetilde{\lambda}^2 + \delta_n)\, \exp(-iE_{+,n}^{(1)}t)\,
 \mathcal{A}_{+,n}^{(1)}+ \tfrac{1}{\sqrt{\mathcal{N}_{n}^{-}}}\,(\widetilde{\lambda}^2 - \delta_n)
 \, \exp(-iE_{-,n}^{(1)}t)\,\mathcal{A}_{-,n}^{(1)}, \nn \\
  \mathcal{B}_{\pm,n}^{(1)} (t)\! \!&=& \! \!  \pm \tfrac{1}{\sqrt{2}}\, \exp(-iE_{0,n}^{(1)}\,t) \, \mathcal{A}_{0,n}^{(1)} +  
 2\, \tfrac{\Delta_n}{\sqrt{\mathcal{N}_{n}^{+}}}\, \exp(-iE_{+,n}^{(1)}\,t)\,  \mathcal{A}_{+,n}^{(1)} 
+ 2\,\tfrac{\Delta_n}{\sqrt{\mathcal{N}_{n}^{-}}} \,\exp(-iE_{-,n}^{(1)}\,t)\, \mathcal{A}_{-,n}^{(1)}.\quad
\label{B-def}
\eea
The off-diagonal elements  of the density operator (\ref{density-1}) carry the correlation functions of the oscillator number states:
\bea
\mathcal{G}_{m n}(\mathcal{X}) &\equiv & \langle m|\mathrm{D}(\mathcal{X})| n \rangle \nn \\
 &=& \begin{cases}
 \exp \lo-\tfrac{|\mathcal{X}|^2}{2} \rc  \mathcal{X}^{m-n} \sqrt{\tfrac{n!}{m!}} \, \, L_n^{m-n}\lo |\mathcal{X}|^2\rc 
 & \forall \,m \geq n,\\
 \exp \lo-\tfrac{|\mathcal{X}|^2}{2} \rc (- \mathcal{X}^{*})^{n-m} \sqrt{\tfrac{m!}{n!}} \, \, L_m^{n-m}\lo |\mathcal{X}|^2\rc & 
 \forall \,m < n.
\end{cases}
\label{G-def}
\eea
The qudit density matrix (\ref{density-1}) maintains the required normalization restriction: 
$\mathrm{Tr} \rho^{(1)}_{\mathcal{Q}} = 1$. The oscillator reduced density matrix for the $s=1$ case may also be extracted by partial tracing of the spin variables  in the  bipartite pure state $\rho^{(s=1)}(t)$: 
\bea
\rho^{(1)}_{_{\mathcal{O}}}(t) \!\!\! \! &=& \!\!\! \! \sum_{n,\widetilde{n}=0}^{\infty} 
\lg  \mathcal{B}_{+,n}^{(1)}(t)
\mathcal{B}_{+,\widetilde{n}}^{(1)}(t)^{*} \ket{n_{1}}\bra{\widetilde{n}_{1}} +
 \mathcal{B}_{0,n}^{(1)}(t)
 \mathcal{B}_{0,\widetilde{n}}^{(1)}(t)^{*} \ket{n_{0}}\bra{\widetilde{n}_{0}} +
\mathcal{B}_{-,n}^{(1)}(t)
\mathcal{B}_{-,\widetilde{n}}^{(1)}(t)^{*} \ket{n_{-1}}\bra{\widetilde{n}_{-1}} \rg, \qquad \quad
\label{spin_1_osc_den_matrix}
\eea
where the normalization reads $\mathrm{Tr} \rho^{(1)}_{_{\mathcal{O}}}(t) = 1$.

\par

Similarly, the basis states $\left|E_{\ell,n}^{\pm, \frac{3}{2}}\right\rangle$ listed in (\ref{3/2-eigenvector}) facilitate the approximate determination of the evolution of the initial state (\ref{state-t0}) for the spin $s=\tfrac{3}{2}$ case:
\beq
\ket{\psi_{\frac{3}{2}}(t)} = \sum_{ \kappa \in \pm}\;\sum_{ \ell \in \pm 1} \sum_{n=0}^{\infty} 
\mathcal{A}^{\kappa, \frac{3}{2}}_{\ell, n} \exp \left( - i E_{\ell,n}^{\kappa, \frac{3}{2}} t \right) 
\left|E_{\ell,n}^{\kappa, \frac{3}{2}}\right\rangle, \;
 \mathcal{A}^{\pm, \frac{3}{2}}_{\ell, n} \equiv  
\Big\langle E_{\ell,n}^{\pm, \frac{3}{2}}\Big| \psi_{\frac{3}{2}}(0)\Big\rangle,
\label{state-3/2-t}
\eeq
which, in turn, furnishes the corresponding pure state bipartite density matrix $\rho^{(s=\frac{3}{2})}(t) \equiv \ket{\psi_{\frac{3}{2}}(t)} \bra{\psi_{\frac{3}{2}}(t)}$.
The explicit evaluation of the coefficients in (\ref{state-3/2-t}) reads 
\bea
\mathcal{A}^{\pm, \frac{3}{2}}_{\ell, n}  \!\!\!\! &=& \!\!\!\! 
\frac{\mathcal{N}_{\frac{3}{2}}}{\mathcal{N}_{\ell,n}^{\pm, \frac{3}{2}} \left( 1+ |\mathfrak{z}|^{2} \right)^{\frac{3}{2}}}
\left\lgroup  \left(  \mathfrak{z}^{3} 
+ \ell(-1)^{n}    \mathrm{c} \right) \exp \left( -i\, \tfrac{3\,\widetilde{ \lambda} }{2} \, \mathrm{Im}( \alpha )\right) \mathcal{S}_{n} \left(\alpha + \tfrac{3\,\widetilde{\lambda}}{2}, \xi \right)\right. \nn \\
&+& \ell \left( 1 - \ell (-1)^{n} \mathrm{c}\, \mathfrak{z}^{3} \right)
\exp \left( i\, \tfrac{3\, \widetilde{ \lambda}}{2}  \, \mathrm{Im}( \alpha )\right) 
\mathcal{S}_{n} \left(\alpha - \tfrac{3\,\widetilde{\lambda}}{2}, \xi \right) + 
\sqrt{3}\, \Gamma_{\ell}^{(\pm)} \left(  \mathfrak{z}^{2} - \ell (-1)^{n} \mathrm{c}\,
\mathfrak{z}  \right) \times \nn \\
&\times & \exp \left(- i\, \tfrac{\, \widetilde{ \lambda}}{2}  \, \mathrm{Im}( \alpha )\right)
\mathcal{S}_{n} \left(\alpha + \tfrac{\,\widetilde{\lambda}}{2} , \xi \right) + \ell\,
\sqrt{3}\, \Gamma_{\ell}^{(\pm)} \left( 
\mathfrak{z} + \ell (-1)^{n} \mathrm{c}\, \mathfrak{z}^{2}  \right ) 
\exp \left(i\, \tfrac{\, \widetilde{ \lambda}}{2}  \, \mathrm{Im}( \alpha )\right)
\times \nn \\
& \times & \!\!\! \!\!\! \left.\mathcal{S}_{n} \left(\alpha - \tfrac{ \widetilde{\lambda}}{2}, \xi \right)\right\rgroup,\quad \ell \in \pm 1.
\label{A-3/2-pm 1}
\eea
The evolution of the state (\ref{state-3/2-t}) now readily yields the spin density matrix for the  
$s= \tfrac{3}{2}$ example:
\beq
\rho^{(\frac{3}{2})}_{_{\mathcal{Q}}}(t) \! = \!\!\! \sum_{n,\widetilde{n}=0}^{\infty} \!\!
\begin{pmatrix}
	\mathrm{B}_{n,\widetilde{n}}^{(2,2)}(t) \delta_{n\widetilde{n}}  & 
	\mathrm{B}_{n,\widetilde{n}}^{(2,1)}(t)
	\mathcal{G}_{\widetilde{n}n}(- \widetilde{\lambda}) &
	\mathrm{B}_{n,\widetilde{n}}^{(2,-1)}(t)
	\mathcal{G}_{\widetilde{n}n}(-2 \widetilde{\lambda}) &
	\mathrm{B}_{n,\widetilde{n}}^{(2,-2)}(t)
	\mathcal{G}_{\widetilde{n}n}(-3 \widetilde{\lambda}) \\
	\mathrm{B}_{n,\widetilde{n}}^{(1,2)}(t)
	\mathcal{G}_{\widetilde{n}n}( \widetilde{\lambda}) &
	\mathrm{B}_{n,\widetilde{n}}^{(1,1)}(t) \delta_{n\widetilde{n}} &
	\mathrm{B}_{n,\widetilde{n}}^{(1,-1)}(t)
	\mathcal{G}_{\widetilde{n}n}(- \widetilde{\lambda}) &
	\mathrm{B}_{n,\widetilde{n}}^{(1,-2)}(t)
	\mathcal{G}_{\widetilde{n}n}(-2 \widetilde{\lambda}) \\
	\mathrm{B}_{n,\widetilde{n}}^{(-1,2)}(t)
	\mathcal{G}_{\widetilde{n}n}(2 \widetilde{ \lambda}) &
	\mathrm{B}_{n,\widetilde{n}}^{(-1,1)}(t) \mathcal{G}_{\widetilde{n}n}( \widetilde{\lambda})  &
	\mathrm{B}_{n,\widetilde{n}}^{(-1,-1)}(t)
	\delta_{n\widetilde{n}} &
	\mathrm{B}_{n,\widetilde{n}}^{(-1,-2)}(t)
	\mathcal{G}_{\widetilde{n}n}(-\widetilde{\lambda}) \\
	\mathrm{B}_{n,\widetilde{n}}^{(-2,2)}(t)
	\mathcal{G}_{\widetilde{n}n}(3 \widetilde{ \lambda}) &
	\mathrm{B}_{n,\widetilde{n}}^{(-2,1)}(t) \mathcal{G}_{\widetilde{n}n}(2 \widetilde{\lambda})  &
	\mathrm{B}_{n,\widetilde{n}}^{(-2,-1)}(t)
	\mathcal{G}_{\widetilde{n}n}( \widetilde{\lambda})  &
	\mathrm{B}_{n,\widetilde{n}}^{(-2,-2)}(t) \delta_{n\widetilde{n}}
\end{pmatrix},
\label{DenMtrx-3/2-t}
\eeq
where the elements  are expressed via the factorized structure 
\beq
\mathrm{B}_{n,\widetilde{n}}^{(\imath,\jmath)}(t) \equiv \mathcal{B}^{(\frac{3}{2})}_{\imath,n}(t) \;\mathcal{B}^{(\frac{3}{2})}_{\jmath,\widetilde{n}}(t)^{*} , \quad \imath,\jmath \in \{ \pm 1, \pm 2 \}
\label{B-n,m}
\eeq
of the following linear combinations 
\bea
\mathcal{B}^{(\frac{3}{2})}_{\pm 2,n}(t) \!\!\! &=& \!\!\! 
\tfrac{\mathcal{A}^{ +,\frac{3}{2}}_{1, n}(t)}{\mathcal{N}_{1,n}^{+,\frac{3}{2}}}+
\tfrac{\mathcal{A}^{ -,\frac{3}{2}}_{1, n}(t)}{\mathcal{N}_{1,n}^{-,\frac{3}{2}}}\pm
\tfrac{\mathcal{A}^{ +,\frac{3}{2}}_{-1, n}(t)}{\mathcal{N}_{-1,n}^{+,\frac{3}{2}}}\pm
\tfrac{\mathcal{A}^{ -,\frac{3}{2}}_{-1, n}(t)}{\mathcal{N}_{-1,n}^{-,\frac{3}{2}}},\nn\\
\mathcal{B}^{(\frac{3}{2})}_{\pm 1,n}(t) \!\!\! &=& \!\!\! 
\tfrac{\mathcal{A}^{ +,\frac{3}{2}}_{1, n}(t) \Gamma_{1}^{(+)}}{\mathcal{N}_{1,n}^{+,\frac{3}{2}}}+
\tfrac{\mathcal{A}^{ -,\frac{3}{2}}_{1, n}(t) \Gamma_{1}^{(-)} }{\mathcal{N}_{1,n}^{-,\frac{3}{2}}}\pm
\tfrac{\mathcal{A}^{ +,\frac{3}{2}}_{-1, n}(t)  \Gamma_{-1}^{(+)}}{\mathcal{N}_{-1,n}^{+,\frac{3}{2}}}\pm
\tfrac{\mathcal{A}^{ -,\frac{3}{2}}_{-1, n}(t) \Gamma_{-1}^{(-)}}{\mathcal{N}_{-1,n}^{-,\frac{3}{2}}}
\eea
containing the time-dependent  phases 
$\mathcal{A}^{\pm, \frac{3}{2}}_{\ell, n}(t) 
\equiv \mathcal{A}^{\pm, \frac{3}{2}}_{\ell, n} \exp \left( - i E_{\ell,n}^{\pm, \frac{3}{2}} t \right),
\; \ell \in \{\pm 1\} $ that reflect the energy eigenvalues. On the other hand the reduced density matrix of the oscillator for the 
instance $s= \tfrac{3}{2}$ is obtained by implementing the partial tracing of the spin degree of freedom on  the bipartite density matrix $\rho^{(s=\frac{3}{2})}(t)$:
\bea
\rho^{(\frac{3}{2})}_{_{\mathcal{O}}}(t) \!\!\! \! &=& \!\!\! \! \sum_{n,\widetilde{n}=0}^{\infty} 
\left\lgroup  
\mathcal{B}_{2,n}^{(\frac{3}{2})}(t)\,
\mathcal{B}_{2,\widetilde{n}}^{(\frac{3}{2})}(t)^{*} \left|n_{\frac{3}{2}}\right\rangle\left\langle \widetilde{n}_{\frac{3}{2}}\right| +
\mathcal{B}_{1,n}^{(\frac{3}{2})}(t)\,
\mathcal{B}_{1,\widetilde{n}}^{(\frac{3}{2})}(t)^{*} \left|n_{\frac{1}{2}}\right\rangle \left\langle \widetilde{n}_{\frac{1}{2}}\right|
\right.\nn \\
& +& \left.
\mathcal{B}_{-1,n}^{(\frac{3}{2})}(t)\,
\mathcal{B}_{-1,\widetilde{n}}^{(\frac{3}{2})}(t)^{*} \left|n_{-\frac{1}{2}}\right\rangle 
\left\langle \widetilde{n}_{-\frac{1}{2}}\right| + \mathcal{B}_{-2,n}^{(\frac{3}{2})}(t)\,
\mathcal{B}_{-2,\widetilde{n}}^{(\frac{3}{2})}(t)^{*} \left|n_{-\frac{3}{2}}\right\rangle 
\left\langle \widetilde{n}_{-\frac{3}{2}}\right|\right\rgroup. 
\label{spin_3by2_osc_den_matrix}
\eea
The density matrices (\ref{DenMtrx-3/2-t}, \ref{spin_3by2_osc_den_matrix}) of both the subsystems obey the normalization requirement:
$\mathrm{Tr}\rho^{(\frac{3}{2})}_{_{\mathcal{Q}}}(t) =1, 
\mathrm{Tr}\rho^{(\frac{3}{2})}_{_{\mathcal{O}}}(t) = 1$.
\section{Phase space representation of the evolving hybrid system}
\label{PWQ-representation}
\setcounter{equation}{0}
To express the phase space quasiprobability densities for the spin variable the author of 
Ref. [\cite{A1981}] introduced the spherical tensor operator
\beq
T_{kq}=\sum_{m, m^{'}}\;(-1)^{s-m}\;\sqrt{2\,k\,+1}\;\left(\begin{array}{clcr}
s & k & s\\
-m & q& m'  \end{array}\right)|s\,m\rangle\langle\,s\,m'|,\quad T_{kq}^{\dag}=\;(-1)^{q}\;T_{k,-q},
\label{T_kq}
\eeq
where the indices read: $k \in (0, 1,\ldots,2 s),\, q \in (-k, -k+1, \ldots, k)$. The Wigner $3j$-coefficient appearing above
follows the standard definition [\cite{Talman1968}]:
\beq
\left(\begin{array}{clcr}
j_{1} & j_{2} & j_{3}\\
m_{1} & m_{2} & m_{3}  \end{array}\right) =\;(-1)^{j_{1}-j_{2}-m_{3}}\;(2\,j_{3}\,+1)^{-\frac{1}{2}}\;
\langle j_{1}\, m_{1}; j_{2} \,m_{2}|j_{3},-m_{3} \rangle.
\label{Wigner-3j}
\eeq
Towards constructing the phase space distributions for the hybrid bipartite system we employ the direct product of  the 
spherical tensor (\ref{T_kq}) and the unit operator acting on the oscillator Hilbert space: 
${\cal T}_{kq}= T_{kq} \otimes \mathbb{I}_{_{\mathcal{O}}}$. The bipartite density matrix $\rho^{(s)} (t)$ given in 
(\ref{DenMaHybrid}) may now be utilized \textit{\`{a} la} [\cite{A1981}]
to acquire the spherical tensor components in the compounded Hilbert space: 
\beq
\varrho_{kq}=\mathrm{Tr}_{_{\mathcal{Q}}}\;\big[\rho^{(s)}(t)\;{\mathcal T}_{kq}^{\dag}\big].
\label{DenMat-kq}
\eeq
In (\ref{DenMat-kq}) the indices referring to the oscillator variable are not explicitly notified. A partial 
tracing on the oscillator degree of freedom in (\ref{DenMat-kq}) readily furnishes  the qudit reduced density matrix in the spherical tensor basis [\cite{A1981}]:  
\beq
\varrho_{kq}^{\mathcal{Q}} \equiv \mathrm{Tr}_{_{\mathcal{O}}}\big[\varrho_{kq}\big] = 
(-1)^{q}\sum_{m, m^{\prime}}(-1)^{s-m}\,\sqrt{2k+1}\,\left(\begin{array}{clcr}
s & {\;\;s} & k\\
m & -m' & q  \end{array}\right)\, \left(\rho_{_{\mathcal{Q}}}\right)_{m^{\prime} m},\quad 
\left(\rho_{_{\mathcal{Q}}}\right)_{m^{\prime} m} = \bra{s\, m'}\rho_{_{\mathcal{Q}}}\ket{s\, m}.
\label{Den-MaQ-SphTn}
\eeq
For the $s=1$ case its structure is obtained via (\ref{density-1}, \ref{Den-MaQ-SphTn}):
\bea
\left.\varrho^{\mathcal{Q}}_{kq}(t)\right|_{s=1} 
\!\!\!\!\! &=& \!\!\!\!\! \sqrt{\tfrac{2k+1}{(2-k)!(3+k)!}} \sum_{n,\widetilde{n}=0}^{\infty} \left\lgroup \delta_{n,\widetilde{n}} \delta_{q,0} \Big( 2 \Big(  \mathcal{B}^{(1)}_{+,n}(t)\, \mathcal{B}^{(1)}_{+,\widetilde{n}}(t)^{*} + (-1)^{k}  \mathcal{B}^{(1)}_{-,n}(t)\, \mathcal{B}^{(1)}_{-,\widetilde{n}}(t)^{*} \Big)\right.  \qquad \qquad \nn \\
& - & \!\!\! \!\!\!  (k^{2}+k-2)\, \mathcal{B}^{(1)}_{0,n}(t)\, \mathcal{B}^{(1)}_{0,\widetilde{n}}(t)^{*} \Big) -\delta_{q,1} \sqrt{2\,\tfrac{(k+1)!}{(k-1)!}} \, \mathcal{G}_{\widetilde{n} n}(-\widetilde{\lambda}) \,\Big(  \mathcal{B}^{(1)}_{+,n}(t)\, \mathcal{B}^{(1)}_{0,\widetilde{n}}(t)^{*}   \nn \\ 
&-& \!\!\! \!\!\! (-1)^{k}  \mathcal{B}^{(1)}_{0,n}(t)\, \mathcal{B}^{(1)}_{-,\widetilde{n}}(t)^{*}\Big) + \delta_{q,-1}\; \sqrt{2\,\tfrac{(k+1)!}{(k-1)!}} \, 
\mathcal{G}_{\widetilde{n} n}(\widetilde{\lambda})\, \Big(  \mathcal{B}^{(1)}_{0,n}(t) \,\mathcal{B}^{(1)}_{+,\widetilde{n}}(t)^{*}  
\nn \\
&-& \!\!\! \!\!\! (-1)^{k} 
\mathcal{B}^{(1)}_{-,n}(t) \,\mathcal{B}^{(1)}_{0,\widetilde{n}}(t)^{*} \Big) +  \sqrt{\tfrac{(k+2)!}{(k-2)!}} 
\Big( \delta_{q,-2}\; \mathcal{G}_{\widetilde{n} n}(2\widetilde{\lambda})\, 
\mathcal{B}^{(1)}_{-,n}(t)\, \mathcal{B}^{(1)}_{+,\widetilde{n}}(t)^{*} \nn \\
&+&
\left.(-1)^{k} \delta_{q,2} \,\mathcal{G}_{\widetilde{n} n}(-2\widetilde{\lambda})\, 
\mathcal{B}^{(1)}_{+,n}(t)\, \mathcal{B}^{(1)}_{-,\widetilde{n}}(t)^{*} \Big)\right\rgroup.
\label{DenMat_kq_s1}
\eea
For the  $s=\tfrac{3}{2}$ example the composition  of the qudit reduced density matrix  in the spherical tensor basis is 
also assembled by employing the construction (\ref{DenMtrx-3/2-t}, \ref{Den-MaQ-SphTn}):
\bea
\left.\varrho^{\mathcal{Q}}_{kq}(t)\right|_{s= \frac{3}{2}} 
\!\!\!\!\! &=& \!\!\!\!\! \sqrt{\tfrac{2k+1}{(3-k)!(4+k)!}} \sum_{n,\widetilde{n}=0}^{\infty}\left\lgroup 
\delta_{n,\widetilde{n}} \delta_{q,0} \Big(6 \left(  \mathrm{B}_{n,\widetilde{n}}^{(2,2)}(t) +  (-1)^{k}  \mathrm{B}_{n,\widetilde{n}}^{(-2,-2)}(t) \right)
- 2 (k^{2}+k-3)\right. \times \quad \quad \quad \nn \\
 & \times & \!\! \!\! \Big( \mathrm{B}_{n,\widetilde{n}}^{(1,1)}(t) + (-1)^{k}\, \mathrm{B}_{n,\widetilde{n}}^{(-1,-1)}(t) \Big) \Big) + 
 \delta_{q,-1} \sqrt{\tfrac{(k+1)!}{(k-1)!}}\, \mathcal{G}_{\widetilde{n}n}( \widetilde{\lambda})\, \Big( 2 \sqrt{3} \,\mathrm{B}_{n,\widetilde{n}}^{(1,2)}(t) \nn \\
 &  - & \!\! \!\! (k^{2}+k-6) \mathrm{B}_{n,\widetilde{n}}^{(-1,1)}(t) - (-1)^{k}\, 2 \sqrt{3} \, \mathrm{B}_{n,\widetilde{n}}^{(-2,-1)}(t)\Big) + \delta_{q,1} \sqrt{\tfrac{(k+1)!}{(k-1)!}}\, \mathcal{G}_{\widetilde{n}n}( -\widetilde{\lambda})\, \times \nn \\
 & \times & \!\! \!\! \Big(-2 \sqrt{3}\, \mathrm{B}_{n,\widetilde{n}}^{(2,1)}(t) - (-1)^{k}\,(k^{2}+k-6)\, 
 \mathrm{B}_{n,\widetilde{n}}^{(1,-1)}(t) + (-1)^{k}\, 2 \sqrt{3}\, \mathrm{B}_{n,\widetilde{n}}^{(-1,-2)}(t) \Big) \nn \\
 &  + & \!\! \!\! \sqrt{3 \tfrac{(k+2)!}{(k-2)!}}   \Big( \delta_{q,-2}\, \mathcal{G}_{\widetilde{n}n}( 2 \widetilde{\lambda}) \Big( \mathrm{B}_{n,\widetilde{n}}^{(-1,2)}(t) 
  + (-1)^{k} \, \mathrm{B}_{n,\widetilde{n}}^{(-2,1)}(t) \Big) \nn \\
&  +& \!\! \!\! \delta_{q,2} \, \mathcal{G}_{\widetilde{n}n}( -2 \widetilde{\lambda}) \Big( \mathrm{B}_{n,\widetilde{n}}^{(2,-1)}(t) 
+ (-1)^{k} \, \mathrm{B}_{n,\widetilde{n}}^{(1,-2)}(t) \Big) \Big) + \sqrt{\tfrac{(k+3)!}{(k-3)!}} \times  \nn \\
& \times & \left.\!\!  \!\! \Big( (-1)^{k}\, \delta_{q,3} \,\mathcal{G}_{\widetilde{n}n}( -3 \widetilde{\lambda}) 
\mathrm{B}_{n,\widetilde{n}}^{(2,-2)}(t) 
 +   \delta_{q,-3} \, \mathcal{G}_{\widetilde{n}n}( 3 \widetilde{\lambda})\, 
\mathrm{B}_{n,\widetilde{n}}^{(-2,2)}(t) \Big) \right\rgroup.
\label{DenMat_kq_s3/2}
\eea

\par

Adapting the formulation in Ref. [\cite{A1981}] for the spin variable and the well known description of the oscillator degree of freedom [\cite{Gerry2005}] we now propose the phase space quasiprobability distributions of the bipartite system via the  decomposition (\ref{DenMat-kq}) of the composite density matrix. In particular, the diagonal 
$\mathrm{P}$-representation of the qudit-oscillator interacting system may be constructed as
\beq
\mathrm{P}(\theta,\phi;\beta,\beta^{*})=\frac{\exp(|\beta|^{2})}{\pi^{2}}\sum_{kq}\;(-1)^{k-q}\;c_{kq}\;
\left\lgroup\int \braket{-\gamma|\,\varrho_{kq}\,|\gamma} \exp(|\gamma|^{2})
\exp(\beta\,\gamma^{*}-\beta^{*}\,\gamma)\;\mathrm{d}^{2}\gamma\right\rgroup\,Y_{kq}(\theta,\phi),
\label{P-bipartite}
\eeq
where the spin coefficient reads [\cite{A1981}]: $c_{kq}=\tfrac{\sqrt{(2\,s-k)!\,(2\,s+k+1)!}}
{\sqrt{4\,\pi}\,(2\,s)!}$. For the sake of completeness we now list the necessary properties of the spherical harmonics appearing in (\ref{P-bipartite}). The spherical functions are expressed [\cite{Talman1968}] via the Legendre polynomials:
\beq
Y_{\ell m}(\theta,\phi)=  \sqrt{\tfrac{2 \ell +1}{4 \pi}} \sqrt{\tfrac{(\ell-m)!}{(\ell + m)!}}\, \exp(i m \phi) \,P_{\ell}^{m}(\cos \theta), \quad P_{\ell}^{m}(x)=(-1)^{m} (1-x^{2})^{\frac{m}{2}} \tfrac{\mathrm{d}^{m}}{\mathrm{d}x^{m}} P_{\ell}(x)\\
\label{Y-lm}
\eeq
that, in turn, may be considered as special cases of the Jacobi polynomials [\cite{AAR1999}]:
\beq
P_{\ell}(x) \equiv  P_{\ell}^{0,0}(x), \qquad P_{\ell}^{\mathsf{a},\mathsf{b}}(x) = \sum_{k} \tbinom{\ell+\mathsf{a}}{\ell-k} \tbinom{\ell+\mathsf{b}}{k} \left(  \tfrac{x-1}{2}\right)^{k} 
\left(  \tfrac{x+1}{2}\right)^{\ell-k}. 
\label{P_alpah, beta}
\eeq
These functions follow the usual orthogonality relation [\cite{Talman1968}]:
$ \int_{0}^{\pi} \int_{0}^{2\pi} Y_{\ell m}(\theta,\phi) Y_{\ell^{'} m^{'}}(\theta,\phi)^{*}  \; \mathrm{d}\Omega =  \delta_{\ell \ell^{'}} \delta_{m m^{'}},\;\mathrm{d}\Omega = \sin \theta \;\mathrm{d}\theta \, \mathrm{d}\phi$.

\par

The diagonal $\mathrm{P}_{\cal{Q}}$-representation for the spin variable [\cite{A1981}] may be procured by integrating the bipartite quasiprobability density (\ref{P-bipartite}) on the oscillator phase space:
\beq
 \mathrm{P}_{\cal{Q}}(\theta,\phi) \equiv \int \mathrm{P}(\theta,\phi;\beta,\beta^{*})\, \mathrm{d}^{2}\beta \;\; \Rightarrow \;\;
\mathrm{P}_{\cal{Q}}(\theta,\phi) = \sum_{kq} (-1)^{k-q}\;c_{kq}\; \varrho_{kq}^{\cal{Q}}\;Y_{kq}(\theta,\phi), \quad \int \mathrm{P}_{\cal{Q}}(\theta,\phi) \mathrm{d}\Omega  =1.
\label{P-spin}
\eeq
Modulo our approximation, the equation (\ref{P-spin}) admits explicit evaluation of the qudit  $\mathrm{P}_{\cal{Q}}(\theta,\phi)$-representation, say for the $s=1$ and $s=\tfrac{3}{2}$  cases, via the substitution  of the corresponding density matrices in the spherical tensor basis  given in (\ref{DenMat_kq_s1}) and (\ref{DenMat_kq_s3/2}), respectively. Unlike its oscillator counterpart the spin $\mathrm{P}_{\cal{Q}}$-representation is 
nonsingular, and, therefore, may be fruitfully applied to observe the phase space structures such as the transient spin kitten states.
We follow this route in our characterization of the spin kitten states in Sec. \ref{nonclassicality}.

\par

The construction of the hybrid bipartite Wigner $\mathrm{W}$-distribution in the product phase space may be similarly 
established. Maintaining the compositions of the individual phase space Wigner functions for the spin and oscillator 
variables, we present the $\mathrm{W}$-distribution for the interacting system as 
\beq
\mathrm{W}(\theta,\phi;\beta,\beta^{*})=\tfrac{1}{\pi^{2}} \sqrt{\tfrac{2s+1}{4 \pi}}\sum_{kq} \;
\left\lgroup \int \mathrm{Tr}_{\cal{O}}\big[\varrho_{kq} \mathrm{D}(\gamma) \big]
\exp(\beta\,\gamma^{*}-\beta^{*}\,\gamma)\;\mathrm{d}^{2}\gamma \right\rgroup\,Y_{kq}(\theta,\phi). 
\label{W-spin-osc}
\eeq
The above Wigner distribution (\ref{W-spin-osc}) may be recast \textit{\`{a} la} [\cite{MK1993}] as an infinite alternating 
series sum of the diagonal matrix elements of the density operator in the displaced oscillator number state basis: 
\beq
\mathrm{W}(\theta,\phi;\beta,\beta^{*})= \tfrac{2}{\pi} \sqrt{\tfrac{2s+1}{4 \pi}}
\sum_{kq} \sum_{n=0}^{\infty} (-1)^{n} \braket{\beta, n |\varrho_{kq}| \beta, n}
Y_{kq}(\theta,\phi), \quad \ket{\beta, n} = \mathrm{D}(\beta) \ket{n}.
\label{W-series-spin-osc}
\eeq
An integration of the bipartite quasiprobability function (\ref{W-spin-osc}) over the oscillator phase space generate the Wigner 
distribution for the spin degree of freedom [\cite{DAS1994}]:
\beq
\mathrm{W}_{\cal{Q}}(\theta,\phi) \equiv \int \mathrm{W}(\theta,\phi;\beta,\beta^{*})\, \mathrm{d}^{2}\beta \;\;\Rightarrow\;\;
\mathrm{W}_{\cal{Q}}(\theta,\phi) =  \sqrt{\tfrac{2s+1}{4 \pi}} \sum_{kq} \varrho_{kq}^{\cal{Q}}\;Y_{kq}(\theta,\phi), \quad \int \mathrm{W}_{\cal{Q}}(\theta,\phi) \; \mathrm{d}\Omega =1.
\label{W-spin}
\eeq

\par

Continuing our description of the bipartite quasiprobability functions on the joint phase space we now constitute
the positive semidefinite Husimi $\mathrm{Q}$-function for the combined spin-oscillator system via the spherical tensor decomposition 
 (\ref{DenMat-kq}) of the compounded density matrix:
\beq
\mathrm{Q}(\theta,\phi; \beta, \beta^{*})=\tfrac{2\,s+1}{4\,\pi^{2}}\,\sum_{kq} (-1)^{k-q}\; (c_{kq})^{-1}\; 
\bra\beta\varrho_{kq}\ket\beta\;Y_{kq}(\theta,\phi).
\label{Q-bipartite}
\eeq
The $Q$-function for the spin degree of freedom [\cite{A1981}] is recovered from the bipartite construction 
(\ref{Q-bipartite}) by an integration over the oscillator phase space:
\beq
\mathrm{Q}_{_{\mathcal{Q}}}(\theta,\phi) \equiv \!\! \int \! \mathrm{Q}(\theta,\phi;\beta,\beta^{*})\, \mathrm{d}^{2}\beta  \Rightarrow  \mathrm{Q}_{_{\mathcal{Q}}}(\theta,\phi)= \tfrac{2\,s+1}{4\,\pi}\!\!\sum_{kq} (-1)^{k-q}\; (c_{kq})^{-1}\! 
\varrho_{kq}^{\cal{Q}}\;Y_{kq}(\theta,\phi), \; \int \!\mathrm{Q}_{\cal{Q}}(\theta,\phi)  \mathrm{d}\Omega  =1.
\label{Q-spin}
\eeq

\par

For the sake of completeness we briefly summarize the recipe for the construction of the oscillator quasiprobability distributions  starting from the bipartite phase space densities. Integration 
of the bipartite distributions over the qudit spherical phase space leads to the corresponding oscillator quasiprobabilities [\cite{Gerry2005}] listed below:
\bea
\int \mathrm{P}(\theta,\phi;\beta,\beta^{*}) \mathrm{d}\Omega \equiv 
\mathrm{P}_{\mathcal{O}}(\beta,\beta^{*}) = 
\frac{\exp(|\beta|^{2})}{\pi^{2}}
\int \braket{-\gamma|\,\rho_{\cal{O}}\,|\gamma} \exp(|\gamma|^{2})
\exp(\beta\,\gamma^{*}-\beta^{*}\,\gamma)\;\mathrm{d}^{2}\gamma,
\label{P-oscillator}
\eea
\bea
\int \mathrm{W}(\theta,\phi;\beta,\beta^{*}) \mathrm{d}\Omega \equiv 
\mathrm{W}_{\cal{O}}(\beta,\beta^{*}) = 
\frac{1}{\pi^{2}}
\int \mathrm{Tr}_{\cal{O}}\big[\rho_{\cal{O}} \mathrm{D}(\gamma) \big]
\exp(\beta\,\gamma^{*}-\beta^{*}\,\gamma)\;\mathrm{d}^{2}\gamma,
\label{W-oscillator}
\eea
\bea
\int \mathrm{Q}(\theta,\phi;\beta,\beta^{*}) \mathrm{d}\Omega \equiv 
\mathrm{Q}_{\cal{O}}(\beta,\beta^{*}) = 
\frac{1}{\pi} \braket{\beta |\rho_{\cal{O}} |\beta}.
\label{Q-oscillator}
\eea
The above oscillator quasiprobability distributions obey [\cite{Gerry2005}] the normalization restriction.  
\subsection{Explicit structures of the hybrid phase space distributions for the $s=1$ case}
The bipartite $\mathrm{P}$-representation (\ref{P-bipartite}) may be explicitly determined by employing the composite 
density matrix elements (\ref{DenMat-kq}). The $s=1$ example reads:  
\bea
\mathrm{P}^{(1)}(\theta,\phi; \beta, \beta^{*}) \!\!\!\! &=& \!\!\!\! \tfrac{1}{16 \pi} \sum_{n,\widetilde{n}=0}^{\infty}  \left \{ 
3(3-4\cos \theta + 5 \cos 2\theta) 
\mathcal{B}^{(1)}_{+,n}(t) \mathcal{B}^{(1)}_{+,\widetilde{n}}(t)^{*} \Lambda^{n,\widetilde{n}}_{1}
-6 (1+ 5 \cos 2\theta) \times \qquad \qquad \right.  \nn \\
&\times & \!\!\!\! 
\; \mathcal{B}^{(1)}_{0,n}(t) \mathcal{B}^{(1)}_{0,\widetilde{n}}(t)^{*} \Lambda^{n,\widetilde{n}}_{0}
+ 3(3+4\cos \theta + 5 \cos 2\theta)  \; \mathcal{B}^{(1)}_{-,n}(t) \mathcal{B}^{(1)}_{-,\widetilde{n}}(t)^{*} \Lambda^{n,\widetilde{n}}_{-1}  \nn \\
&+& \!\!\!\! \left. \sum_{k=0}^{\infty} \left[ 12 \sqrt{2} \sin \theta (1-5 \cos \theta) \, \mathrm{Re} \left \lgroup \exp \left(i \phi \right) 
\mathcal{B}^{(1)}_{+,n}(t) \mathcal{B}^{(1)}_{0,\widetilde{n}}(t)^{*}\,\mathcal{G}_{kn}{(-\widetilde{\lambda})} \Lambda^{k,\widetilde{n}}_{0}  \right \rgroup \right.  \right. \nn \\
&+& \!\!\!\! \left.
12 \sqrt{2} \sin \theta (1+5 \cos \theta)\, \mathrm{Re} \left \lgroup  \exp \left(-i \phi \right) 
\; \mathcal{B}^{(1)}_{-,n}(t) \mathcal{B}^{(1)}_{0,\widetilde{n}}(t)^{*}\,
\mathcal{G}_{kn}{(\widetilde{\lambda})} \Lambda^{k,\widetilde{n}}_{0}  \right \rgroup \right] \nn \\
&+& \!\!\!\!  \sum_{k,\ell=0}^{\infty}  60 \sin^{2} \theta \;
\mathrm{Re} \left \lgroup \exp \left( 2 i \phi \right) 
\mathcal{B}^{(1)}_{+,n}(t) \mathcal{B}^{(1)}_{-,\widetilde{n}}(t)^{*}
\mathcal{G}_{kn}{(-\widetilde{\lambda})} \;
\mathcal{G}_{\widetilde{n}\ell}{(-\widetilde{\lambda}) \Lambda ^{k,\ell}_{0} } \right \rgroup \Big \}, 
\label{P-bipartite-1}
\eea
where the weighted distribution is structured as 
$\Lambda^{n,\widetilde{n}}_{m} = \tfrac{1}{\sqrt{n!\widetilde{n}!}} \exp(|\beta_{m}|^{2}) 
\left( -\tfrac{\partial}{\partial \beta_{m}} \right)^{n}
\left( -\tfrac{\partial}{\partial \beta_{m}^{*}} \right)^{\widetilde{n}}
\delta^{(2)}(\beta_{m})$, and  the spin dependent displaced coordinate  is given by $\beta_{m}=\beta + m \widetilde{\lambda}$.
 The oscillator phase
space integral of the composite quasiprobability density (\ref{P-bipartite-1}) provides, \textit{\`{a} la} 
(\ref{P-spin}), the qudit $s=1$ diagonal $\mathrm{P}_{\mathcal{Q}}$-representation:
\bea
\mathrm{P}^{(1)}_{\mathcal{Q}}(\theta,\phi) \!\!\!\!\! &=& \!\!\!\!\! \tfrac{1}{16 \pi } \!\! \sum_{n,\widetilde{n}=0}^{\infty} 
\!\!  \left \{ \! 3\,(3-4\cos \theta + 5 \cos 2\theta) 
\mathcal{B}^{(1)}_{+,n}(t) \mathcal{B}^{(1)}_{+,\widetilde{n}}(t)^{*}\, \delta_{n,\widetilde{n}}-6 \,(1+ 5 \cos 2\theta) \;
\mathcal{B}^{(1)}_{0,n}(t) \mathcal{B}^{(1)}_{0,\widetilde{n}}(t)^{*}\, \delta_{n,\widetilde{n}} \right.  \nn  \\
& + & \!\!\!\! 
3\,(3+4\cos \theta + 5 \cos 2\theta) 
\mathcal{B}^{(1)}_{-,n}(t) \mathcal{B}^{(1)}_{-,\widetilde{n}}(t)^{*} \, \delta_{n,\widetilde{n}} 
+ 12 \,\sqrt{2} \sin \theta \left(1-5 \cos \theta \right)  \mathrm{Re}\lg \exp \left(i  \phi  \right) \right. \times \nn \\
& \times & \!\!\!\! \left. \mathcal{B}^{(1)}_{+,n}(t) \mathcal{B}^{(1)}_{0,\widetilde{n}}(t)^{*} \mathcal{G}_{\widetilde{n} n}(-\widetilde{\lambda}) \right \rgroup 
+12 \sqrt{2} \sin \theta \left(1+5 \cos \theta \right) 
\mathrm{Re} \!\!  \left \lgroup \exp \left(-i  \phi  \right) 
\mathcal{B}^{(1)}_{-,n}(t) \mathcal{B}^{(1)}_{0,\widetilde{n}}(t)^{*} 
\mathcal{G}_{\widetilde{n} n}(\widetilde{\lambda}) \right \rgroup  \nn \\ 
& + & \!\!\!\! \left.
60 \,\sin^{2}\theta \; \mathrm{Re} \!\! \lg \exp \left(2i  \phi  \right) \mathcal{B}^{(1)}_{+,n}(t) 
\mathcal{B}^{(1)}_{-,\widetilde{n}}(t)^{*} \mathcal{G}_{\widetilde{n} n}(-2\widetilde{\lambda}) \rg  \right \}.
\label{P_spin_1}
\eea
Alternately (\ref{P_spin_1}) may be directly computed via the corresponding spin density matrix  (\ref{DenMat_kq_s1}) in 
the spherical tensor basis  and the construction [\cite{A1981}] of the $\mathrm{P}_{\cal{Q}}(\theta,\phi)$-representation 
appearing at the first equality in (\ref{P-spin}). This provides a consistency check on the structure of the hybrid 
$\mathrm{P}$-representation (\ref{P-bipartite}).  Similarly, the composition (\ref{W-series-spin-osc}) of the bipartite Wigner 
$\mathrm{W}$-distribution provides its explicit evaluation for the $s=1$ case: 
\bea
\mathrm{W}^{(1)}(\theta,\phi; \beta, \beta^{*}) \!\!\!\! &=& \!\!\!\! \tfrac{1}{16 \pi^{2} } \sum_{n,\widetilde{n}=0}^{\infty}
 \left \{\left( 8+ \sqrt{10} + 12 \sqrt{2} \cos \theta + 3 \sqrt{10} \cos 2\theta \right) \mathcal{H}^{n,\widetilde{n}}_{1,1}(\beta , \beta^{*}) 
\mathcal{B}^{(1)}_{+,n}(t) \mathcal{B}^{(1)}_{+,\widetilde{n}}(t)^{*}
 \right.   \nn \quad \quad \\
&+ & \!\!\!\! \left( 8+4 \sqrt{10} -12 \sqrt{10} \cos^{2}\theta \right)
 \mathcal{H}^{n,\widetilde{n}}_{0,0}(\beta , \beta^{*}) \mathcal{B}^{(1)}_{0,n}(t) \mathcal{B}^{(1)}_{0,\widetilde{n}}(t)^{*} +
 \left( 8+ \sqrt{10} - 12 \sqrt{2} \cos \theta \right.
\nn \\
& + & \!\!\!\! \left.
   3 \sqrt{10} \cos 2\theta \right) 
\mathcal{H}^{n,\widetilde{n}}_{-1,-1}(\beta , \beta^{*}) 
 \mathcal{B}^{(1)}_{-,n}(t) \mathcal{B}^{(1)}_{-,\widetilde{n}}(t)^{*}
 +24(1+\sqrt{5} \cos \theta) \sin \theta \times
   \nn \\
& \times & \!\!\!\!\!   \mathrm{Re}  \left \lgroup \!
\mathcal{H}^{n,\widetilde{n}}_{1,0}(\beta , \beta^{*}) \exp \left(  i\left( \phi + \widetilde{\lambda} \mathrm{Im}(\beta)\right)\right)
\mathcal{B}^{(1)}_{+,n}(t) \mathcal{B}^{(1)}_{0,\widetilde{n}}(t)^{*} \! \right \rgroup \!\!
+  24(1-\sqrt{5} \cos \theta) \sin \theta  \times\nn \\
& \times &  \!\!\!\!\!\! \,
\mathrm{Re}  \left \lgroup \! \mathcal{H}^{n,\widetilde{n}}_{-1,0}(\beta , \beta^{*})
  \exp \left( i\left( \phi + \widetilde{\lambda} \mathrm{Im}(\beta)\right)\right) \mathcal{B}^{(1)}_{-,n}(t) \mathcal{B}^{(1)}_{0,\widetilde{n}}(t)^{*} \!\! \right \rgroup +  12 \sqrt{10} \; \sin ^{2}\theta \times  \nn \\
& \times & \!\!\!\!\!\! \left. 
\mathrm{Re} \lg \!  \mathcal{H}^{n,\widetilde{n}}_{1,-1}(\beta , \beta^{*}) 
\exp \left( 2i\left( \phi + \widetilde{\lambda} \mathrm{Im}(\beta)\right)\right)
\mathcal{B}^{(1)}_{+,n}(t) \mathcal{B}^{(1)}_{-,\widetilde{n}}(t)^{*} \rg \!\!\right \},
\label{W_spin_1_bipartite}
\eea
where the complex Gaussian structure stands as
\beq
\mathcal{H}^{n,\widetilde{n}}_{k,\ell}(\beta , \beta^{*}) 
= \tfrac{1}{\sqrt{n!\widetilde{n}!}} 
\left(\beta^{*}_{k} + \beta^{*}_{\ell} \right)^{n} 
\left(\beta_{k} + \beta_{\ell} \right)^{\widetilde{n}}
\exp \left( - \tfrac{1}{2} \left( |\beta_{k}|^{2} + |\beta_{\ell}|^{2}  \right)- 
\beta_{k} \beta^{*}_{\ell} \right) 
{_2}F_{0}\left( -n,-\widetilde{n} ; \phantom{}_{-} ; -\tfrac{1}{|\beta_{k} + \beta_{\ell}|^{2}}\right)\nn\\
\eeq
and the  hypergeometric sum  is given by [\cite{AAR1999}]
$  {}_2F_0(\mathsf{x},\mathsf{y};\phantom{}_{-}; \tau)
	= \sum_{k=0}^{\infty} (\mathsf {x})_{k} (\mathsf{y})_{k}\, \frac{\tau^{k}}{k!},\,
	(\mathsf {x})_{k} = \prod_{\ell = 0}^{k - 1}(\mathsf {x} + \ell) $.
Choosing  negative integers as numerator coefficients the function  ${}_2F_0$ may be  expressed [\cite{M1939}] via the Charlier polynomial: $\mathrm{c}_{k} (\ell; \tau) = {}_{2}F_0\left(-k,-\ell;\phantom{}_{-} ;
-\frac{1}{\tau}\right)\,\forall \tau > 0$. 
To derive the bipartite $\mathrm{W}$-distribution (\ref{W_spin_1_bipartite}) we utilize an identity [\cite{M1939}]  that readily follows  from the bilinear generating function of the Charlier polynomials:
\bea
\sum_{k=0}^{\infty} \tfrac{(-1)^{k} \tau^{k}}{k!}\, {}_2F_0\Big( -n,-k;\phantom{}_{-} ; -\tfrac{1}{\mathsf{x}}\Big) 
\,{}_2F_0\Big( -k,-m;\phantom{}_{-} ; -\tfrac{1}{\mathsf{y}}\Big) 
&=& 
 \left( 1+ \tfrac{\tau}{\mathsf{x}}\right)^{n} \left( 1+ \tfrac{\tau}{\mathsf{y}}\right)^{m} \exp(-\tau) \times \nn \\
 & & \times {}_2F_0\Big( -n,-m;\phantom{}_{-} ; -\tfrac{\tau}{(\mathsf{x}+\tau)(\mathsf{y}+\tau)}\Big).
\label{IdentityHyper}
\eea
  A reduction of the phase space via the integration over the complex plane given in (\ref{W-spin}) now procures the qudit 
  ${\mathrm{W}}_{\mathcal{Q}}$-distribution concretely for the $s=1$ case:
\bea
\mathrm{W}^{(1)}_{\mathcal{Q}}(\theta,\phi) \!\!\!\! &=& \!\!\!\! \tfrac{1}{32 \pi} \!\! \sum_{n,\widetilde{n}=0}^{\infty}
\left \{ \left( 8+ \sqrt{10} + 12 \sqrt{2} \cos \theta + 3 \sqrt{10} \cos 2\theta \right) \mathcal{B}^{(1)}_{+,n}(t) \mathcal{B}^{(1)}_{+,\widetilde{n}}(t)^{*} \; \delta_{n,\widetilde{n}} +
  \left( 8+4 \sqrt{10}
 \right. \right. \nn \qquad \qquad \qquad \\
& - & \!\!\!\! \left. 
12 \sqrt{10} \cos^{2}\theta \right) \mathcal{B}^{(1)}_{0,n}(t) \mathcal{B}^{(1)}_{0,\widetilde{n}}(t)^{*} \; \delta_{n,\widetilde{n}} 
 + 
 \left( 8+ \sqrt{10} - 12 \sqrt{2} \cos \theta + 3 \sqrt{10} \cos 2\theta \right)  \times \nn
 \\  
& \times &  \!\!\!\! 
\mathcal{B}^{(1)}_{-,n}(t) \mathcal{B}^{(1)}_{-,\widetilde{n}}(t)^{*} \; \delta_{n,\widetilde{n}} 
+ 24(1+\sqrt{5} \cos \theta) \sin \theta \;
\mathrm{Re} \lg \exp(i \phi) \;
\mathcal{B}^{(1)}_{+,n}(t) \mathcal{B}^{(1)}_{0,\widetilde{n}}(t)^{*} \mathcal{G}_{\widetilde{n} n}(-\widetilde{\lambda}) \rg
\nn \\
&+&  \!\!\!\! 24(1-\sqrt{5} \cos \theta) \sin \theta \; \mathrm{Re} \left \lgroup \exp(-i \phi) \;
\mathcal{B}^{(1)}_{-,n}(t) \mathcal{B}^{(1)}_{0,\widetilde{n}}(t)^{*} \mathcal{G}_{\widetilde{n} n}(\widetilde{\lambda}) \right \rgroup + 12 \sqrt{10} \; \sin ^{2}\theta  \nn \\
& \times & \!\!\!\! \left.
 \mathrm{Re} \lg \exp(2i\phi) \;
 \mathcal{B}^{(1)}_{+,n}(t) \mathcal{B}^{(1)}_{-,\widetilde{n}}(t)^{*} \mathcal{G}_{\widetilde{n} n}(-2\widetilde{\lambda}) \rg
 \right \}.
\label{W_spin_1_function}
\eea
The above expression  may also be directly obtained by applying our evaluation of the spin density matrix 
(\ref{DenMat_kq_s1}) in a spherical basis, and utilizing  the construction [\cite{A1981}] of the 
$\mathrm{P}_{\cal{Q}}(\theta,\phi)$-representation realized  in (\ref{W-spin}). This implements a consistency check on the validity of the hybrid $\mathrm{W}$-representation (\ref{W-spin-osc}) advanced here. We also proceed with a similar 
demonstration of the bipartite $\mathrm{Q}$-function given in (\ref{Q-bipartite}). The compounded $\mathrm{Q}$-function for the 
qudit-oscillator system may be composed by employing the corresponding hybrid density matrix (\ref{DenMat-kq}). The $s=1$ example is quoted below:  
\bea
\mathrm{Q}^{(1)}(\theta,\phi; \beta, \beta^{*}) \!\!\!\! &=& \!\!\!\! \tfrac{3}{4 \pi^{2} } \sum_{n,\widetilde{n}=0}^{\infty}
\left \{ \sin ^{4}\left( \tfrac{\theta}{2}\right) \mathcal{Y}^{n,\widetilde{n}}_{1,1}(\beta,\beta^{*})
\mathcal{B}^{(1)}_{+,n}(t) \mathcal{B}^{(1)}_{+,\widetilde{n}}(t)^{*} 
+ \tfrac{\sin ^{2}\theta}{2}  \mathcal{Y}^{n,\widetilde{n}}_{0,0}(\beta,\beta^{*}) 
\mathcal{B}^{(1)}_{0,n}(t) \mathcal{B}^{(1)}_{0,\widetilde{n}}(t)^{*} \right.  \nn  \\
& +& \!\!\!\! 
\cos ^{4}\left( \tfrac{\theta}{2}\right)  \mathcal{Y}^{n,\widetilde{n}}_{-1,-1}(\beta,\beta^{*}) 
\mathcal{B}^{(1)}_{-,n}(t) \mathcal{B}^{(1)}_{-,\widetilde{n}}(t)^{*}
+ \sqrt{2} \sin \theta \sin ^{2}\left( \tfrac{\theta}{2}\right)  
\mathrm{Re}\left \lgroup 
\mathcal{Y}^{n,\widetilde{n}}_{1,0}(\beta,\beta^{*}) 
 \right. \!\! \times \nn \\
&\times & \!\!\!\! \left. \exp \left(i \left( \phi + \widetilde{\lambda} 
\mathrm{Im}(\beta) \right) \right)
 \mathcal{B}^{(1)}_{+,n}(t) \mathcal{B}^{(1)}_{0,\widetilde{n}}(t)^{*}  \right \rgroup
+  \sqrt{2} \sin \theta \cos ^{2}\left( \tfrac{\theta}{2}\right) 
 \mathrm{Re}\left \lgroup \mathcal{Y}^{n,\widetilde{n}}_{-1,0}(\beta,\beta^{*})
  \right. \times \nn \\
&\times & \!\!\!\! \left.  
\exp \left(-i \left( \phi + \widetilde{\lambda} \mathrm{Im}(\beta) \right) \right)
 \mathcal{B}^{(1)}_{-,n}(t) \mathcal{B}^{(1)}_{0,\widetilde{n}}(t)^{*} \right \rgroup + 
\tfrac{\sin^{2}\theta}{2} \,  \mathrm{Re}\left \lgroup 
\mathcal{Y}^{n,\widetilde{n}}_{1,-1}(\beta,\beta^{*}) \right. \times \nn \\
&\times & \!\!\!\! \left. \left.
\exp \left(2i \left( \phi + \widetilde{\lambda} \mathrm{Im}(\beta) \right) \right) 
 \mathcal{B}^{(1)}_{+,n}(t) \mathcal{B}^{(1)}_{-,\widetilde{n}}(t)^{*}
  \right \rgroup \right \}.  \quad \quad
\label{Q_spin_1_bipartite}
\eea
In the expression (\ref{Q_spin_1_bipartite}) we have employed the  notation
$\mathcal{Y}^{n,\widetilde{n}}_{k,\ell}(\beta,\beta^{*}) =
 \tfrac{1}{\sqrt{n!\widetilde{n}!}} \beta^{*n}_{k} \beta^{\widetilde{n}}_{\ell}
 \exp\left(-\tfrac{1}{2} (|\beta_{k}|^{2} +|\beta_{\ell}|^{2})\right)$.
A further integration on the oscillator variables \textit{\`{a} la} (\ref{Q-spin}) now generates the $s=1$ qudit $\mathrm{Q}^{(1)}_{\mathcal{Q}}$-function that may alternately be established starting from  the spin density matrix  (\ref{DenMat_kq_s1}) in the spherical basis set and implementing the construction given in [\cite{A1981}]: 
\bea
\mathrm{Q}^{(1)}_{\mathcal{Q}}(\theta,\phi) \!\!\!\! &=& \!\!\!\! \tfrac{3}{4 \pi } \sum_{n,\widetilde{n}=0}^{\infty} 
\left \{ \sin ^{4}\left( \tfrac{\theta}{2}\right)  
\mathcal{B}^{(1)}_{+,n}(t) \mathcal{B}^{(1)}_{+,\widetilde{n}}(t)^{*} \delta_{n,\widetilde{n}} + \tfrac{1}{2} \sin ^{2}\theta \;
\mathcal{B}^{(1)}_{0,n}(t) \mathcal{B}^{(1)}_{0,\widetilde{n}}(t)^{*} \delta_{n,\widetilde{n}} \right. \nn  \\
& + & \!\!\!\! 
\cos ^{4}\left( \tfrac{\theta}{2}\right)
\mathcal{B}^{(1)}_{-,n}(t) \mathcal{B}^{(1)}_{-,\widetilde{n}}(t)^{*}  \delta_{n,\widetilde{n}} 
+  \sqrt{2} \sin \theta \sin ^{2}\left( \tfrac{\theta}{2}\right) \mathrm{Re}\left \lgroup \exp \left(i  \phi  \right) \mathcal{B}^{(1)}_{+,n}(t) \mathcal{B}^{(1)}_{0,\widetilde{n}}(t)^{*} \mathcal{G}_{\widetilde{n} n}(-\widetilde{\lambda}) \right \rgroup  \nn \\
&+ & \!\!\!\! 
  \sqrt{2} \sin \theta \cos ^{2}\left( \tfrac{\theta}{2}\right) 
\mathrm{Re}\left \lgroup \exp \left(-i  \phi  \right) \mathcal{B}^{(1)}_{-,n}(t) \mathcal{B}^{(1)}_{0,\widetilde{n}}(t)^{*} \mathcal{G}_{\widetilde{n} n}(\widetilde{\lambda}) \right \rgroup + 
\tfrac{\sin^{2}\theta}{2}  \times  \nn \\
&\times & \!\!\!\! \left. \mathrm{Re}\left \lgroup \exp \left(2i  \phi  \right) \mathcal{B}^{(1)}_{+,n}(t) \mathcal{B}^{(1)}_{-,\widetilde{n}}(t)^{*} \mathcal{G}_{\widetilde{n} n}(-2\widetilde{\lambda}) \right \rgroup \right \}.  \quad \quad
\label{Q_spin_1}
\eea
The preceding discussion allows us to view the hybrid bipartite phase space quasiprobability distributions as the 
underlying structure that 
produces the appropriate qudit phase space functions following a dimensional reduction. 

\par

The integral representations specified in (\ref{P-oscillator}-\ref{Q-oscillator}) now allow us to extract the oscillator quasiprobability 
distributions for the spin $s=1$ case:
\bea
\mathrm{P}^{(1)}_{\cal{O}}(\beta,\beta^*)&=&\sum_{n,\widetilde{n}=0}^{\infty}
\lg\mathcal{B}^{(1)}_{+,n}(t)\;\mathcal{B}^{(1)}_{+,\widetilde{n}}(t)^*\,\Lambda^{n,\widetilde{n}}_{1}
+\mathcal{B}^{(1)}_{0,n}(t)\;\mathcal{B}^{(1)}_{0,\widetilde{n}}(t)^*\,\Lambda^{n,\widetilde{n}}_{0}
+\mathcal{B}^{(1)}_{-,n}(t)\;\mathcal{B}^{(1)}_{-,\widetilde{n}}(t)^*\,\Lambda^{n,\widetilde{n}}_{-1}\rg,\nn\\
\mathrm{W}^{(1)}_{\cal{O}}(\beta,\beta^*) \!\!\!\! &=& \!\!\!\! \tfrac{2}{\pi}\sum_{n,\widetilde{n}=0}^{\infty} \!\!
\lg \mathcal{H}^{n,\widetilde{n}}_{1,1}(\beta , \beta^{*}) 
\mathcal{B}^{(1)}_{+,n}(t)\;\mathcal{B}^{(1)}_{+,\widetilde{n}}(t)^{*}
+ \mathcal{H}^{n,\widetilde{n}}_{0,0}(\beta , \beta^{*})
\mathcal{B}^{(1)}_{0,n}(t)\;\mathcal{B}^{(1)}_{0,\widetilde{n}}(t)^{*} \right. \nn \\
&+& \!\!\!\!\!\! \left. \mathcal{H}^{n,\widetilde{n}}_{-1,-1}(\beta , \beta^{*})
\mathcal{B}^{(1)}_{-,n}(t)\;\mathcal{B}^{(1)}_{-,\widetilde{n}}(t)^{*}\rg,\nn\\
\mathrm{Q}^{(1)}_{\cal{O}}(\beta,\beta^*) \!\!\!\! &=& \!\!\!\! \tfrac{1}{\pi}\sum_{n,\widetilde{n}=0}^{\infty} \!\!
\lg\mathcal{Y}^{n,\widetilde{n}}_{1,1}(\beta , \beta^{*}) 
\mathcal{B}^{(1)}_{+,n}(t)\;\mathcal{B}^{(1)}_{+,\widetilde{n}}(t)^{*}
+\mathcal{Y}^{n,\widetilde{n}}_{0,0}(\beta , \beta^{*})
\mathcal{B}^{(1)}_{0,n}(t)\;\mathcal{B}^{(1)}_{0,\widetilde{n}}(t)^{*} \right. \nn \\
&+& \!\!\!\!\!\! \left. \mathcal{Y}^{n,\widetilde{n}}_{-1,-1}(\beta , \beta^{*})
\mathcal{B}^{(1)}_{-,n}(t)\;\mathcal{B}^{(1)}_{-,\widetilde{n}}(t)^{*}\rg.
\label{PWQ-oscillator-1}
\eea
\subsection{Hybrid phase space distributions for the $s=\tfrac{3}{2}$ case}

We now concisely list the analytic  expressions for the quasiprobability densities  in the phase space for the spin  
$s=\tfrac{3}{2}$ case with the following objective in mind.
The increasing complexity of the transitory Schr\"{o}dinger spin kitten states endowed with higher values of $s$ is a consequence of 
 realization of approximately pure qudit states reflecting a relatively more extensive structure of superposition that may
exist in a larger Hilbert space. To explore
the said increment in complexity  we proceed by explicitly constructing the time evolution of the hybrid quasiprobability distributions (\ref{P-bipartite}, \ref{W-spin-osc}, \ref{Q-bipartite}) for the $s=\tfrac{3}{2}$ example:
\newpage
\bea
\mathrm{P}^{(\frac{3}{2})}(\theta,\phi; \beta, \beta^{*}) \!\!\!\! &=& \!\!\!\! \tfrac{1}{32 \pi} \sum_{n,\widetilde{n}=0}^{\infty}  \left \{
(18-45\cos \theta + 30 \cos 2\theta-35\cos 3\theta) 
\mathrm{B}_{n,\widetilde{n}}^{(2,2)}(t) \Lambda^{n,\widetilde{n}}_{\frac{3}{2}} 
+ (18+45\cos \theta
\right.   \nn \\
&+ & \!\!\!\!  30 \cos 2\theta+35\cos 3\theta) 
\mathrm{B}_{n,\widetilde{n}}^{(-2,-2)}(t) \Lambda^{n,\widetilde{n}}_{-\frac{3}{2}} +
(-2+55\cos \theta - 30 \cos 2\theta+105\cos 3\theta) \times \nn \\
& \times & \!\!\!\!  
\mathrm{B}_{n,\widetilde{n}}^{(1,1)}(t) \Lambda^{n,\widetilde{n}}_{\frac{1}{2}}
- (2+55\cos \theta + 30 \cos 2\theta+105\cos 3\theta) 
\mathrm{B}_{n,\widetilde{n}}^{(-1,-1)}(t) \Lambda^{n,\widetilde{n}}_{-\frac{1}{2}}  
\nn \\
& + & \!\!\!\! \!\!  
\sum_{k,\ell=0}^{\infty}  \left[
10 \sqrt{3} (3 \sin \theta - 4 \sin 2 \theta + 7 \sin 3 \theta ) \;
\mathrm{Re} \!\!  \left \lgroup \exp \left( i \phi \right) 
\mathrm{B}_{n,\widetilde{n}}^{(2,1)}(t)
\mathcal{G}_{kn}{(-\tfrac{3\widetilde{\lambda}}{2})} 
\mathcal{G}_{\widetilde{n}\ell}{(\tfrac{\widetilde{\lambda}}{2})}
\Lambda^{k,\ell}_{0}   \right \rgroup \right. \nn \\
& + & \!\!\!\!   
40 \sqrt{3} (1-7 \cos \theta) \sin^{2} \theta  
\; \mathrm{Re} \!\!  \left \lgroup \exp \left(2 i \phi \right) 
\mathrm{B}_{n,\widetilde{n}}^{(2,-1)}(t) 
\mathcal{G}_{kn}{(-\tfrac{3\widetilde{\lambda}}{2})} 
\mathcal{G}_{\widetilde{n} \ell}{(-\tfrac{\widetilde{\lambda}}{2})} 
\Lambda^{k,\ell}_{0} \right \rgroup
\nn \\
& + & \!\!\!\!   
280 \sin^{3} \theta \;
\mathrm{Re} \left \lgroup \exp \left(3 i \phi \right) \mathrm{B}_{n,\widetilde{n}}^{(2,-2)}(t) 
\mathcal{G}_{kn}{(-\tfrac{3\widetilde{\lambda}}{2})} 
\mathcal{G}_{\widetilde{n} \ell}{(-\tfrac{3\widetilde{\lambda}}{2})} \Lambda^{k,\ell}_{0} \right \rgroup 
\!\! -10 (\sin \theta + 21 \sin 3 \theta) \times \nn \\
& \times & \!\!\!\!   
\mathrm{Re} \!\!  \left \lgroup \exp \left(i \phi \right) \mathrm{B}_{n,\widetilde{n}}^{(1,-1)}(t) 
\mathcal{G}_{kn}{(-\tfrac{\widetilde{\lambda}}{2})} 
\mathcal{G}_{\widetilde{n} \ell}{(-\tfrac{\widetilde{\lambda}}{2})} \Lambda^{k,\ell}_{0} \right \rgroup 
+ 40 \sqrt{3} (1+7 \cos \theta) \sin^{2} \theta \times \nn \\
& \times & \!\!\!\!  \mathrm{Re} \!\!  \left \lgroup \exp \left(2i \phi \right)
\mathrm{B}_{n,\widetilde{n}}^{(1,-2)}(t) 
\mathcal{G}_{kn}{(-\tfrac{\widetilde{\lambda}}{2})} 
\mathcal{G}_{\widetilde{n} \ell}{(-\tfrac{3\widetilde{\lambda}}{2})} 
\Lambda^{k,\ell}_{0} \rg 
\!\!  + \! 10 \sqrt{3} (3 \sin \theta + 4 \sin 2 \theta + 7 \sin 3 \theta) \times \nn \\
& \times & \!\!\!\!\!\!  \left. \left.
\mathrm{Re} \!\!  \lg \exp \left(i \phi \right)
\mathrm{B}_{n,\widetilde{n}}^{(-1,-2)}(t) \mathcal{G}_{kn}{(\tfrac{\widetilde{\lambda}}{2})} 
\mathcal{G}_{\widetilde{n} \ell}{(-\tfrac{3\widetilde{\lambda}}{2})} \Lambda^{k,\ell}_{0}
\rg \right]\right \},\nn\\
W^{(\frac{3}{2})}(\theta,\phi;\beta, \beta^{*}) \!\!\!\! &=& \!\!\!\!\!\! \tfrac{1}{80\pi^{2}} \!\!
\sum_{n,\widetilde{n}=0}^{\infty} \!\!   \left \{ 
\left(3(8\sqrt{15}+\sqrt{35})\cos \theta + 5(8+2\sqrt{5}+6\sqrt{5} \cos{2 \theta }+\sqrt{35} \cos 3\theta) \right) \times  \right.  \nn \\ 
&\times & \!\!\!\!\!\! \mathcal{H}^{n,\widetilde{n}}_{\frac{3}{2},\frac{3}{2}}(\beta , \beta^{*}) 
\mathrm{B}_{n,\widetilde{n}}^{(2,2)}(t) + \left(-3(8\sqrt{15} + \sqrt{35}) 
\cos \theta + 5(8+2\sqrt{5}+6\sqrt{5} \cos{2 \theta } \right.  \nn \\
&-& \!\!\!\!\!\! \left. \sqrt{35} \cos 3\theta) \right)
\mathcal{H}^{n,\widetilde{n}}_{-\frac{3}{2},-\frac{3}{2}}(\beta , \beta^{*}) \mathrm{B}_{n,\widetilde{n}}^{(-2,-2)}(t)  +
 \left((8\sqrt{15}-9\sqrt{35})\cos \theta 
 - 5(-8+2\sqrt{5} \right.
  \nn \\
&+& \!\!\!\!\!\! \left.  6\sqrt{5} \cos{2 \theta }+3\sqrt{35} \cos 3\theta) \right)
\mathcal{H}^{n,\widetilde{n}}_{\frac{1}{2},\frac{1}{2}}(\beta , \beta^{*}) \mathrm{B}_{n,\widetilde{n}}^{(1,1)}(t) +\left((-8\sqrt{15}+9\sqrt{35}) \cos \theta  \right.
 \nn \\
&+ & \!\!\!\!\!\!  \left.
 5(8-2\sqrt{5} -6\sqrt{5} \cos{2 \theta }+3\sqrt{35} \cos 3\theta) \right)
 \mathcal{H}^{n,\widetilde{n}}_{-\frac{1}{2},-\frac{1}{2}}(\beta , \beta^{*})
\mathrm{B}_{n,\widetilde{n}}^{(-1,-1)}(t) + 4 \sqrt{5} \times \nn \\
&\times & \!\!\!\!\!\! (3(4+\sqrt{21})+20 \sqrt{3} \cos \theta + 5 \sqrt{21} \cos 2\theta)
\sin \theta \, \mathrm{Re} \!\! \left \lgroup \!
\mathcal{H}^{n,\widetilde{n}}_{\frac{3}{2},\frac{1}{2}}(\beta , \beta^{*})
\exp\left(i (\phi + \widetilde{\lambda} \mathrm{Im}(\beta)) \right) \right. \!\!
\times
\nn \\
&\times & \!\!\!\!\!\! \left. \mathrm{B}_{n,\widetilde{n}}^{(2,1)}(t) \right \rgroup \!\!
 + 40 \sqrt{15} (1+\sqrt{7} \cos \theta) 
\sin^{2}\theta  \,
\mathrm{Re} \! \left \lgroup 
\mathcal{H}^{n,\widetilde{n}}_{\frac{3}{2},-\frac{1}{2}}(\beta , \beta^{*})
\exp \left(2i (\phi + \widetilde{\lambda} \mathrm{Im}(\beta)) \right) \times \right.   \!\! \nn \\ 
& \times & \!\!\!\!\!\! \left. \mathrm{B}_{n,\widetilde{n}}^{(2,-1)}(t) \right \rgroup \!\!
+ 40 \sqrt{35} \sin^{3} \theta \, \mathrm{Re} \!\! \left \lgroup \!
\mathcal{H}^{n,\widetilde{n}}_{\frac{3}{2},-\frac{3}{2}}(\beta , \beta^{*})
\exp\left(3i (\phi + \widetilde{\lambda} \mathrm{Im}(\beta)) \right) \mathrm{B}_{n,\widetilde{n}}^{(2,-2)}(t) \right \rgroup
\nn \\
&+ & \!\!\!\!\!\!  
  8 \sqrt{5} (4\sqrt{3}+3\sqrt{7}
- 15\sqrt{7} \cos^{2} \theta) \sin \theta \, \mathrm{Re} \!\! \left \lgroup \!
 \mathcal{H}^{n,\widetilde{n}}_{\frac{1}{2},-\frac{1}{2}}(\beta , \beta^{*})
 \exp\left(i (\phi + \widetilde{\lambda} \mathrm{Im}(\beta)) \right) \right. \times
 \!\!\!    \nn \\
& \times & \!\!\!\!\!\! \left. \mathrm{B}_{n,\widetilde{n}}^{(1,-1)}(t) \right \rgroup \!\! + 40 \sqrt{15}(1-\sqrt{7} \cos \theta) \! \sin^{2}\theta \,
\mathrm{Re} \!\! \left \lgroup \!
\mathcal{H}^{n,\widetilde{n}}_{\frac{1}{2},-\frac{3}{2}}(\beta , \beta^{*}) 
\exp\left(2i (\phi + \widetilde{\lambda} \mathrm{Im}(\beta)) \right) \times
\right.  \nn \\
&\times & \!\!\!\!\!\! \left. 
\mathrm{B}_{n,\widetilde{n}}^{(1,-2)}(t) \right \rgroup  \!\! + 4 \sqrt{5}(3(4+\sqrt{21})
-  20\sqrt{3} \cos \theta + 5 \sqrt{21} \cos 2 \theta) 
\sin \theta  \times \nn \\
&\times & \!\!\!\!\!\! \left.
\mathrm{Re} \!\! \lg 
\mathcal{H}^{n,\widetilde{n}}_{-\frac{1}{2},-\frac{3}{2}}(\beta , \beta^{*}) 
  \exp\left(i (\phi + \widetilde{\lambda} \mathrm{Im}(\beta)) \right)   
\mathrm{B}_{n,\widetilde{n}}^{(-1,-2)}(t) \rg
\right \}, \quad \quad\nn\\
\mathrm{Q}^{(\frac{3}{2})}(\theta,\phi; \beta, \beta^{*}) \!\!\!\! &=& \!\!\!\! \tfrac{1}{8 \pi^{2}} \sum_{n,\widetilde{n}=0}^{\infty}
\left \{ 8 \sin^{6}\left( \tfrac{\theta}{2}\right) 
\mathcal{Y}^{n,\widetilde{n}}_{\frac{3}{2},\frac{3}{2}}(\beta, \beta^{*}) \mathrm{B}_{n,\widetilde{n}}^{(2,2)}(t) + 24 \sin^{4}\left( \tfrac{\theta}{2}\right)
\cos^{2}\left( \tfrac{\theta}{2}\right)
\mathcal{Y}^{n,\widetilde{n}}_{\frac{1}{2},\frac{1}{2}}(\beta, \beta^{*}) \mathrm{B}_{n,\widetilde{n}}^{(1,1)}(t) \right.  \nn \\
& +& \!\!\!\! 
24 \cos^{4}\left( \tfrac{\theta}{2}\right)
\sin^{2}\left( \tfrac{\theta}{2}\right) 
\mathcal{Y}^{n,\widetilde{n}}_{-\frac{1}{2},-\frac{1}{2}}(\beta, \beta^{*})
\mathrm{B}_{n,\widetilde{n}}^{(-1,-1)}(t)
  + 8 \cos^{6}\left( \tfrac{\theta}{2}\right)
  \mathcal{Y}^{n,\widetilde{n}}_{-\frac{3}{2},-\frac{3}{2}}(\beta, \beta^{*}) 
 \mathrm{B}_{n,\widetilde{n}}^{(-2,-2)}(t) \nn \\
 &+& \!\!\!\!  8\sqrt{3} \sin \theta
\sin^{4}\left( \tfrac{\theta}{2}\right) 
\mathrm{Re} \! \left \lgroup    \mathcal{Y}^{n,\widetilde{n}}_{\frac{1}{2},\frac{3}{2}}(\beta, \beta^{*}) 
\exp \left( -i\left( \phi + \widetilde{\lambda} \mathrm{Im}(\beta)\right) 
- 2 \widetilde{\lambda} \mathrm{Re}(\beta)\right) \mathrm{B}_{n,\widetilde{n}}^{(1,2)}(t) \right \rgroup \nn \\
& + &  \!\!\!\! 8\sqrt{3} \sin \theta \cos^{4}\left( \tfrac{\theta}{2}\right) 
\mathrm{Re} \! \left \lgroup 
\mathcal{Y}^{n,\widetilde{n}}_{-\frac{3}{2},-\frac{1}{2}}(\beta, \beta^{*})
\exp \left( -i\left( \phi + \widetilde{\lambda} \mathrm{Im}(\beta) \right) 
+ 2 \widetilde{\lambda} \mathrm{Re}(\beta) \right)
  \mathrm{B}_{n,\widetilde{n}}^{(-2,-1)}(t) \right \rgroup \nn \\
& + &  \!\!\!\! 
 4 \sqrt{3} \sin ^{2} \theta \sin ^{2}\left( \tfrac{\theta}{2}\right)
 \mathrm{Re} \! \left \lgroup \mathcal{Y}^{n,\widetilde{n}}_{-\frac{1}{2},\frac{3}{2}}(\beta, \beta^{*})
\exp \left( -2i\left( \phi + \widetilde{\lambda} \mathrm{Im}(\beta) \right) 
- \widetilde{\lambda} \mathrm{Re}(\beta)  \right) \mathrm{B}_{n,\widetilde{n}}^{(-1,2)}(t) \right \rgroup
\nn \\
& + & \!\!\!\!  4 \sqrt{3} \sin ^{2} \theta \cos ^{2}\left( \tfrac{\theta}{2}\right) 
\mathrm{Re} \left \lgroup \mathcal{Y}^{n,\widetilde{n}}_{-\frac{3}{2},\frac{1}{2}}(\beta, \beta^{*}) 
\exp \left( -2i\left( \phi + \widetilde{\lambda} \mathrm{Im}(\beta) \right) 
+ \widetilde{\lambda} \mathrm{Re}(\beta) \right) \mathrm{B}_{n,\widetilde{n}}^{(-2,1)}(t) \right \rgroup \nn \\
& + & \!\!\!\! 6 \sin ^{3} \theta \,
\mathrm{Re} \left \lgroup \mathcal{Y}^{n,\widetilde{n}}_{-\frac{1}{2},\frac{1}{2}}(\beta, \beta^{*}) 
\exp \left( -i\left( \phi + \widetilde{\lambda} \mathrm{Im}(\beta) \right) \right) \mathrm{B}_{n,\widetilde{n}}^{(-1,1)}(t) \right \rgroup \nn \\
& + & \left. \!\!\!\! 2 \sin ^{3} \theta \,
\mathrm{Re} \left \lgroup \mathcal{Y}^{n,\widetilde{n}}_{-\frac{3}{2},\frac{3}{2}}(\beta, \beta^{*}) 
\exp \left( -3i\left( \phi + \widetilde{\lambda} \mathrm{Im}(\beta) \right) \right) \mathrm{B}_{n,\widetilde{n}}^{(-2,2)}(t) \right \rgroup \right \}.
\label{PWQ_spin_bipartite_3/2}
\eea
In the derivation of the hybrid $\mathrm{W}^{(\frac{3}{2})}$-distribution (\ref{PWQ_spin_bipartite_3/2}) the identity  
(\ref{IdentityHyper}) has been used. The above quasiprobability densities on the composite phase space lead, in turn, to the 
construction of the relevant spin distributions on the sphere via the integration of the oscillator degree of freedom given 
in equations (\ref{P-spin}), (\ref{W-spin}) and (\ref{Q-spin}), respectively:
\bea
\mathrm{P}^{(\frac{3}{2})}_{\mathcal{Q}}(\theta,\phi) \!\!\!\! &=& \!\!\!\! \dfrac{1}{32 \pi} \sum_{n,\widetilde{n}=0}^{\infty}  \left \{
 (18-45\cos \theta + 30 \cos 2\theta-35\cos 3\theta) \, \mathrm{B}_{n,\widetilde{n}}^{(2,2)}(t) \, \delta_{n,\widetilde{n}}  + (18+45\cos \theta  \right.  \nn \\
&+ & \!\!\!\! 30 \cos 2\theta+35\cos 3\theta)   \mathrm{B}_{n,\widetilde{n}}^{(-2,-2)}(t) 
\delta_{n,\widetilde{n}} +(-2+55\cos \theta - 30 \cos 2\theta+105\cos 3\theta) \times \nn \\
& \times & \!\!\!\!  \mathrm{B}_{n,\widetilde{n}}^{(1,1)}(t)\,\delta_{n,\widetilde{n}}
- (2+55\cos \theta + 30 \cos 2\theta+105\cos 3\theta) \, \mathrm{B}_{n,\widetilde{n}}^{(-1,-1)}(t) \,
\delta_{n,\widetilde{n}} + 10 \sqrt{3} (3 \sin \theta  \nn \\
& - & \!\!\!\! 4 \sin 2 \theta +7 \sin 3 \theta ) 
\mathrm{Re} \left \lgroup \exp \left( i \phi \right) 
\mathrm{B}_{n,\widetilde{n}}^{(2,1)}(t) \mathcal{G}_{mn}{(-\widetilde{\lambda})}
  \right \rgroup + 40 \sqrt{3} (1-7 \cos \theta) \sin^{2} \theta \times \nn \\
& \times & \!\!\!\! 
\mathrm{Re} \left \lgroup \exp \left(2 i \phi \right) 
 \mathrm{B}_{n,\widetilde{n}}^{(2,-1)}(t) \mathcal{G}_{mn}{(-2\widetilde{\lambda})} \right \rgroup 
 +  280 \sin^{3} \theta \,
 \mathrm{Re} \left \lgroup \exp \left(3 i \phi \right) \mathrm{B}_{n,\widetilde{n}}^{(2,-2)}(t) \right. \times \nn \\
& \times & \!\!\!\! \left.  
\mathcal{G}_{mn}{(-3\widetilde{\lambda})} \right \rgroup 
-10 (\sin \theta + 21 \sin 3 \theta) \,
\mathrm{Re} \left \lgroup \exp \left(i \phi \right)  
\mathrm{B}_{n,\widetilde{n}}^{(1,-1)}(t) \mathcal{G}_{mn}{(-\widetilde{\lambda})} 
\right \rgroup \nn \\
& + & \!\!\!\!
40 \sqrt{3} (1+7 \cos \theta) \sin^{2} \theta \,
\mathrm{Re} \left \lgroup \exp \left(2i \phi \right)
\mathrm{B}_{n,\widetilde{n}}^{(1,-2)}(t) \mathcal{G}_{mn}{(-2\widetilde{\lambda})} \right \rgroup 
+ 10 \sqrt{3} (3 \sin \theta   \nn \\
&+ & \!\!\!\! \left. 4 \sin 2 \theta + 7 \sin 3 \theta) 
\mathrm{Re} \left \lgroup \exp \left(i \phi \right)
 \mathrm{B}_{n,\widetilde{n}}^{(-1,-2)}(t) \mathcal{G}_{mn}{(-\widetilde{\lambda})} \right \rgroup 
 \right \},\nn\\
W^{(\frac{3}{2})}_{\mathcal{Q}}(\theta,\phi) \!\!\!\! &=& \!\!\!\!\!\! \tfrac{1}{160\pi} \!\!
\sum_{n,\widetilde{n}=0}^{\infty} \!\!   \left \{ 
\left(3(8\sqrt{15}+\sqrt{35})\cos \theta + 5(8+2\sqrt{5}+6\sqrt{5} \cos{2 \theta }+\sqrt{35} \cos 3\theta) \right) \times  \right.  
\nn \\ 
&\times & \!\!\!\!\!\! \left.
\mathrm{B}_{n,\widetilde{n}}^{(2,2)}(t) \delta_{n,\widetilde{n}} + \left(-3(8\sqrt{15} + \sqrt{35}) \cos \theta 
+ 5(8+2\sqrt{5}+6\sqrt{5} \cos{2 \theta }-\sqrt{35} \cos 3\theta) \right) \right. \times  \nn \\
&\times & \!\!\!\!\!\! \left. 
\mathrm{B}_{n,\widetilde{n}}^{(-2,-2)}(t) \delta_{n,\widetilde{n}} + \left((8\sqrt{15}-9\sqrt{35})\cos \theta 
- 5(-8+2\sqrt{5} + 6\sqrt{5} \cos{2 \theta }  \right. \right.  \nn \\ 
&+& \!\!\!\!\!\! \left. 3\sqrt{35} \cos 3\theta) \right)  \mathrm{B}_{n,\widetilde{n}}^{(1,1)}(t) \delta_{n,\widetilde{n}} + \left((-8\sqrt{15}+9\sqrt{35}) \cos \theta 
+ 5(8-2\sqrt{5} -  6\sqrt{5} \cos{2 \theta }\right. \nn \\
&+& \!\!\!\!\!\! \left.
3\sqrt{35} \cos 3\theta) \right) \mathrm{B}_{n,\widetilde{n}}^{(-1,-1)}(t) 
\delta_{n,\widetilde{n}}  + 4 \sqrt{5} 
\left(3(4+\sqrt{21})+20 \sqrt{3} \cos \theta + 5 \sqrt{21} \cos 2\theta \right) \times
  \nn \\
&\times & \!\!\!\!\!\! \sin \theta \, 
 \mathrm{Re} \left \lgroup 
 \exp(i \phi) \mathrm{B}_{n,\widetilde{n}}^{(2,1)}(t) 
 \mathcal{G}_{\widetilde{n}n}(-\widetilde{\lambda}) \right \rgroup 
 + 40 \sqrt{15} (1+\sqrt{7} \cos \theta) 
 \sin^{2}\theta \times \nn \\
&\times & \!\!\!\!\!\! 
\mathrm{Re} \!\! \left \lgroup 
\exp(2i \phi) \mathrm{B}_{n,\widetilde{n}}^{(2,-1)}(t) 
\mathcal{G}_{\widetilde{n}n}(-2\widetilde{\lambda}) \right \rgroup \!\! + 40 \sqrt{35} \sin^{3} \theta \,  \mathrm{Re} \!\! \left \lgroup \!
\exp(3i \phi) \mathrm{B}_{n,\widetilde{n}}^{(2,-2)}(t) 
\mathcal{G}_{\widetilde{n}n}(-3\widetilde{\lambda}) \! \right \rgroup  \nn \\ 
&+ & \!\!\!\!\!\!  8 \sqrt{5} (4\sqrt{3}+3\sqrt{7}
- 15\sqrt{7} \cos^{2} \theta) \sin \theta \, 
\mathrm{Re} \!\! \left \lgroup 
\exp(i \phi) \mathrm{B}_{n,\widetilde{n}}^{(1,-1)}(t) 
\mathcal{G}_{\widetilde{n}n}(-\widetilde{\lambda}) \right \rgroup
 \nn \\
&+ & \!\!\!\!\!\! 40 \sqrt{15}(1-\sqrt{7} \cos \theta) \sin^{2}\theta \,
\mathrm{Re} \!\! \left \lgroup 
\exp(2i \phi) \mathrm{B}_{n,\widetilde{n}}^{(1,-2)}(t) 
\mathcal{G}_{\widetilde{n}n}(-2\widetilde{\lambda}) \right \rgroup 
+4 \sqrt{5}(3(4+\sqrt{21}) \nn \\
&- & \!\!\!\!\!\! \left. 20\sqrt{3} \cos \theta + 5 \sqrt{21} \cos 2 \theta) 
\sin \theta \,  \mathrm{Re} \!\! \left \lgroup 
\exp(i \phi) \mathrm{B}_{n,\widetilde{n}}^{(-1,-2)}(t) 
\mathcal{G}_{\widetilde{n}n}(-\widetilde{\lambda}) \right \rgroup  
\right \},\nn\\
\mathrm{Q}^{(\frac{3}{2})}_{\mathcal{Q}}(\theta,\phi) \!\!\!\! &=& \!\!\!\!\!\! \tfrac{1}{8 \pi} \sum_{n,\widetilde{n}=0}^{\infty}
\!\! \left \{ 8 \sin^{6}\left( \tfrac{\theta}{2}\right) 
 \mathrm{B}_{n,\widetilde{n}}^{(2,2)}(t) \delta_{n,\widetilde{n}} + 24 \sin^{4}\left( \tfrac{\theta}{2}\right)
\cos^{2}\left( \tfrac{\theta}{2}\right) \mathrm{B}_{n,\widetilde{n}}^{(1,1)}(t) \delta_{n,\widetilde{n}} 
+24 \cos^{4}\left( \tfrac{\theta}{2}\right) \sin^{2}\left( \tfrac{\theta}{2} \right)  \right. \!\! \times  \nn \\
& \times & \!\!\!\! \mathrm{B}_{n,\widetilde{n}}^{(-1,-1)}(t) \delta_{n,\widetilde{n}} +
8 \cos^{6}\left( \tfrac{\theta}{2}\right) 
 \mathrm{B}_{n,\widetilde{n}}^{(-2,-2)}(t) \delta_{n,\widetilde{n}} 
 + 8\sqrt{3} \sin \theta \sin^{4}\left( \tfrac{\theta}{2}\right) 
 \mathrm{Re} \left \lgroup 
 \exp \left( -i \phi \right) \times \right.
 \nn \\
& \times &  \!\!\!\! \left. \mathrm{B}_{n,\widetilde{n}}^{(1,2)}(t) \mathcal{G}_{\widetilde{n}n}(\widetilde{\lambda}) \right \rgroup + 8\sqrt{3} \sin \theta \cos^{4}\left( \tfrac{\theta}{2}\right) 
\mathrm{Re} \left \lgroup  
\exp \left( -i\phi \right) \mathrm{B}_{n,\widetilde{n}}^{(-2,-1)}(t) \mathcal{G}_{\widetilde{n}n}(\widetilde{\lambda}) \right \rgroup   \nn \\
& +&  \!\!\!\! 
 4 \sqrt{3} \sin ^{2} \theta \sin ^{2}\left( \tfrac{\theta}{2}\right) 
 \mathrm{Re} \left \lgroup 
 \exp \left( -2i \phi \right) \mathrm{B}_{n,\widetilde{n}}^{(-1,2)}(t) \mathcal{G}_{\widetilde{n}n}(2\widetilde{\lambda}) \right \rgroup +
  4 \sqrt{3} \sin ^{2} \theta \cos ^{2}\left( \tfrac{\theta}{2}\right) \times
\nn \\
& \times & \!\!\!\! 
\mathrm{Re} \left \lgroup  
\exp \left( -2i \phi \right) \mathrm{B}_{n,\widetilde{n}}^{(-2,1)}(t) \mathcal{G}_{\widetilde{n}n}(2\widetilde{\lambda}) \right \rgroup + 6 \sin ^{3} \theta \,
\mathrm{Re} \left \lgroup  
\exp \left( -i \phi \right) \mathrm{B}_{n,\widetilde{n}}^{(-1,1)}(t) \mathcal{G}_{\widetilde{n}n}(\widetilde{\lambda})  \right \rgroup \nn \\
& + & \left. \!\!\!\! 2 \sin ^{3} \theta \,
\mathrm{Re} \left \lgroup 
\exp \left( -3i \phi \right) \mathrm{B}_{n,\widetilde{n}}^{(-2,2)}(t) \mathcal{G}_{\widetilde{n}n}(3\widetilde{\lambda})  \right \rgroup \right \}.
\label{PWQ_spin_3/2}
\eea
As already noted the spin phase space densities (\ref{PWQ_spin_3/2}) may be alternately  constructed following the 
recipe developed in [\cite{A1981}] and utilizing our derivation of the spherical components of the density matrix 
(\ref{DenMat_kq_s3/2}). This confirms the validity of the approach considered here that starts from the composite phase space quasiprobability distributions. Our description of the spin kitten states for the $s=\tfrac{3}{2}$ example given in 
Sec. \ref{nonclassicality} is based on the structure (\ref{PWQ_spin_3/2}) enunciated above.

\par

The integral representations given in (\ref{P-oscillator}-\ref{Q-oscillator}) permit us to obtain the oscillator phase space
quasiprobability densities for the present example of spin $s=\tfrac{3}{2}$:
\bea
 \mathrm{P}^{(\frac{3}{2})}_{\cal{O}}(\beta,\beta^*) \!\!\!\! &=& \!\!\!\! \sum_{n,\widetilde{n}=0}^{\infty}
\lg \mathrm{B}^{(2,2)}_{n,\widetilde{n}}(t)\,\Lambda^{n,\widetilde{n}}_{\frac{3}{2}}
+\mathrm{B}^{(1,1)}_{n,\widetilde{n}}(t)\,\Lambda^{n,\widetilde{n}}_{\frac{1}{2}}
+\mathrm{B}^{(-1,-1)}_{n,\widetilde{n}}(t)\,\Lambda^{n,\widetilde{n}}_{-\frac{1}{2}}
+\mathrm{B}^{(-2,-2)}_{n,\widetilde{n}}(t)\,\Lambda^{n,\widetilde{n}}_{-\frac{3}{2}}
\rg ,\nn\\
 \mathrm{W}^{(\frac{3}{2})}_{\cal{O}}(\beta,\beta^*) \!\!\!\! &=& \!\!\!\! \tfrac{2}{\pi} 
\sum_{n,\widetilde{n}=0}^{\infty}
\lg
\mathcal{H}^{n,\widetilde{n}}_{\frac{3}{2},\frac{3}{2}}(\beta , \beta^{*})
\mathrm{B}^{(2,2)}_{n,\widetilde{n}}(t)
+
\mathcal{H}^{n,\widetilde{n}}_{\frac{1}{2},\frac{1}{2}}(\beta , \beta^{*})
\mathrm{B}^{(1,1)}_{n,\widetilde{n}}(t)
+
\mathcal{H}^{n,\widetilde{n}}_{-\frac{1}{2},-\frac{1}{2}}(\beta , \beta^{*})
\mathrm{B}^{(-1,-1)}_{n,\widetilde{n}}(t) \right. \nn \\
&+& \!\!\!\!\!\!
\left.
\mathcal{H}^{n,\widetilde{n}}_{-\frac{3}{2},-\frac{3}{2}}(\beta , \beta^{*})
 \mathrm{B}^{(-2,-2)}_{n,\widetilde{n}}(t)
\rg, \nn\\
\mathrm{Q}^{(\frac{3}{2})}_{\cal{O}}(\beta,\beta^*) \!\!\!\! &=& \!\!\!\! \tfrac{1}{\pi} 
\sum_{n,\widetilde{n}=0}^{\infty}
\lg
\mathcal{Y}^{n,\widetilde{n}}_{\frac{3}{2},\frac{3}{2}}(\beta , \beta^{*})
\mathrm{B}^{(2,2)}_{n,\widetilde{n}}(t)
+
\mathcal{Y}^{n,\widetilde{n}}_{\frac{1}{2},\frac{1}{2}}(\beta , \beta^{*})
\mathrm{B}^{(1,1)}_{n,\widetilde{n}}(t)
+
\mathcal{Y}^{n,\widetilde{n}}_{-\frac{1}{2},-\frac{1}{2}}(\beta , \beta^{*})
\mathrm{B}^{(-1,-1)}_{n,\widetilde{n}}(t) \right. \nn \\
&+& \!\!\!\!\!\!
\left.
\mathcal{Y}^{n,\widetilde{n}}_{-\frac{3}{2},-\frac{3}{2}}(\beta , \beta^{*})
\mathrm{B}^{(-2,-2)}_{n,\widetilde{n}}(t)
\rg.
\label{PWQ-oscillator-3/2}
\eea
\section{Nonclassical features in the phase space: spin kitten states}
\label{nonclassicality}
\setcounter{equation}{0}
\label{spin-kitten}
To explore  the emergence of transitory spin kitten states during the evolution generated by the bipartite Hamiltonian (\ref{H-Sp}) 
at moderately strong coupling $\widetilde{\lambda} \sim \mathrm {O}\lo 10^{-2}\rc$ we study   the  qudit entropy  given by  
\beq
S(\rho_{\mathcal{Q}}) = - \mathrm{Tr} \left[\rho_{\mathcal{Q}}\, \log \rho_{\mathcal{Q}}\right],
\label{entropy}
\eeq
where  we have omitted explicit reference to the spin quantum number: $s$. As our  bipartite system inhabits a pure state, 
the entropies of two individual subsystems are 
equal [\cite{AL1970}]. The entropy associated with the oscillator degree of freedom is, therefore, identical to that of the spin variable $S(\rho_{\mathcal{Q}})$ given in (\ref{entropy}). This elementary feature has an interesting consequence on the experimental realization of the spin kitten states. We will comment on this later. The entropy of the individual subsystems may be viewed as the entanglement entropy of the bipartite hybrid system. The characteristic time scale that governs the said  appearance of the kitten states  may be understood  as follows. The order of  terms in the
 expansion of the Laguerre polynomial  in the off-diagonal elements of the Hamiltonian for the spin $s=1,\tfrac{3}{2}$ 
cases given in (\ref{H-n-1}) and (\ref{H-n-3/2}), respectively, engenders the corresponding time scales inherent to the process. In the analysis presented here we explore the first nontrivial  dimensionless time scale   $\omega\, t_{\mathsf{long}} 
\sim \mathrm{O}\lg\lo\tfrac{\Delta}{\omega}  \exp\lo-\frac{\widetilde{\lambda}^2}{2}\rc \widetilde{\lambda}^2\rc^{-1}\rg$ that originates as a consequence of the linear term appearing in the Laguerre polynomial.   For the set of parameters considered here the above time scale pertinent to the realization of the kitten states is given by $\omega\, t_{\mathsf{long}} \sim \mathrm{O}\lo 10^{6}\rc$. 
To proceed, we consider the entropy (\ref{entropy})
for the spin $s = 1, \frac{3}{2}$ cases in Figs. \ref{entropy-1-tL}{\sf (a)} and \ref{entropy-3/2-tL}{\sf (a)}, respectively. One general feature evident in these two diagrams is the existence of a quasiperiodicity of the system that renders the entropy reducing to approximately zero periodically, which, in turn, imparts a near-factorizability to the bipartite composite state. Following previous argument this quasiperiod may be quantified as 
\beq
\omega \mathrm{T}|_{\mathsf{quasiperiod}} = 2 \pi \lg\lo\tfrac{\Delta}{\omega}  \exp\lo-\tfrac{\widetilde{\lambda}^2}{2}\rc \widetilde{\lambda}^2\rc^{-1}\rg.
\label{quasi-period}
\eeq
Our later discussions in the context of Figs. \ref{entropy-1-tL} and \ref{entropy-3/2-tL} regarding the evolution of specific states  will support the remarkable  accuracy of this description.
Towards checking the true periodicity of the system we evaluate the Hilbert-Schmidt distance [\cite{DMMW2000}] between the initial qudit state and the evolving state under investigation. Between any two arbitrary density matrices $\lo \lo\rho_{\mathcal{Q}} \rc_{1}, \lo\rho_{\mathcal{Q}} \rc_{2}\rc$ the Hilbert-Schmidt distance is defined as follows:
\beq
\left[\mathrm{d_{HS}} \lo\lo \rho_{\mathcal{Q}} \rc_{1}, \lo \rho_{\mathcal{Q}} \rc_{2}\rc\right]^{2} \equiv  \mathrm{Tr} \left[ \lg \left( \rho_{\mathcal{Q}} \right)_{1}- 
\left( \rho_{\mathcal{Q}}  \right)_{2} \rg ^{2} \right] = \tfrac{4 \pi}{2 s +1}\, \int 
\lg \left( \mathrm{W}_{\mathcal{Q}}(\theta,\phi) \right)_{1} 
- \left( \mathrm{W}_{\mathcal{Q}}(\theta,\phi) \right)_{2} \rg ^{2} \mathrm{d}\Omega.
\label{dHS}
\eeq
The  equality in (\ref{dHS}) expresses the metric on the Hilbert space  $\mathrm{d_{HS}} \lo\lo \rho_{\mathcal{Q}} \rc_{1}, \lo \rho_{\mathcal{Q}} \rc_{2}\rc$ via the  corresponding spin Wigner $\mathrm{W}_{\mathcal{Q}}(\theta,\phi)$-distributions. The distance measures 
between the initial state and various states in question are quoted in Figs. \ref{entropy-1-tL}{\sf (a)} and 
\ref{entropy-3/2-tL}{(\sf a)}. In these illustrations we notice that the system returns, in a time scale, say, $\mathrm {T_{rev}}$, close to its initial state after an integral number of its arrivals to the approximately
zero entropy configurations. Roughly speaking, we obtain
\beq 
\mathrm {T_{rev}} \approx \mathfrak{n} \mathrm{T}|_{\mathsf{quasiperiod}},
\label{T-rev}
\eeq
where $\mathfrak{n}$ is a small  positive integer. The near reproduction of the original state in the phase space may be considered as a spin analog of the quantum revival of a wave packet [\cite{RS2007}].  Another important feature noticed in the Figs. \ref{entropy-1-tL}{\sf(a)} and 
\ref{entropy-3/2-tL}{\sf(a)} is the almost 
periodic manifestation of the \textit{nonzero locally  minimum entropy} configurations. These structures are crucial for the study of the spin kitten states.  For the configurations depicted in Figs. \ref{kitten-1} and 
\ref{kitten-3/2} the  initial states are, as a consequence of the choice $\mathrm{c} = 0$ in  (\ref{state-t0}), 
\textit{completely 
 factorizable, and, therefore, do not contain any coherent superposition of states. Consequently, the said short-lived spin kitten states are generated due to the dynamical spin-photon interaction}.  For the spin $s=1$ case the nonzero locally minimum entropy states occur at times $\lo t_{\mathsf{D}}, t_{\mathsf{E}}, t_{\mathsf{F}}, t_{\mathsf{G}}\rc$ in Fig. \ref{entropy-1-tL}{\sf (a)}, when the approximate Schr\"{o}dinger $2$-kitten  states are observed. The diagonal spin 
 $\mathrm{P}_{\mathcal{Q}}$-representation of these  states are depicted in 
 diagrams {\sf (b, c, d, e)}, respectively, of Fig. \ref {kitten-1}. As the spin increases, more complex kitten states start appearing due to the feasibility of increased quantum coherence among the spin wave function components. For the spin $s=\tfrac{3}{2}$ case  we investigate the evolving states at times  $\lo t_{\mathsf{C}}, t_{\mathsf{D}}, t_{\mathsf{E}}, t_{\mathsf{F}}\rc $ considered in 
 Fig.  \ref {entropy-3/2-tL}{\sf (a)}. The
 spin  $\mathrm{P}_{\mathcal{Q}}$-representation corresponding to these times are given in illustrations {\sf(b, c, d, e)}, respectively, in Fig. \ref {kitten-3/2}. While Figs. \ref {kitten-3/2}{\sf (b, c)} represent $3$-kitten states at respective times $\lo t_{\mathsf{C}}, t_{\mathsf{D}}\rc$, the Fig. \ref {kitten-3/2}{\sf (d)} embodies a $2$-kitten state  at time $t_{\mathsf{E}}$. Moreover, the Fig. \ref {kitten-3/2}{\sf (e)} constitutes the $4$-kitten state at time $t_{\mathsf{F}}$. The formation of spin kitten states coinciding with 
 nonzero locally minimum entropy configurations has close correspondence with the fractional revival  of the wave packets [\cite{RS2007}]. Akin to the system of wave packets, the present bipartite system also maintains the time of manifestation of the quantum fractional revivals as 
 $\tfrac{k}{\ell}\,\mathrm {T_{rev}}$, where $(k, \ell)$ are coprime integers [\cite{AP1989}].  In this regard a distinguishing feature between the  Figs. \ref{entropy-1-tL}{\sf (a)} and \ref{entropy-3/2-tL}{(\sf a)} is that the latter, owing to the higher dimensionality of its Hilbert space contains more locally minimum entropy structures reflecting a wider possibility of formations of coherent superposition of states. This translates into the transitory formation of more complex kitten states for the $s= \tfrac{3}{2}$ case. Lastly, the equality of the entropy of both the subsystems discussed following (\ref{entropy}) provides a cue 
 towards experimental detection of the spin kitten states. The concurrent appearance of local entropy minima for both sectors now indicates \textit{simultaneous} 
realizations of spin and  oscillator Schr\"{o}dinger kitten states. The plots of oscillator Wigner distribution $\mathrm{W}_{\cal{O}}(\beta,\beta^*)$ given in  (\ref{PWQ-oscillator-1}, \ref{PWQ-oscillator-3/2}) substantiate this property. We do not reproduce these diagrams here. 
 
\par
 
In contrast to the above discussion, the ultrastrong coupling domain $\widetilde{\lambda} \sim \mathrm {O}(1)$ incorporates
a large number of interaction modes and their harmonics with a wide range of characteristic time scales $ \mathrm{O}\!\!\!\lg\!\!\lo\tfrac{\Delta}{\omega}  \exp\!\lo-\frac{\widetilde{\lambda}^2}{2}\rc \widetilde{\lambda}^{2n}\!\rc^{-1}\!\rg$, where  $n\in (0, 1,2,\ldots)$. As a consequence the phase correlations among the interacting modes in the system are lost causing the disappearance of any collapse and revival pattern. The randomization of the phases induces the quasiperiodicity of the qudit entropy to vanish, and the system \textit{does not return} close to its initial configuration in a finite time. A  stabilization of the value of  entropy
(Figs. \ref{entropy-1-tL}{\sf (b)}, \ref{entropy-3/2-tL}{\sf (b)}) sets in, while a random stochastic fluctuation  around 
the stabilized value of entropy is also observed.

\subsection{Second order spin correlation function}
Towards further study on the coherence properties of the spin kitten states we use the normalized second order spin correlation function [\cite{ABNV1977}, \cite{GG1997}]  defined as 
\beq
g_{s}(t)=\tfrac{\braket{S_{+}^{2}S_{-}^{2}}} {\braket{S_{+}S_{-}}^{2}},\qquad 
\braket{\Theta} \equiv \mathrm{Tr}\left[ \rho_{_{\mathcal{Q}}}(t) \, \Theta \right].
\label{def_corr}
\eeq
The correlation function (\ref{def_corr}) determines [\cite{ABNV1977}] possible existence of antibunching effect in the emission spectrum of photons. It  has also been employed [\cite{GG1997}] to signal the contrast between the properties of  spin kitten states with that of a spin coherent state. For the spin coherent state (\ref{s-polar}) emerging in  (\ref{state-t0}) with the  $c=0$ choice, the  correlation function (\ref{def_corr}) reads [\cite{W1984}] 
\beq
g_{s}(t=0)= \left( \frac{2s-1}{s} \right) \left(2s + \tan ^{2} \!\left( \tfrac{\widetilde{\theta}}{2}\right)  \right) ^{-2}
\left( \tan ^{4} \!\left( \tfrac{\widetilde{\theta}}{2}\right) + 2(2s-1)
\tan ^{2} \!\left( \tfrac{\widetilde{\theta}}{2}\right) + s(2s-1) \right).
\label{second-spin}
\eeq

\par

\begin{figure}
\begin{center}
\captionsetup[subfigure]{labelfont={sf}} 
\subfloat[]{\includegraphics[scale=0.43]{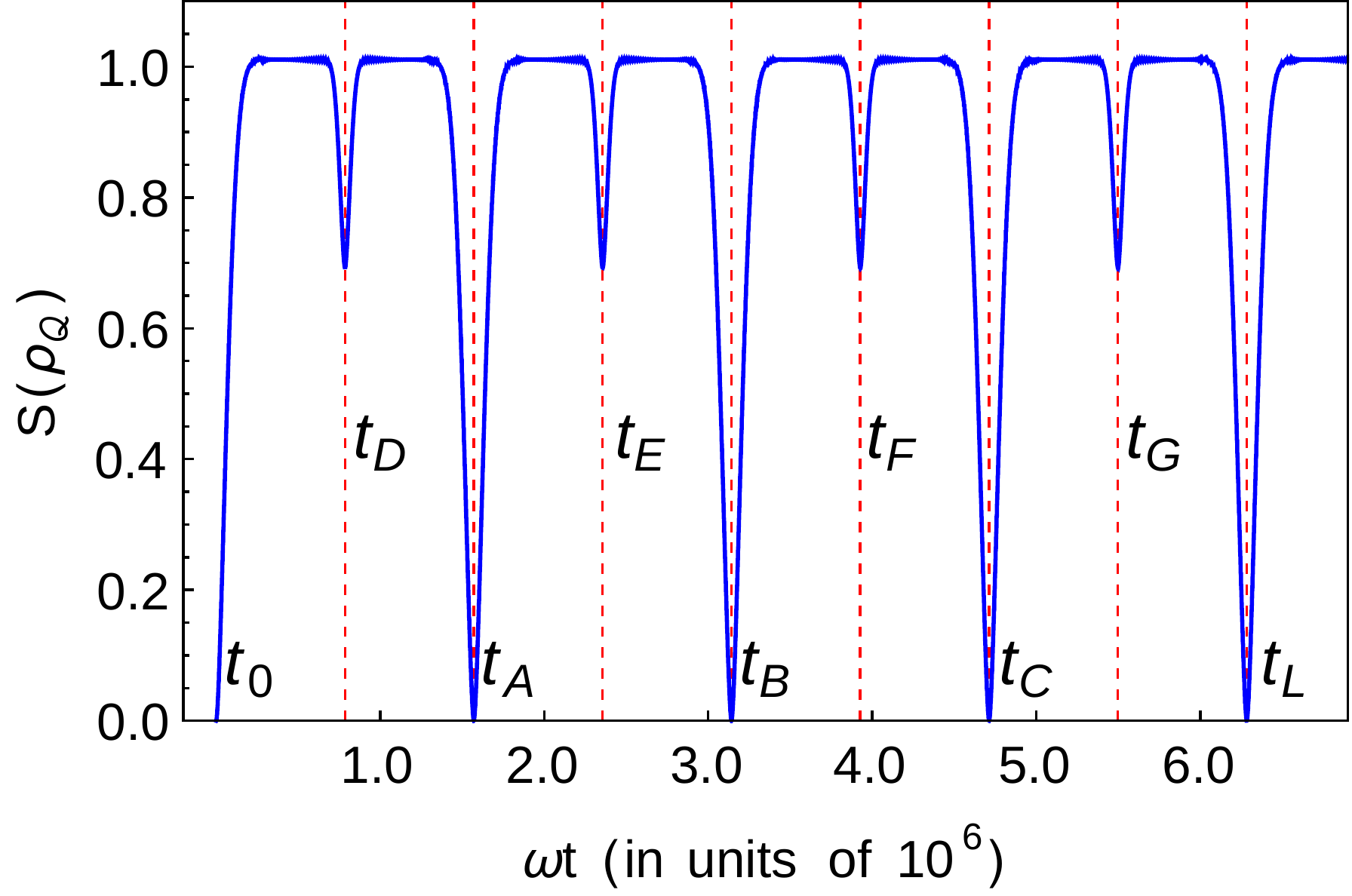}}
\hspace*{0.3cm}
\subfloat[]{\includegraphics[scale=0.43]{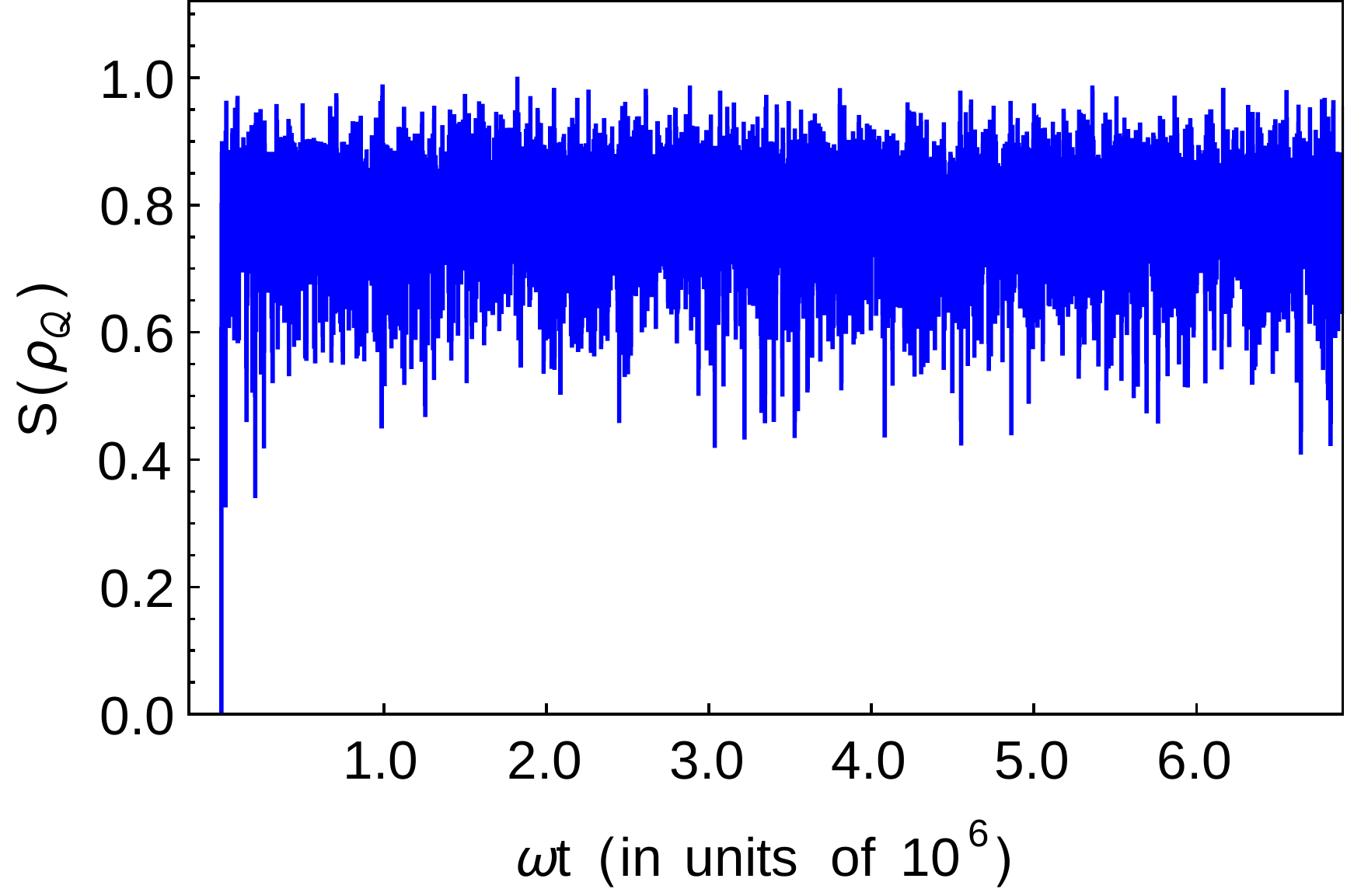}}\\
\caption{For the \textit{factorized initial state} (\ref{state-t0}) with $\mathrm{c}=0$ the time evolution of the entropy for the  $s=1$ case is 
studied. {\sf (a)}: For the strong coupling regime $(\widetilde{\lambda}=0.005)$ the parametric values are as follows:  $\Delta =0.16,\,\mathfrak{z}=0.1051,\,\alpha =3,\,r=0.2$. To explore the near-recurrence of the initial state during the evolution process we compute the Hilbert-Schmidt distances between the initial 
state at $t_{0}= 0$ and the states at 
$t_{\mathrm{A}}=1.571170\times 10^6,  t_{\mathrm{B}}=3.142120\times 10^6, t_{\mathrm{C}}=4.712820\times 10^6, t_{\mathrm{L}}=6.284030 \times 10^{6}$. Here and elsewhere all times are specified in the scale $\omega^{-1}$, and for all numerical work we use the unit 
$\omega = 1$. The relevant distances read
$\mathrm{d}_{HS}|_{t_{\mathrm{A}}}=0.526507,\,\mathrm{d}_{HS}|_{t_{\mathrm{B}}}=0.951675,\,\mathrm{d}_{HS}|_{t_{\mathrm{C}}}=0.514655,\,
\mathrm{d}_{HS}|_{t_{\mathrm{L}}}=0.009092$. Thereby it is manifest that the system achieves near-reproduction of its initial state at 
$t_{\mathrm{L}} \equiv \mathrm {T_{rev}}$. The quasiperiod  (\ref{quasi-period}) of the time evolution of the entropy for the present set of parameters equals $1.570816 \times 10^{6}$, which, very accurately, may be identified with $t_{\mathrm{A}}$. The full revival time 
(\ref{T-rev}) now corresponds to $\mathfrak{n} = 4$. The kitten states realized at $\lo t_{\mathrm{D}}, t_{\mathrm{E}}, t_{\mathrm{F}}, t_{\mathrm{G}}\rc$ are studied in Fig. \ref{kitten-1}. {\sf (b)}: For the ultrastrong coupling regime we chose $\widetilde{\lambda}= 0.2$ while all other parameters 
remain identical to those in diagram {\sf (a)}. Large coupling leads to the generation of many interaction modes and their harmonics. As the phase relationships between these modes are randomized, the quasiperiodicity of the system disappears leaving a stabilized value of  entropy around which stochastic fluctuations develop. }
\label{entropy-1-tL}
\end{center}
\end{figure}

\begin{figure}
\begin{center}
\captionsetup[subfigure]{labelfont={sf}} 
\subfloat[]{\includegraphics[scale=0.4]{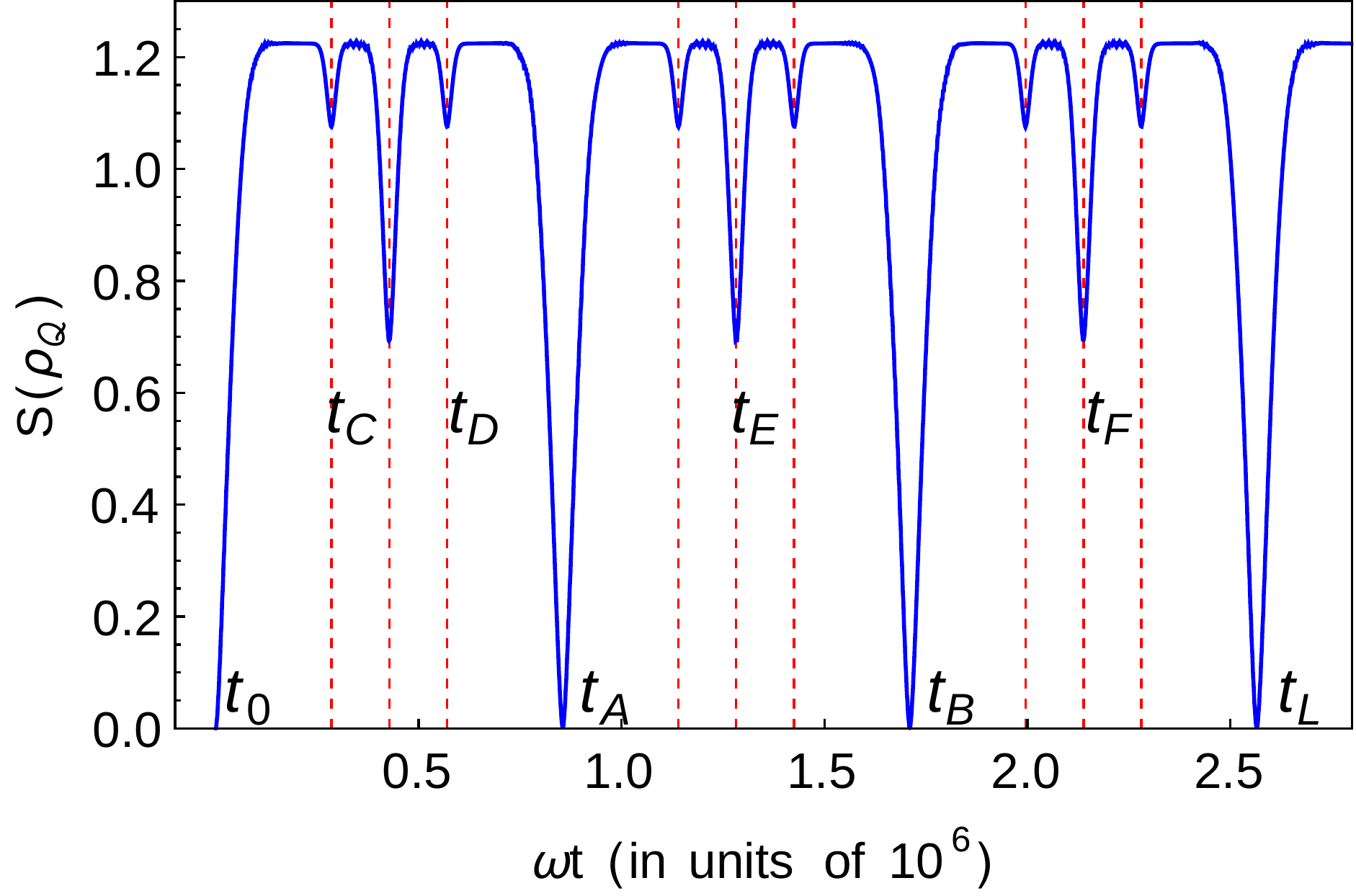}}
\hspace*{0.4cm}
\subfloat[]{\includegraphics[scale=0.4]{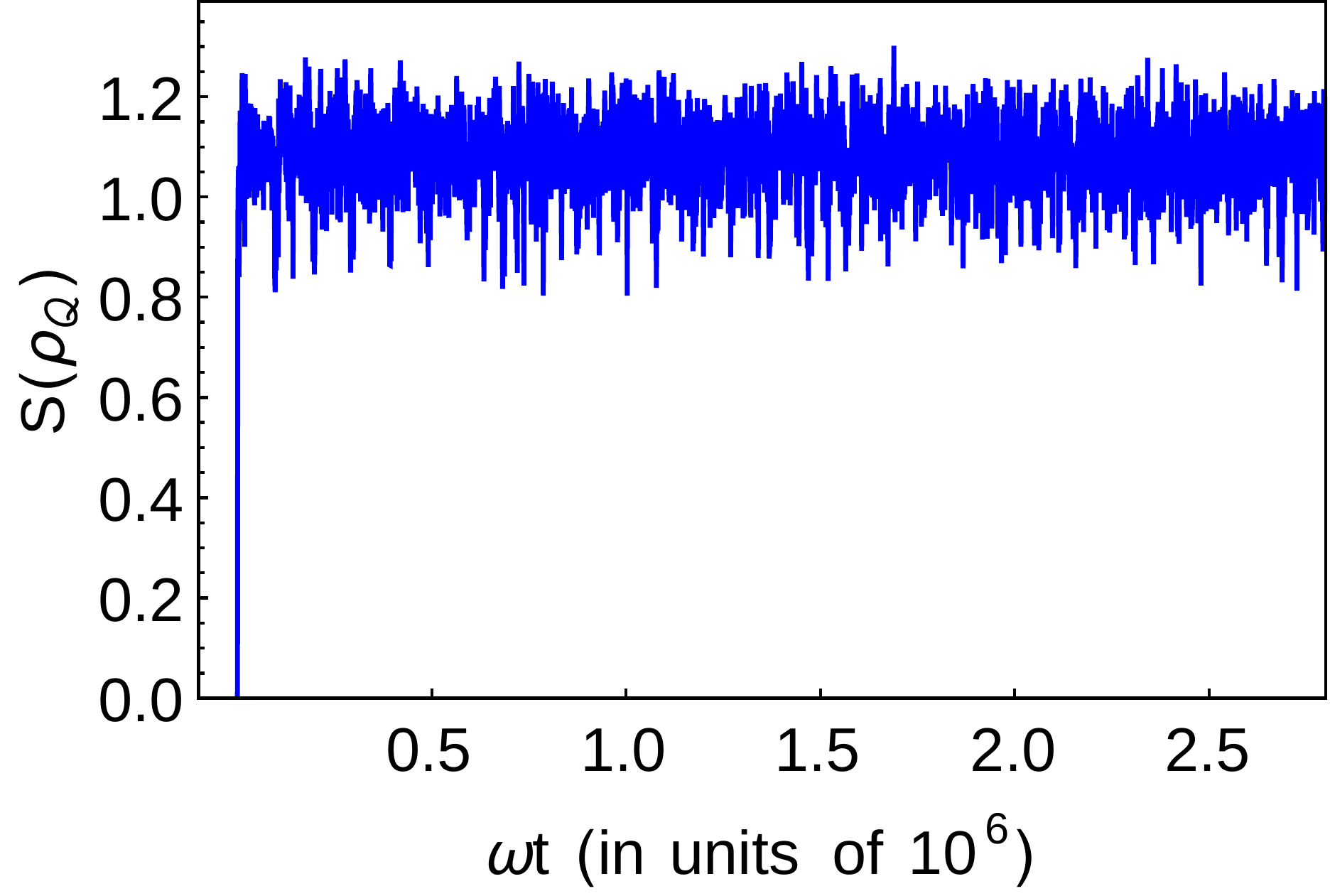}}\\
\caption{The time evolution of the entropy for the  $s=\tfrac{3}{2}$ case is produced for the \textit{factorized initial state} (\ref{state-t0}) with $\mathrm{c}=0$. {\sf (a)}: For the quasiperiodic evolution the chosen parametric values are $\Delta =0.15,\,\mathfrak{z}=0.1051,\,\alpha =3,\,r=0.2$, and the coupling constant equals $\widetilde{\lambda}=0.007$. The Hilbert-Schmidt distance is studied to infer the closeness of the evolving state with its initial ($t_{0}= 0$) counterpart. At times $t_{\mathrm{A}}=0.854819\times10^6, t_{\mathrm{B}}=1.710040\times10^6, t_{\mathrm{L}}=2.564985 \times 10^{6}$ the state achieves zero entropy configuration.
The Hilbert-Schmidt distances between the initial state and the qudit states at these times $\mathrm{d}_{HS}|_{t_{\mathrm{A}}}=0.917149,\,\mathrm{d}_{HS}|_{t_{\mathrm{B}}}=0.924257,\,\mathrm{d}_{HS}|_{t_{\mathrm{L}}}=0.019044$ suggest near-duplication of the initial state at time $t_{\mathrm{L}}  \equiv \mathrm {T_{rev}}$.
For the current set of parameters the quasiperiod  (\ref{quasi-period}) of the near-null value of the entropy stands as 
$0.854876 \times 10^{6}$, which closely  equals $t_{\mathrm{A}}$. The corresponding full revival time (\ref{T-rev}) is given by
  $\mathfrak{n} = 3$.
The  observed transitory kitten states, say, at times  $\lo t_{\mathrm{C}}, t_{\mathrm{D}}, t_{\mathrm{E}}, t_{\mathrm{F}}\rc$, when the entropy reduces to nonzero local minimal values, are specified in Fig. \ref{kitten-3/2}. {\sf (b)}: We observe the evolution of the entropy by increasing the coupling to the ultrastrong regime $\widetilde{\lambda}=0.2$, while retaining all other parameters equal to their values considered in {\sf (a)}. As a consequence of creation of large number of 
modes, the quasiperiodicity of the evolving state is lost. Random fluctuations occur around a steady state of entropy.}
\label{entropy-3/2-tL}
\end{center}
\end{figure}

In the strong coupling regime here we explore
 the time dependence of the correlation function (\ref{def_corr}) to distinguish the temporal spin kitten states 
from the remainder characterized by high entropy configurations evident in 
Figs.  \ref{entropy-1-tL}{\sf (a)} and \ref{entropy-3/2-tL}{\sf (a)}.
The qudit density matrices for  $s=1$ and $s=\frac{3}{2}$ examples given in  (\ref{density-1}) and 
(\ref{DenMtrx-3/2-t}), respectively, now furnish the corresponding time dependent spin correlation function (\ref{def_corr}). The  function $g_{s}(t)$ for the  examples considered here may be expressed via the  ensemble averages listed below:
\bea
\braket{S_{+}S_{-}}_{s=1} &=& 2 \sum_{n=0}^{\infty}
 \left( \lvert  \mathcal{B}_{+,n}^{(1)}(t) \rvert ^{2} + \lvert  \mathcal{B}_{0,n}^{(1)}(t) \rvert ^{2} \right), \quad 
 \braket{S_{+}^{2}S_{-}^{2}}_{s=1} = 4 \sum_{n=0}^{\infty}  \lvert  \mathcal{B}_{+,n}^{(1)}(t) \rvert ^{2}, \\
\braket{S_{+}S_{-}}_{s=\frac{3}{2}} &=&  \sum_{n=0}^{\infty} \left(
 3 \left(  \mathrm{B}_{n,n}^{(2,2)}(t)  +  \mathrm{B}_{n,n}^{(-1,-1)}(t)  \right) +
 4\mathrm{B}_{n,n}^{(1,1)}(t)  \right), \nn \\
\braket{S_{+}^{2}S_{-}^{2}}_{s=\frac{3}{2}} &=& 12 \sum_{n=0}^{\infty} 
 \left(  \mathrm{B}_{n,n}^{(2,2)}(t)  +  \mathrm{B}_{n,n}^{(1,1)}(t)  \right).
 \label{2nd-spin-corr}
\eea

\begin{figure}
\begin{center}
\captionsetup[subfigure]{labelfont={sf}} 
\hspace*{-1cm}
\subfloat[]{\includegraphics[scale=0.4]{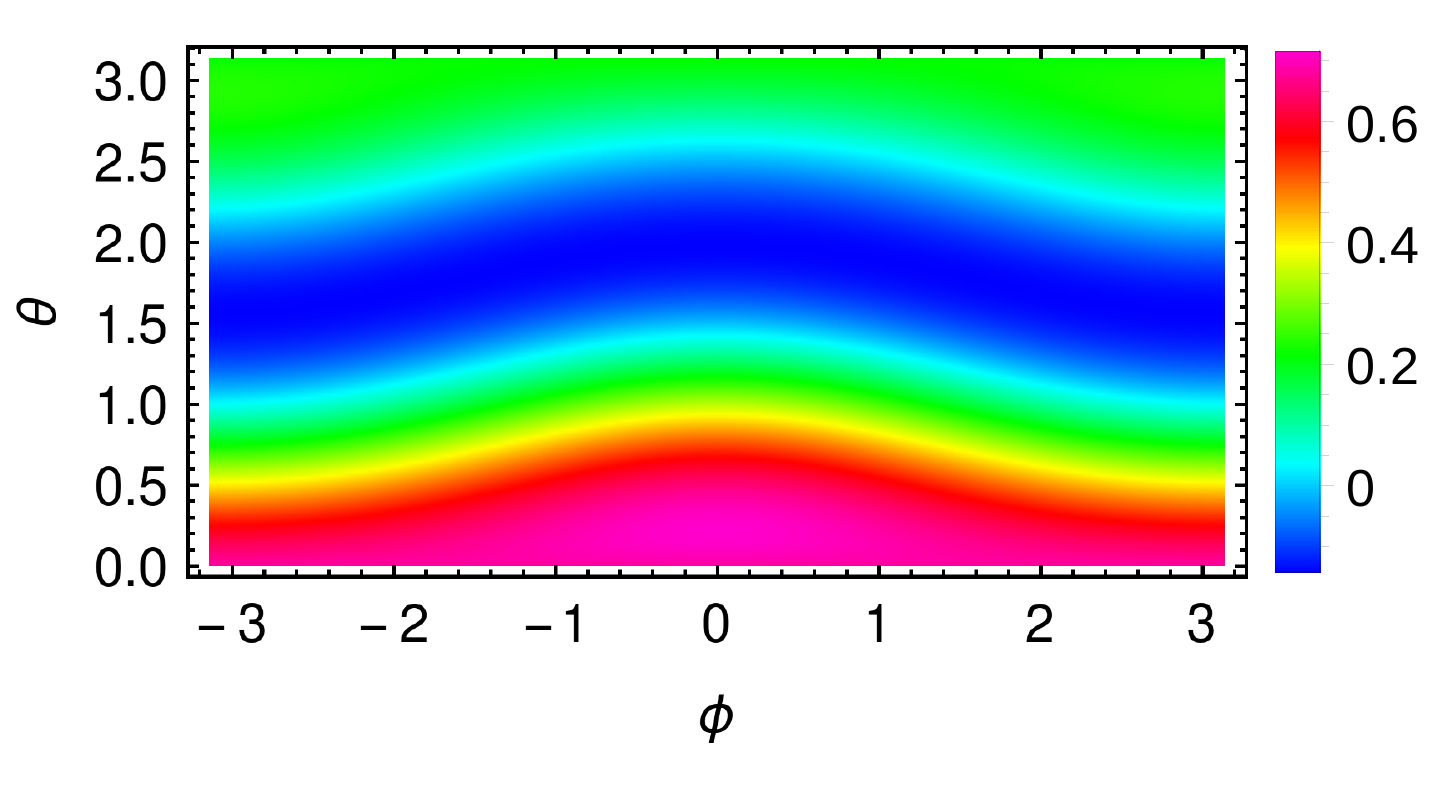}}
\subfloat[]{\includegraphics[scale=0.4]{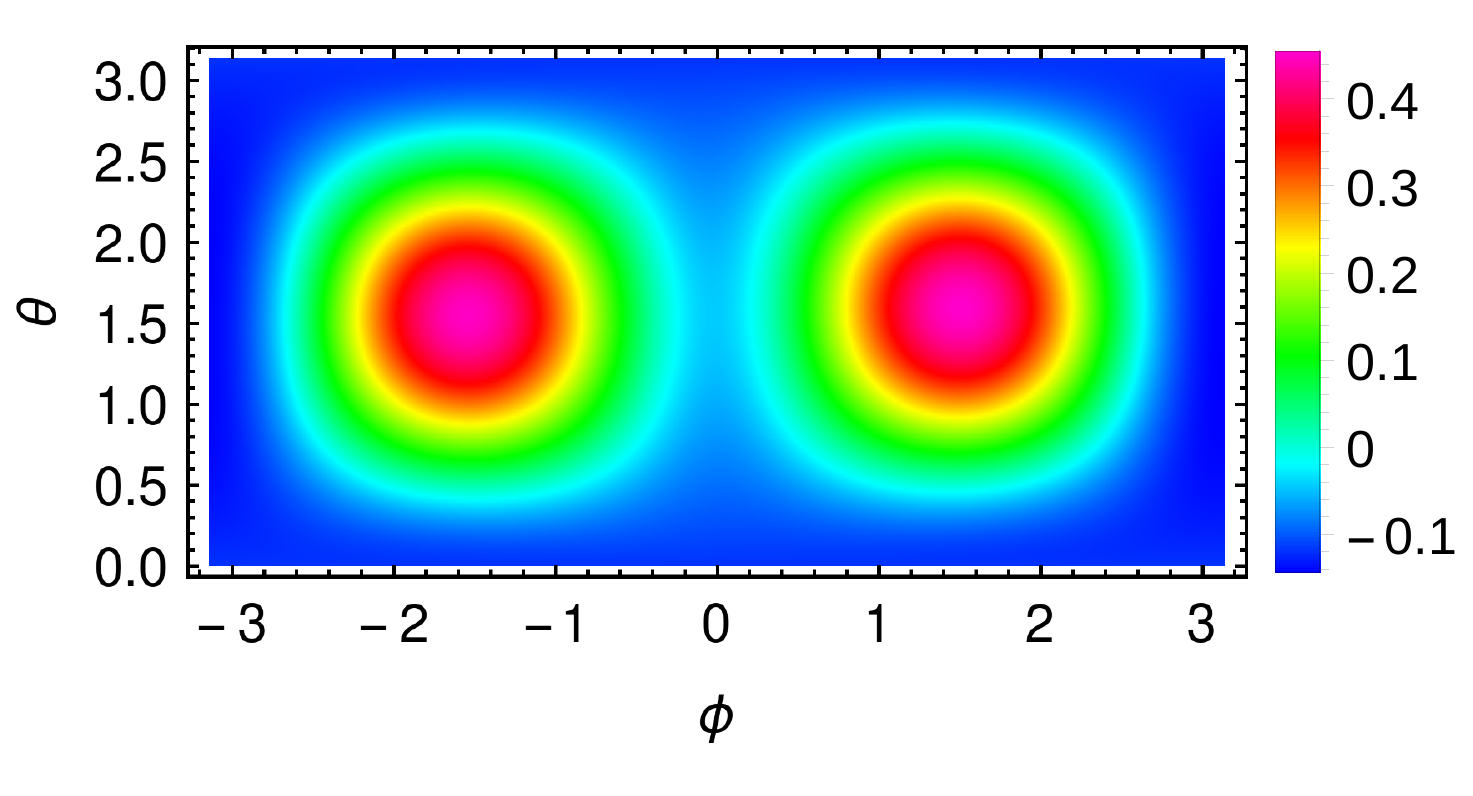}}
\subfloat[]{\includegraphics[scale=0.4]{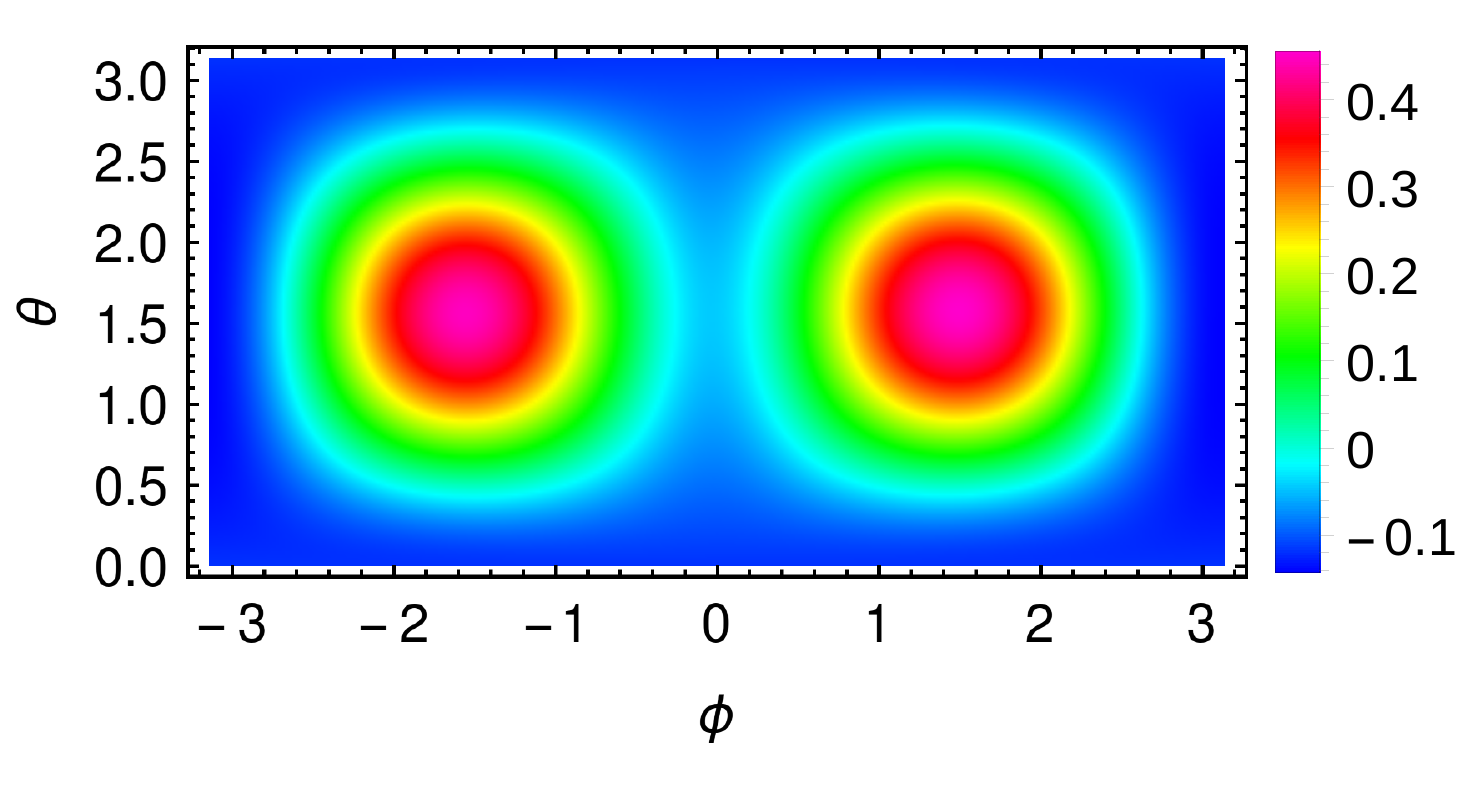}}\\
\hspace*{-1cm}
\subfloat[]{\includegraphics[scale=0.4]{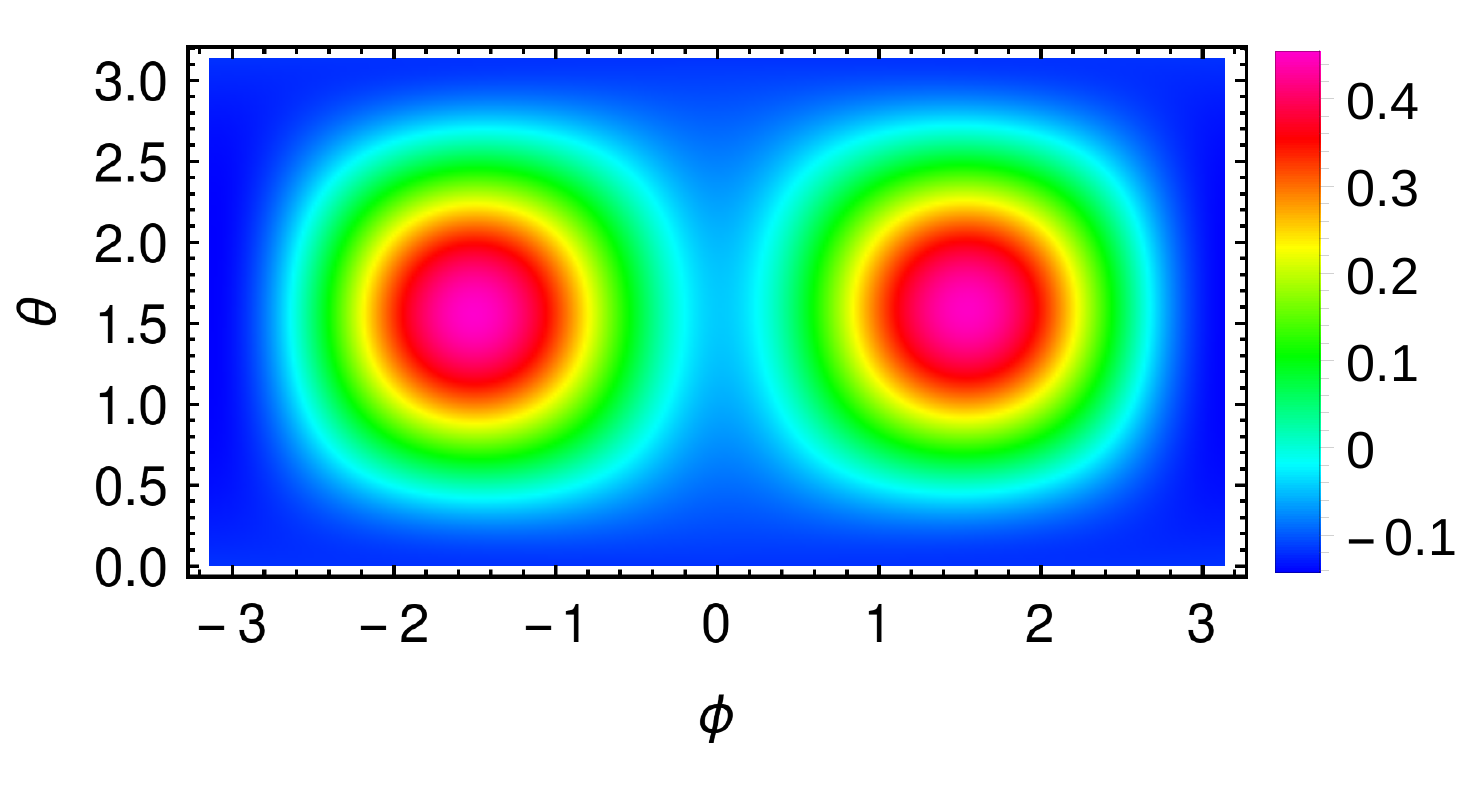}}
\subfloat[]{\includegraphics[scale=0.4]{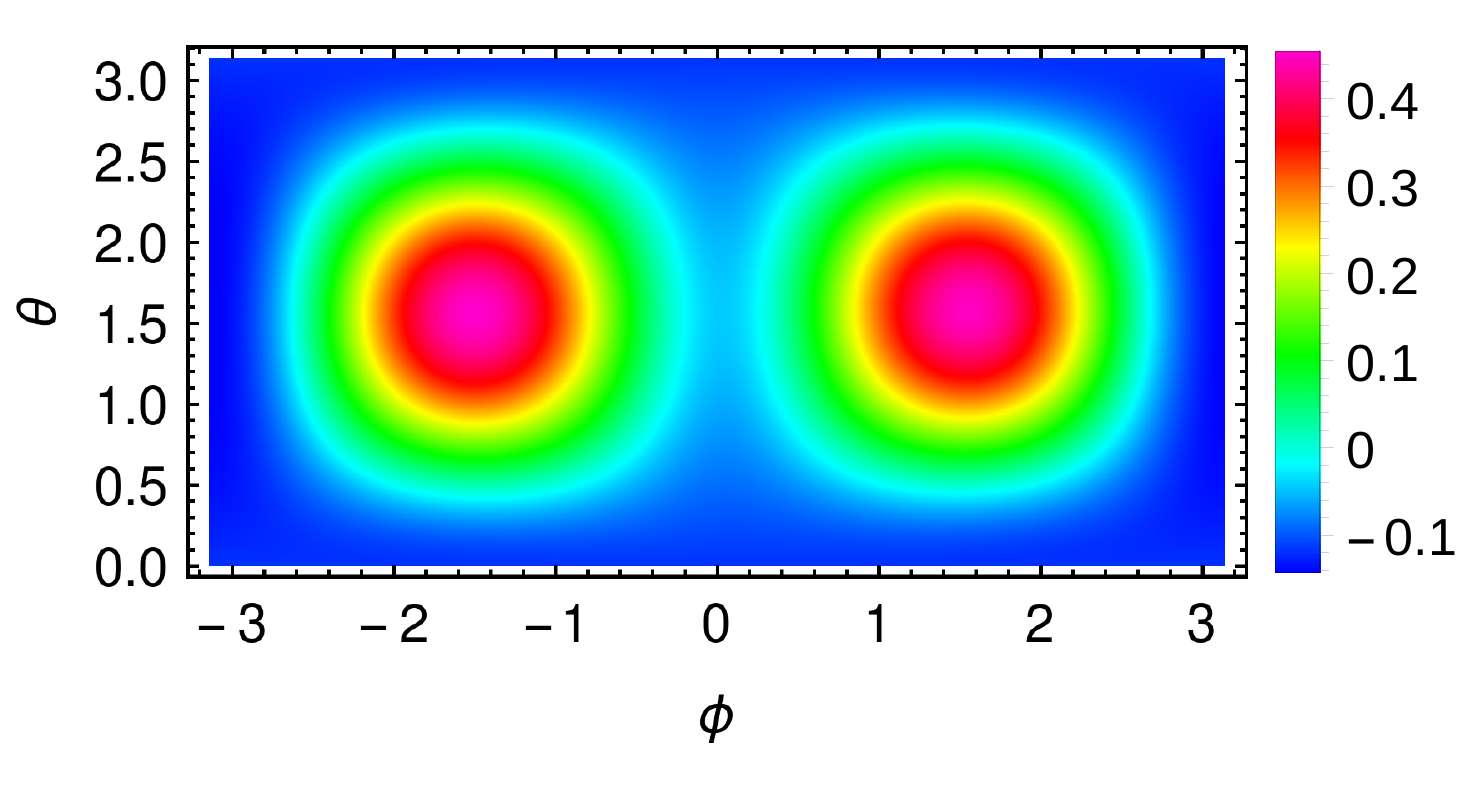}}
\subfloat[]{\includegraphics[scale=0.4]{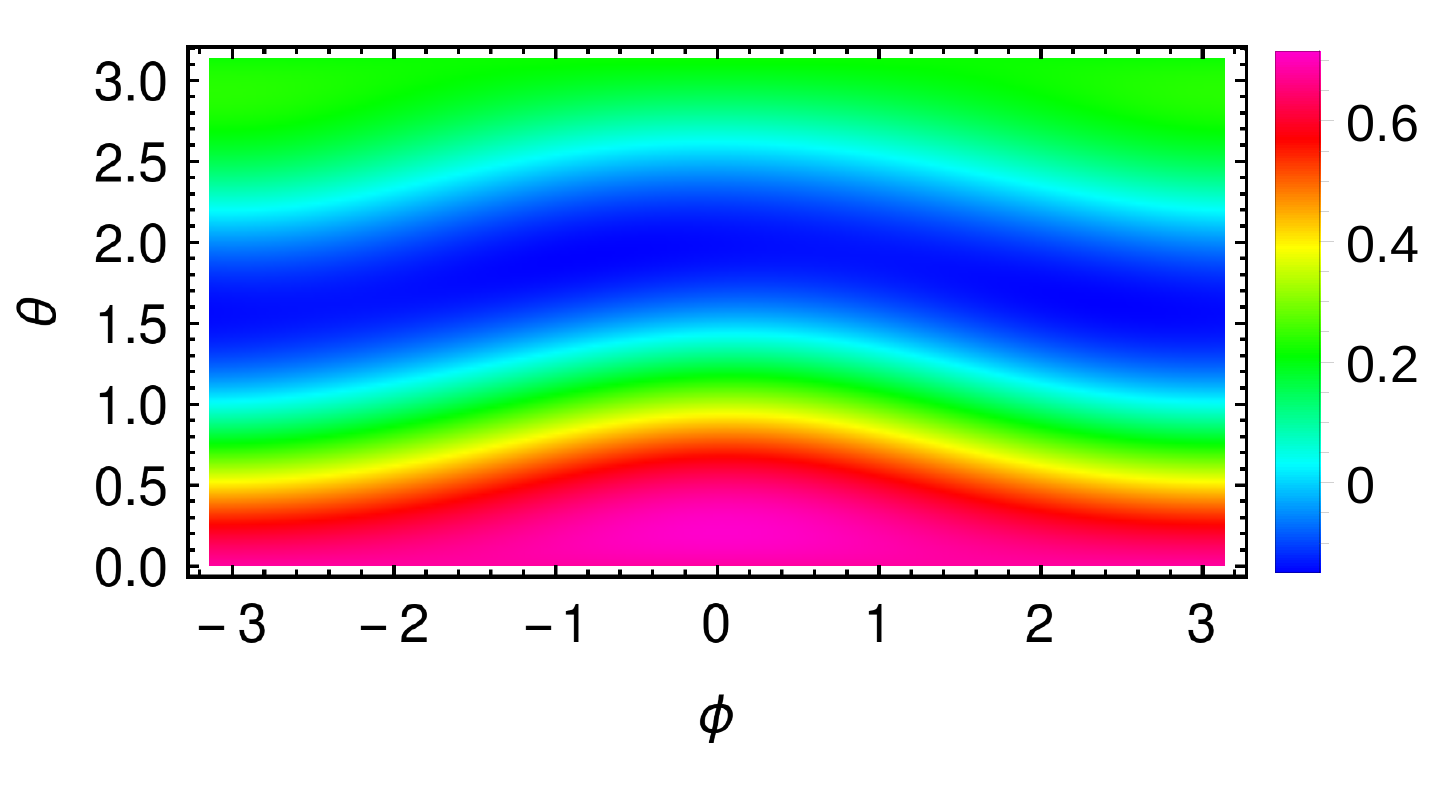}}
\caption{We display the construction of the spin diagonal $\mathrm{P}_{\mathcal{Q}}(\theta, \phi)$-representation 
for the $s=1$ example studied in 
Fig. \ref{entropy-1-tL} at the initial and other times marked therein. The diagram {\sf (a-f)} refer, successively,
to the times  $t_{0}=0, t_{\mathrm{D}}=0.785496 \times 10^{6}, t_{\mathrm{E}}=2.355998\times 10^{6}, t_{\mathrm{F}}=3.927010 \times 10^{6}, t_{\mathrm{G}}=5.498003 \times 10^{6},  t_{\mathrm{L}}=6.284030 \times 10^{6}$. The parametric choices here are identical to those in Fig. \ref{entropy-1-tL}{\sf (a)}. The diagrams {\sf (a)} and {\sf (f)} indicate that a close repetition of the initial state occurs at $t_{\mathrm{L}}$, when an almost complete revival of the system is manifest. The spin kitten states, evident in the illustrations {\sf (b-e)}, are realized  at respective times $\lo t_{\mathrm{D}}, t_{\mathrm{E}}, t_{\mathrm{F}}, t_{\mathrm{G}}\rc$ specified above. The formation of spin kitten states may be regarded as fractional revivals in the hybrid system.} 
\label{kitten-1}
\end{center}
\end{figure}

\begin{figure}
\begin{center}
\captionsetup[subfigure]{labelfont={sf}} 
\hspace*{-1cm}
\subfloat[]{\includegraphics[scale=0.4]{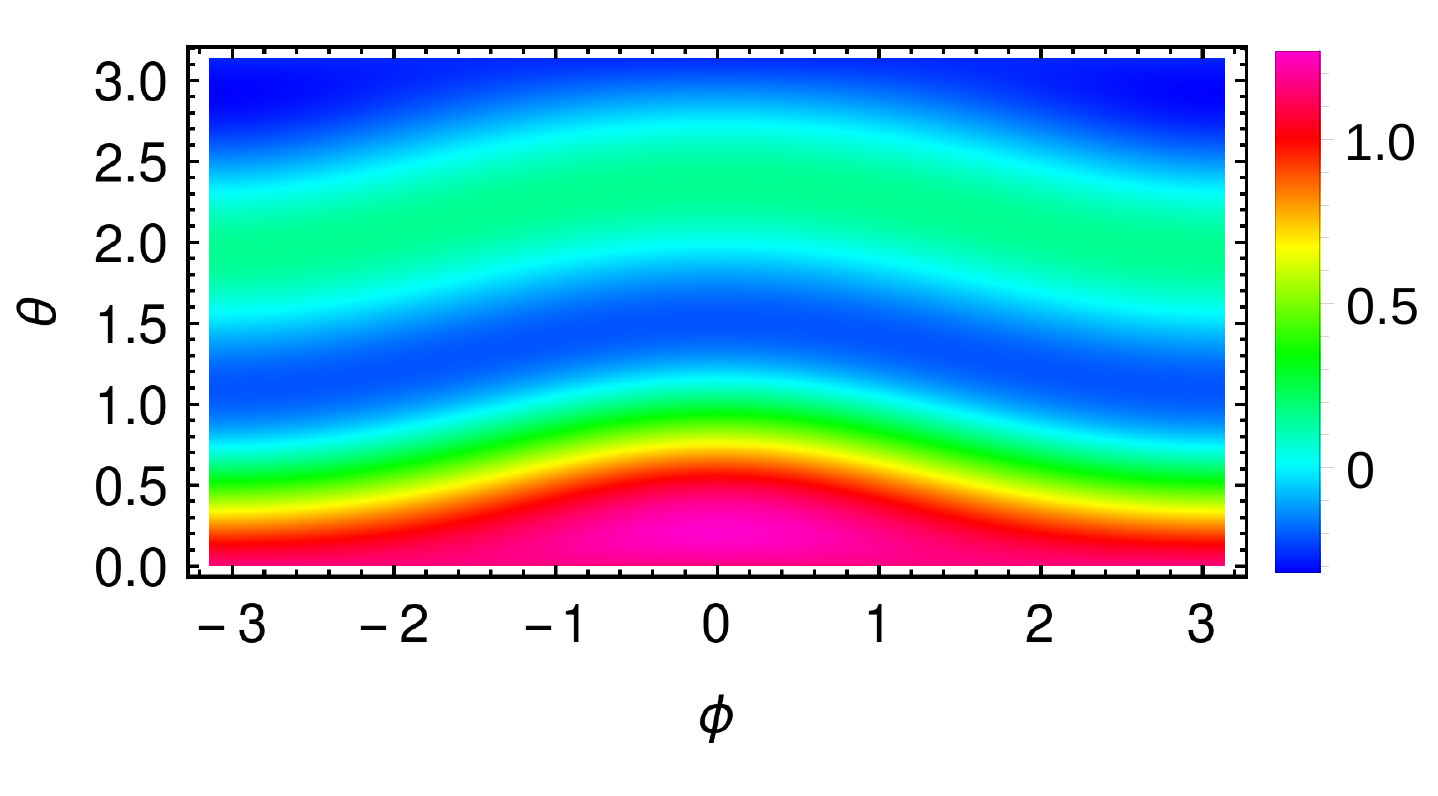}}
\subfloat[]{\includegraphics[scale=0.4]{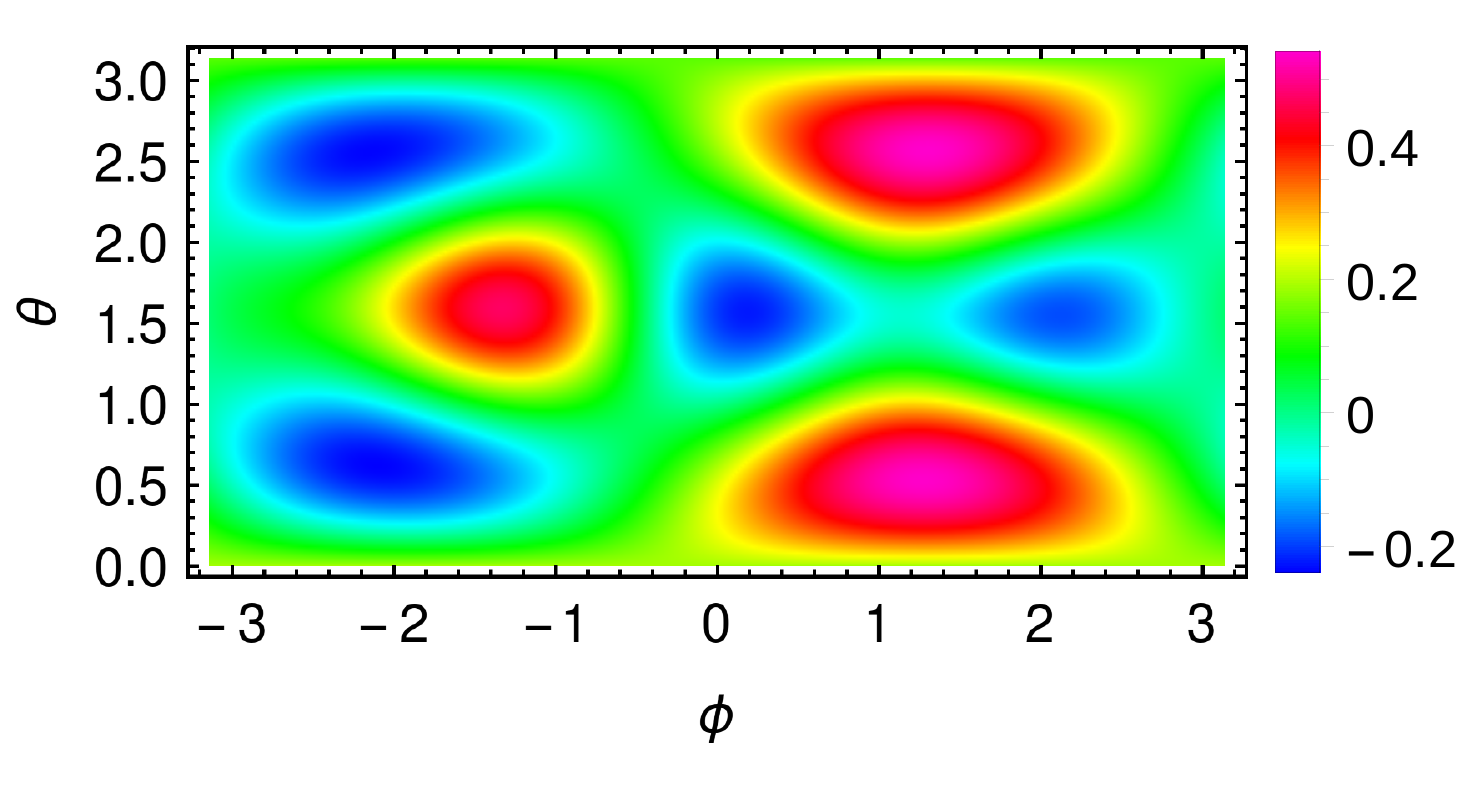}}
\subfloat[]{\includegraphics[scale=0.4]{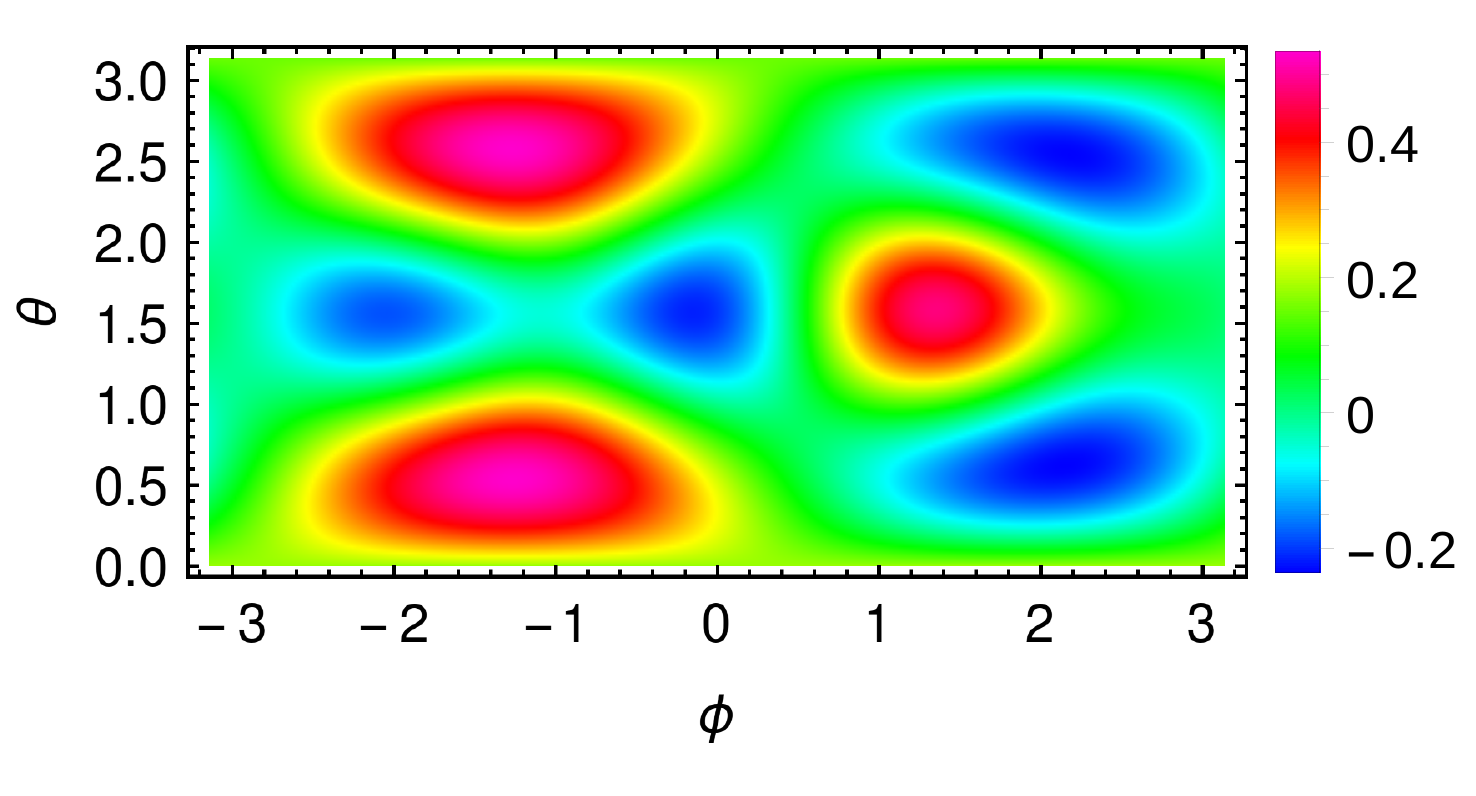}}\\
\hspace*{-1cm}
\subfloat[]{\includegraphics[scale=0.4]{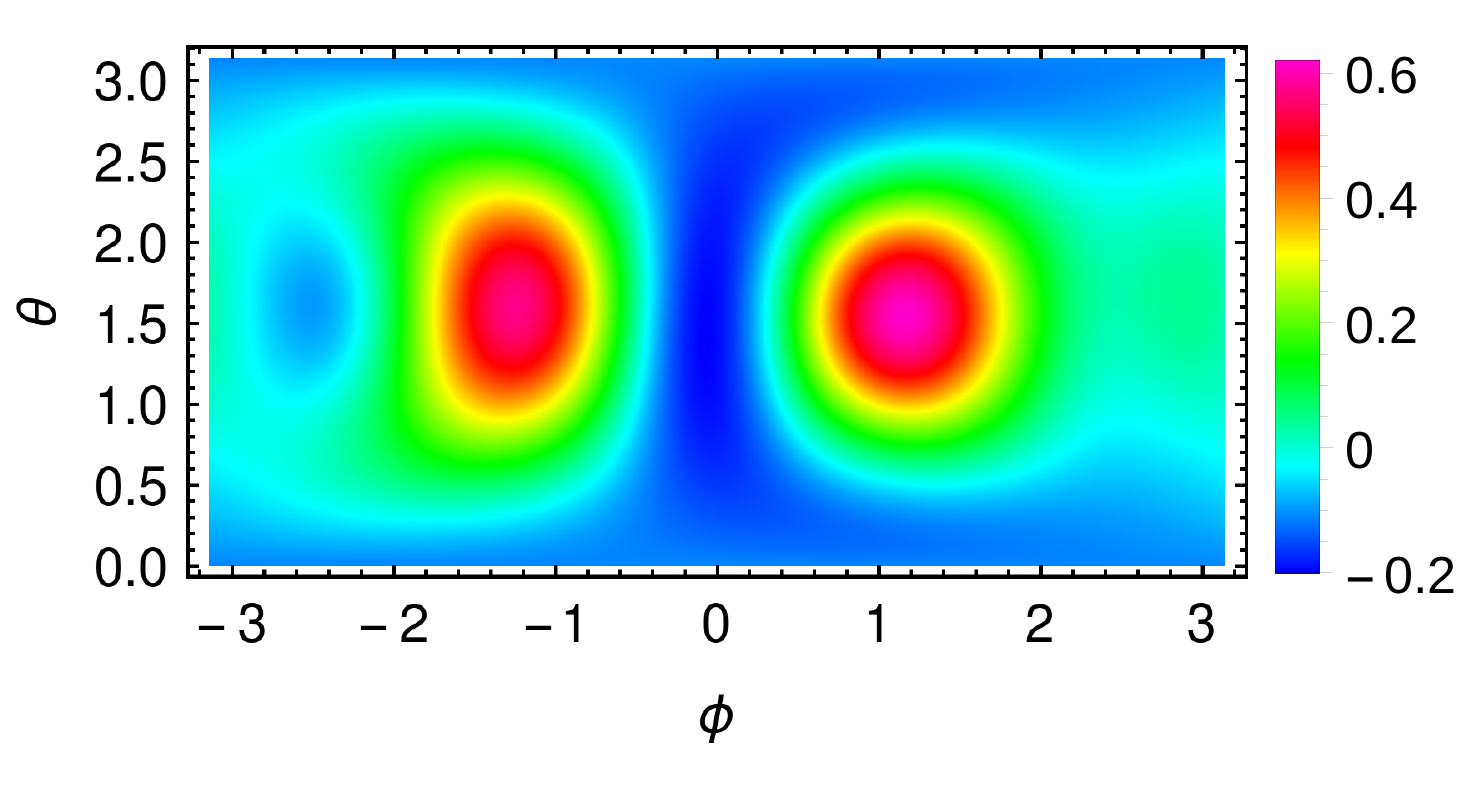}}
\subfloat[]{\includegraphics[scale=0.4]{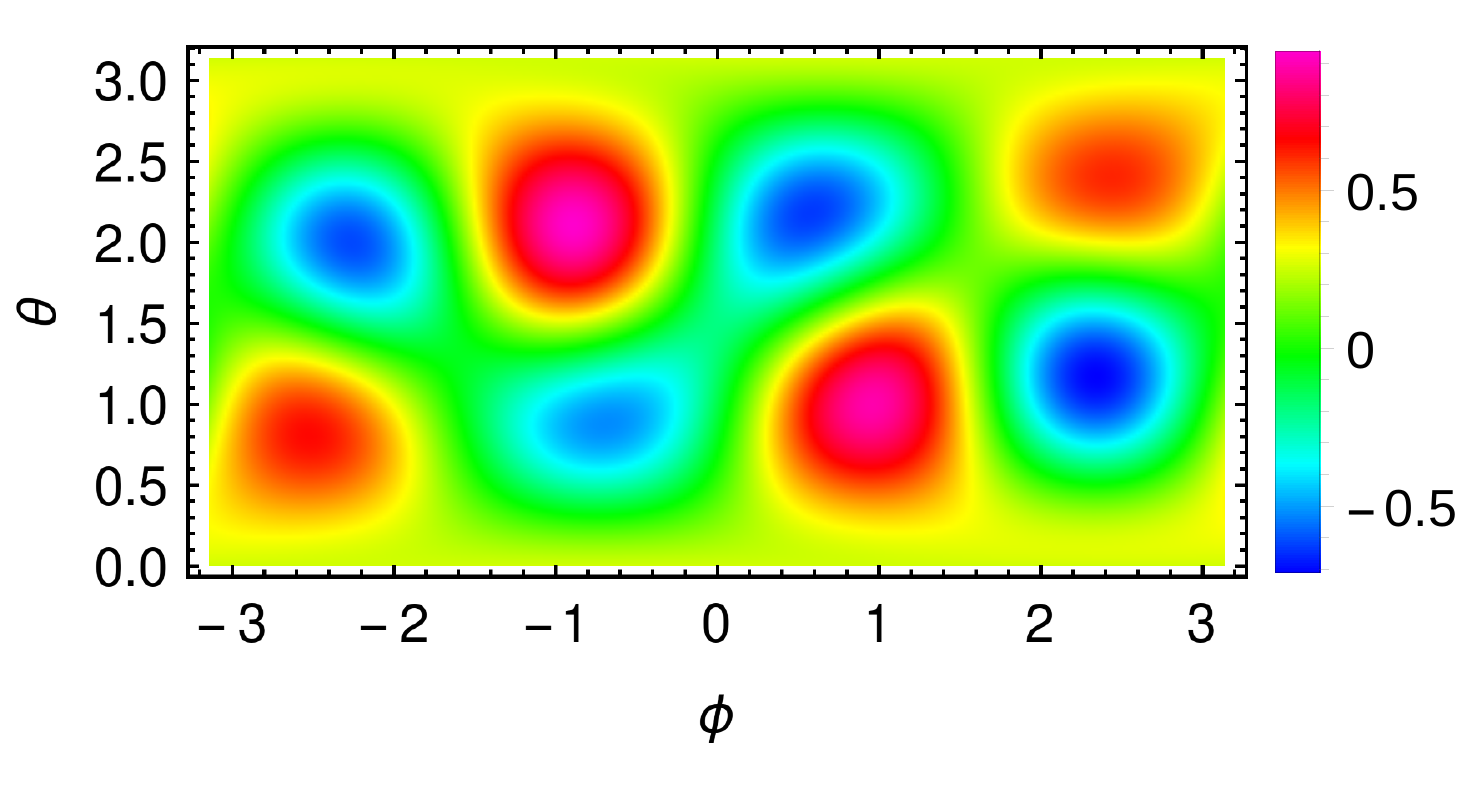}}
\subfloat[]{\includegraphics[scale=0.4]{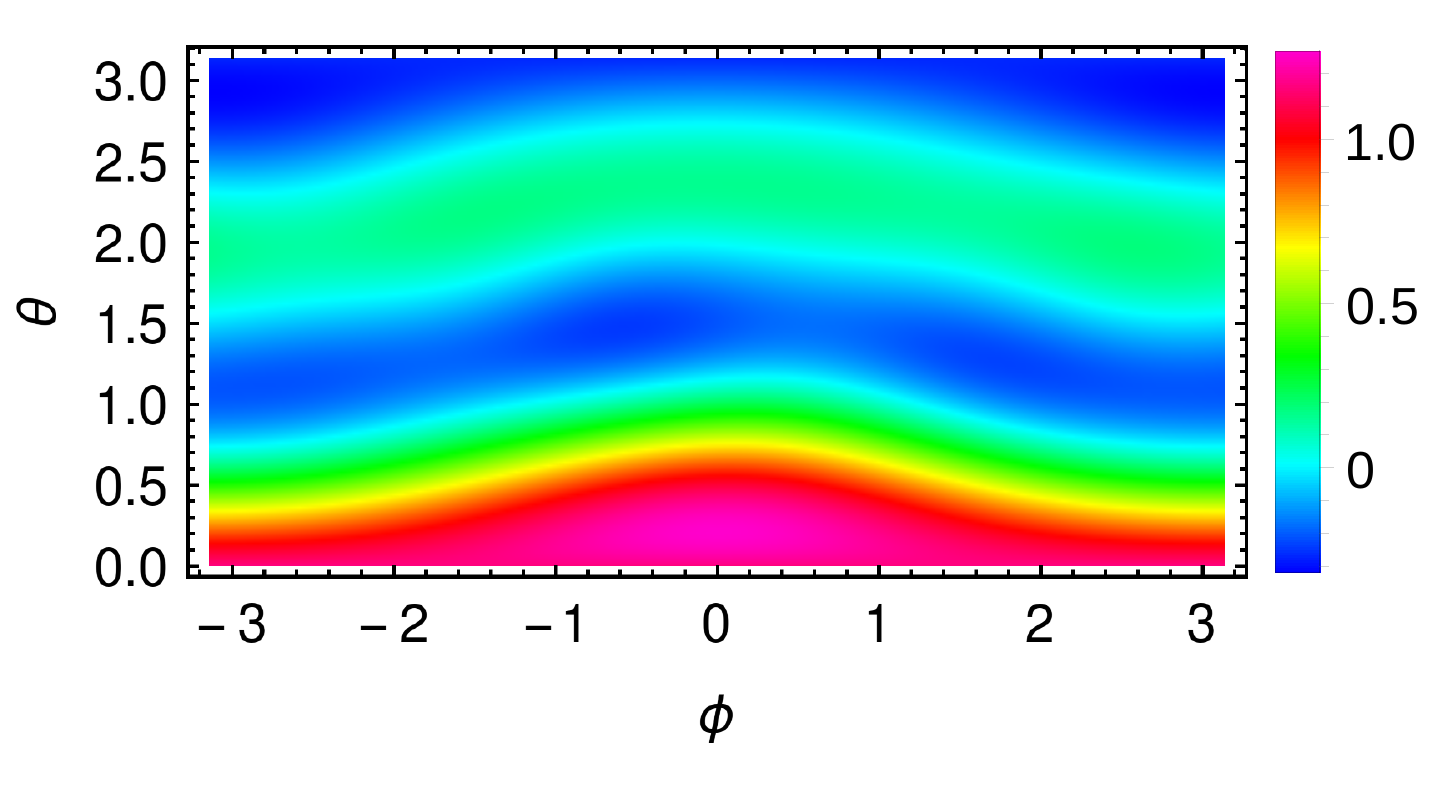}}
\caption{For the choice of the factorized initial state (\ref{state-t0}) with $\mathrm{c}=0$ we produce the  spin quasiprobability 
$\mathrm{P}_{\mathcal{Q}}(\theta, \phi)$-representation for the $s=\tfrac{3}{2}$ case at various times  considered in
Fig. \ref{entropy-3/2-tL}{\sf(a)}. The
parametric choices here are identical to those in Fig. \ref{entropy-3/2-tL}{\sf(a)}. The diagrams {\sf (a, f)} refer to the initial time  $t_{0}=0$ and the time $t_{\mathrm{L}}=2.564985 \times 10^6$, when the system returns close to the initial state. The $3$-kitten states arise at times $t_{\mathrm{C}}=0.284981 \times 10^{6}$ and, for instance, $t_{\mathrm{D}}=0.570000 \times 10^{6}$, where the locally minimum entropy configurations are produced. These are illustrated in diagrams {\sf (b, c)}, respectively. The $2$-kitten state  formed at $t_{\mathrm{E}} = 1.282021 \times 10^{6}$ is depicted in diagram {\sf (d)}. Finally we observe $4$-kitten state at time $t_{\mathrm{F}}=     2.137999  \times 10^{6}$. This is given in diagram  {\sf (e)}. In contrast to the $s=1$ case, the larger size of the Hilbert space of the qudit $\big(s=\tfrac{3}{2}\big)$  allows more
quantum correlation to be present within the system. This, for instance, produces the $3$ and $4$-kitten states in the present example.
}
\label{kitten-3/2}
\end{center}
\end{figure}

\begin{figure}[h]
\begin{center}
\captionsetup[subfigure]{labelformat=empty}
\hspace*{-1cm}
\subfloat[{\sf (a$_{1}$)}]{\includegraphics[scale=0.37]{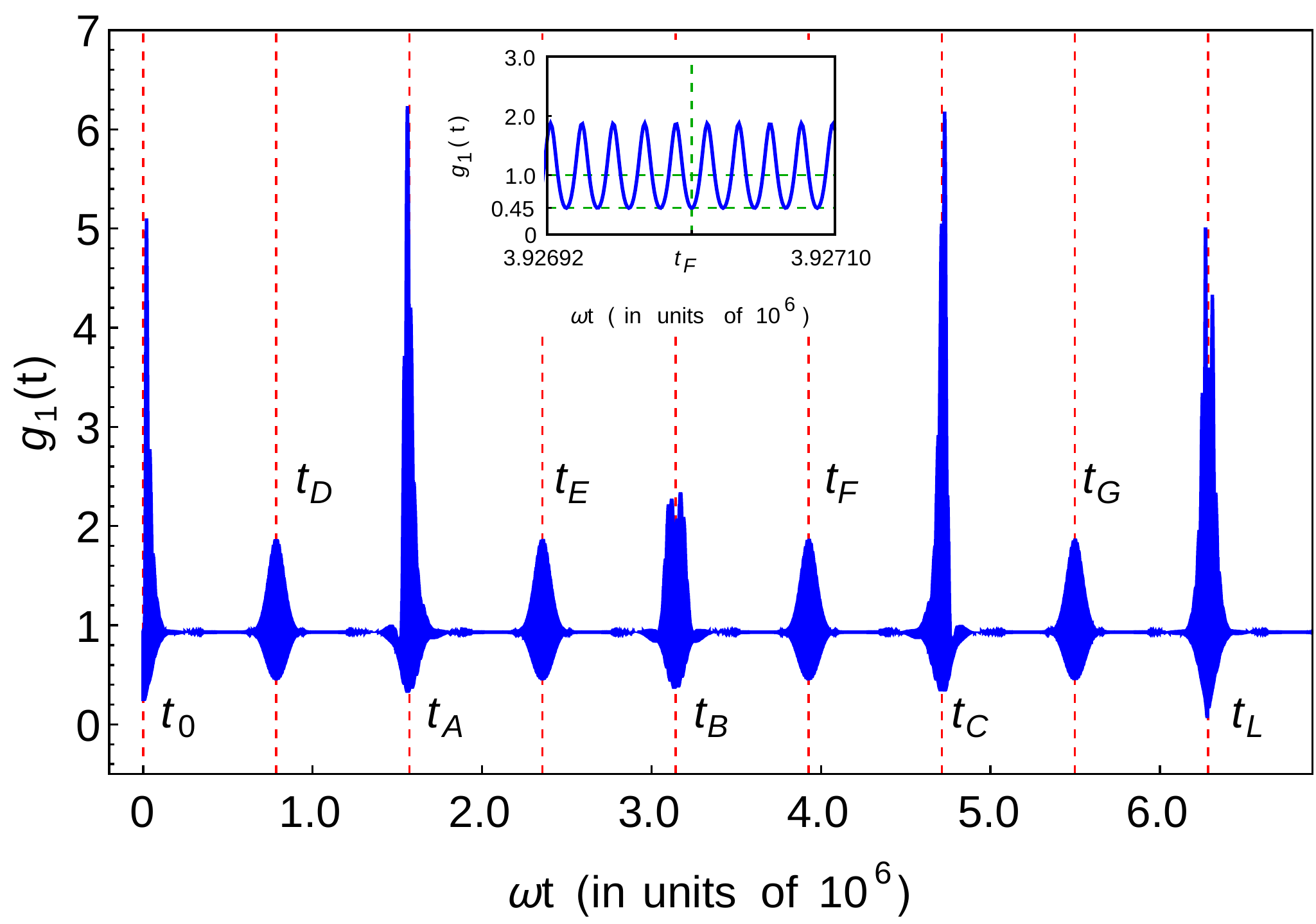}}
\hspace*{0.5cm}
\subfloat[{\sf (a$_{2}$)}]{\includegraphics[scale=0.38]{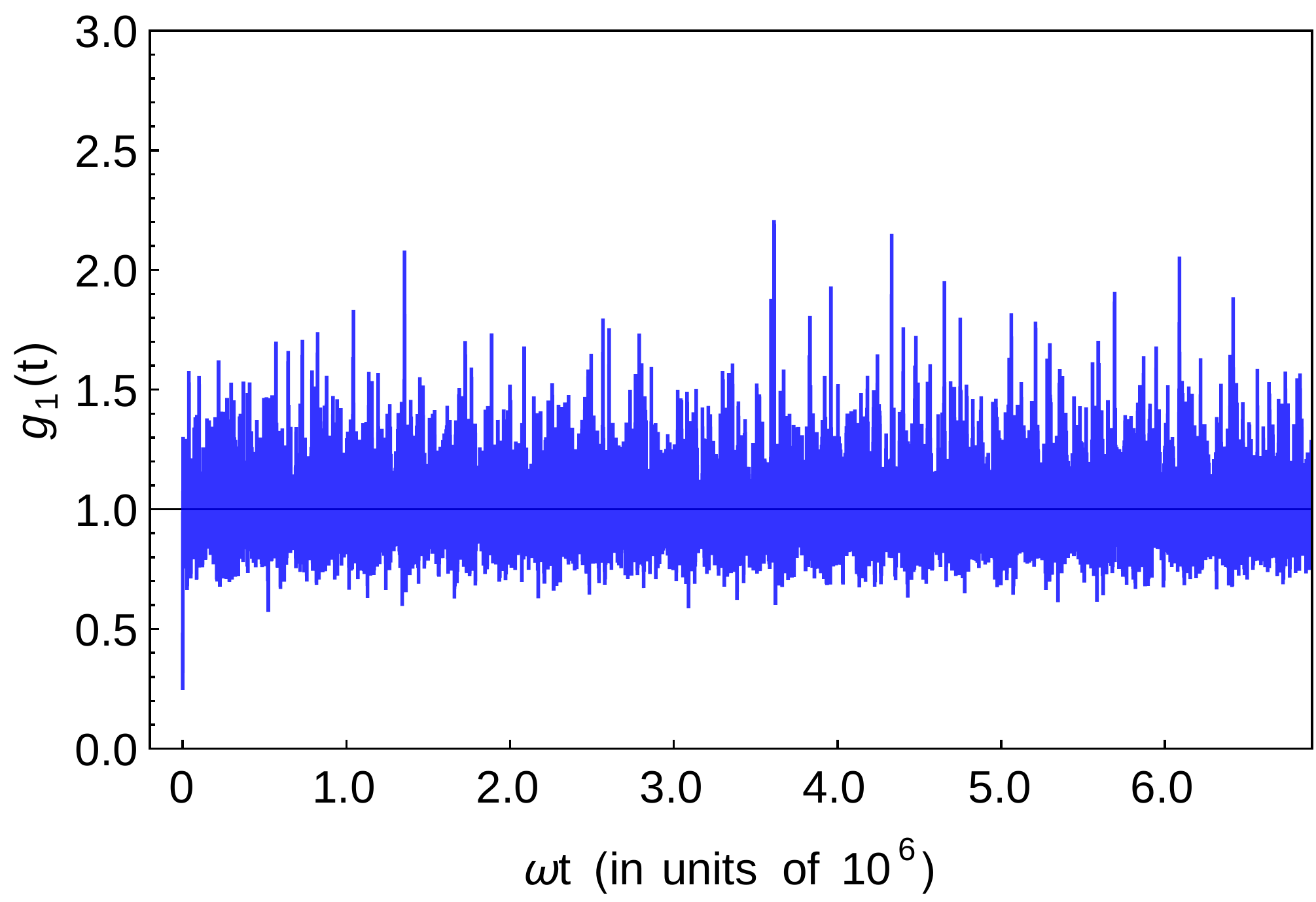}}\\
\hspace*{-1cm}
\subfloat[{\sf (b$_{1}$)}]{\includegraphics[scale=0.31]{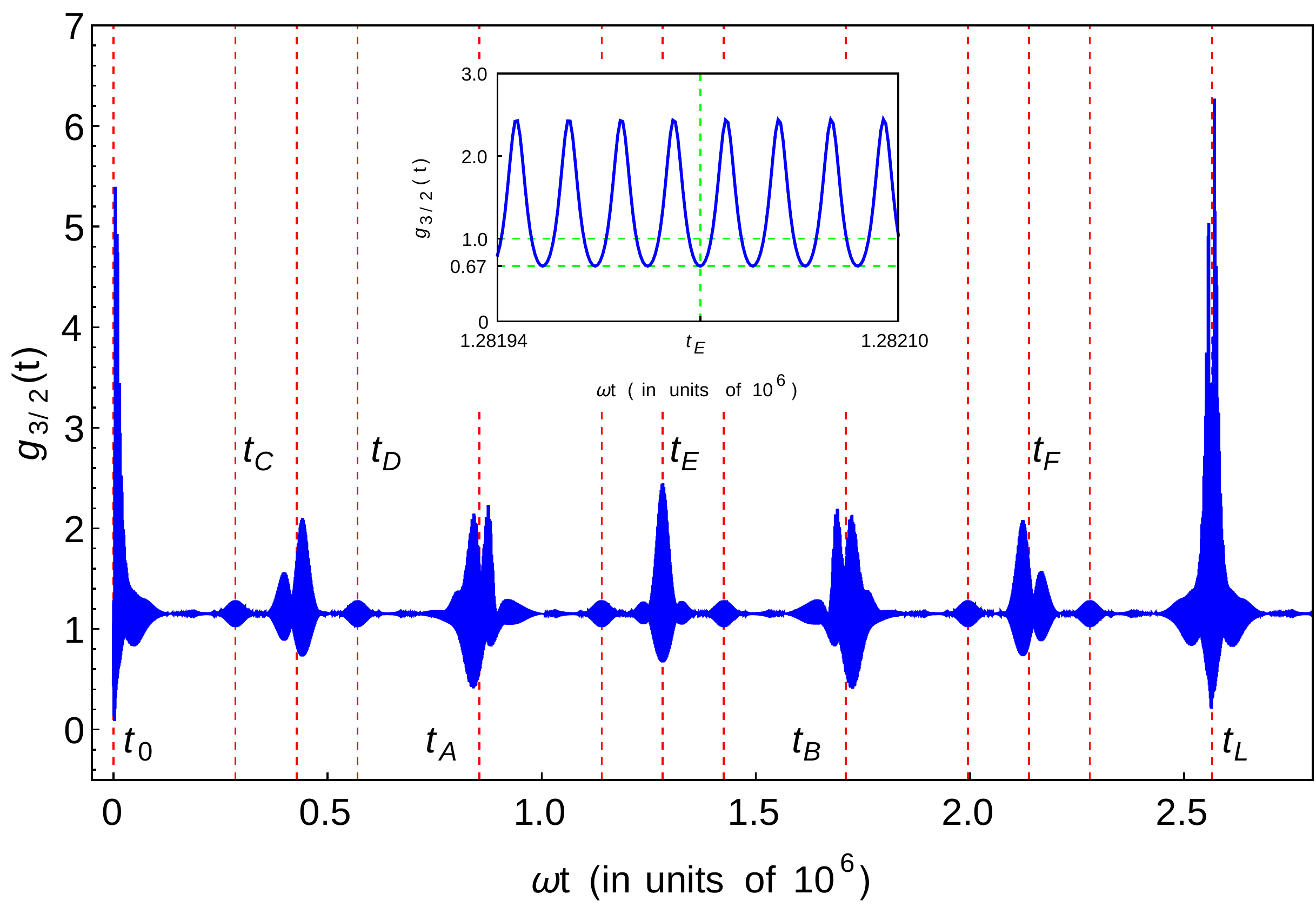}}
\hspace*{0.5cm}
\subfloat[{\sf (b$_{2}$)}]{\includegraphics[scale=0.51]{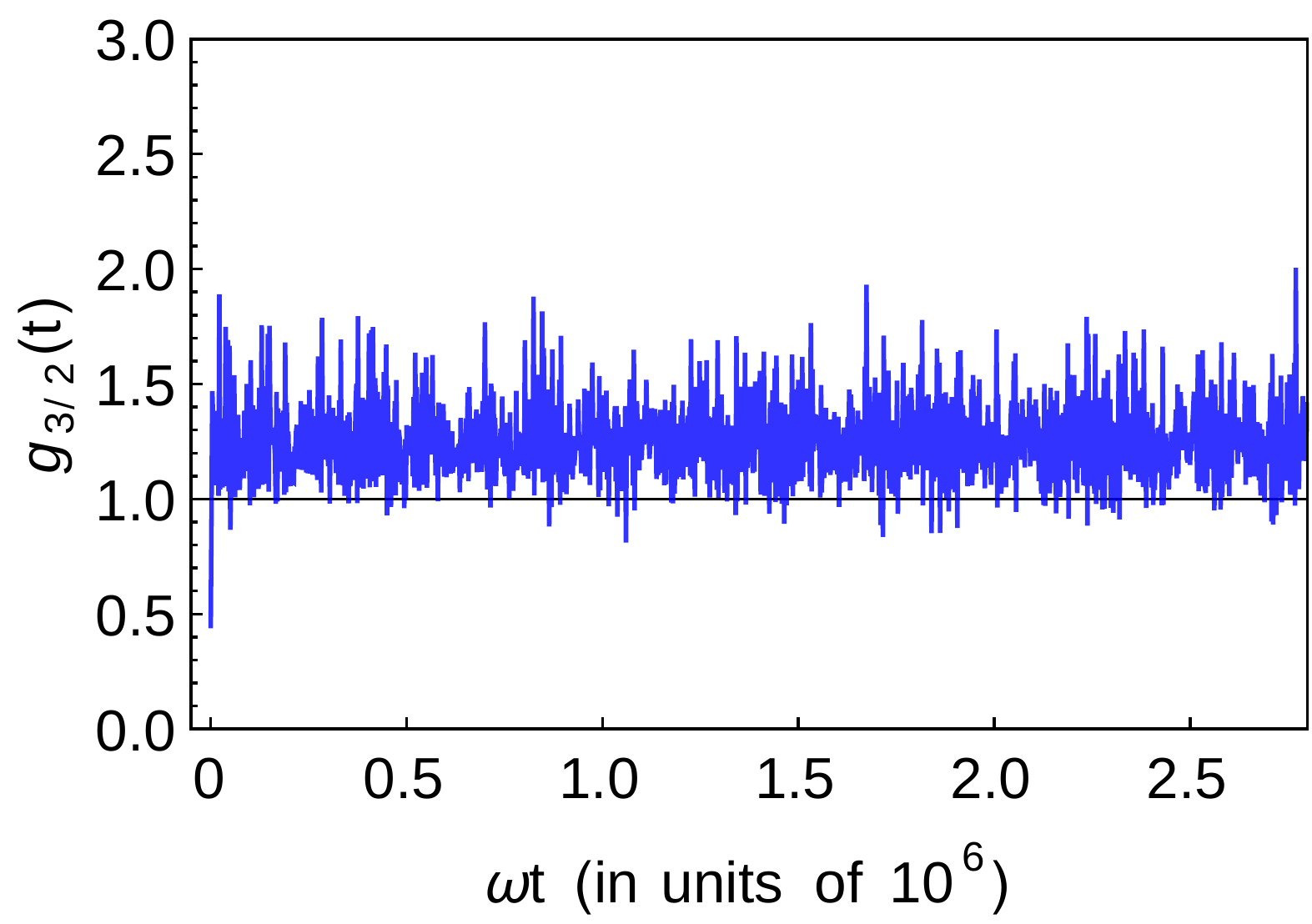}}
\caption{For the factorized initial state (\ref{state-t0}) with $\mathrm{c}=0$ the time evolution of the second order correlation function $g_{s}(t)$ is studied.  For the  $s=1$ case in the strong coupling regime considered in the diagram $\mathsf{(a_1)}$, 
 the coupling strength, other parameters as well as the marked  times of  revivals are taken to be identical to those in 
 Fig. \ref{entropy-1-tL}$\mathsf{(a)}$. Similarly the diagram  
$\mathsf{(b_1)}$ depicts the correlation function for the example $s=\frac{3}{2}$ where the coupling strength, other parametric 
values, and the 
times of total as well as fractional revivals are exactly same as those in Fig. \ref{entropy-3/2-tL}$\mathsf{(a)}$. Compared to the other revivals, the fluctuations in $g_{s}(t)$ as observed for $3$-kitten states at times $t_{\mathrm C}, t_{\mathrm D}$ in $\mathsf{(b_1)}$ are  lesser as the corresponding dips in the entropy (Fig.  \ref{entropy-3/2-tL}$\mathsf{(a)}$) are marginal. This diminishes the pure state component in the $3$-kitten density matrices. A comparison of the insets in Figs. $\mathsf{(a_1)}$ and $\mathsf{(b_1)}$ reveals that the condition of antibunching of the emitted photons $g_{s}(t)<1$ is more strongly satisfied for the lesser spin $s=1$ case. Diagrams $\mathsf{(a_2)}$ and 
$\mathsf{(b_2)}$ study the incoherent chaotic behavior of $g_{s}(t)$ at the ultrastrong coupling regime $\widetilde{\lambda}=0.2$. Other parametric values of $\mathsf{(a_1)/(b_1)}$ are retained in $\mathsf{(a_2)/(b_2)}$. Here also the validity of the antibunching condition occurs far more frequently for the lower spin  $s=1$ example than its higher spin analog. }
\label{fig_corr}
	\end{center}
	\end{figure}

\begin{figure}[h]
\begin{center}
\captionsetup[subfigure]{labelformat=empty}
\hspace*{0cm}
\subfloat[{\sf (a)}]{\includegraphics[scale=0.55]{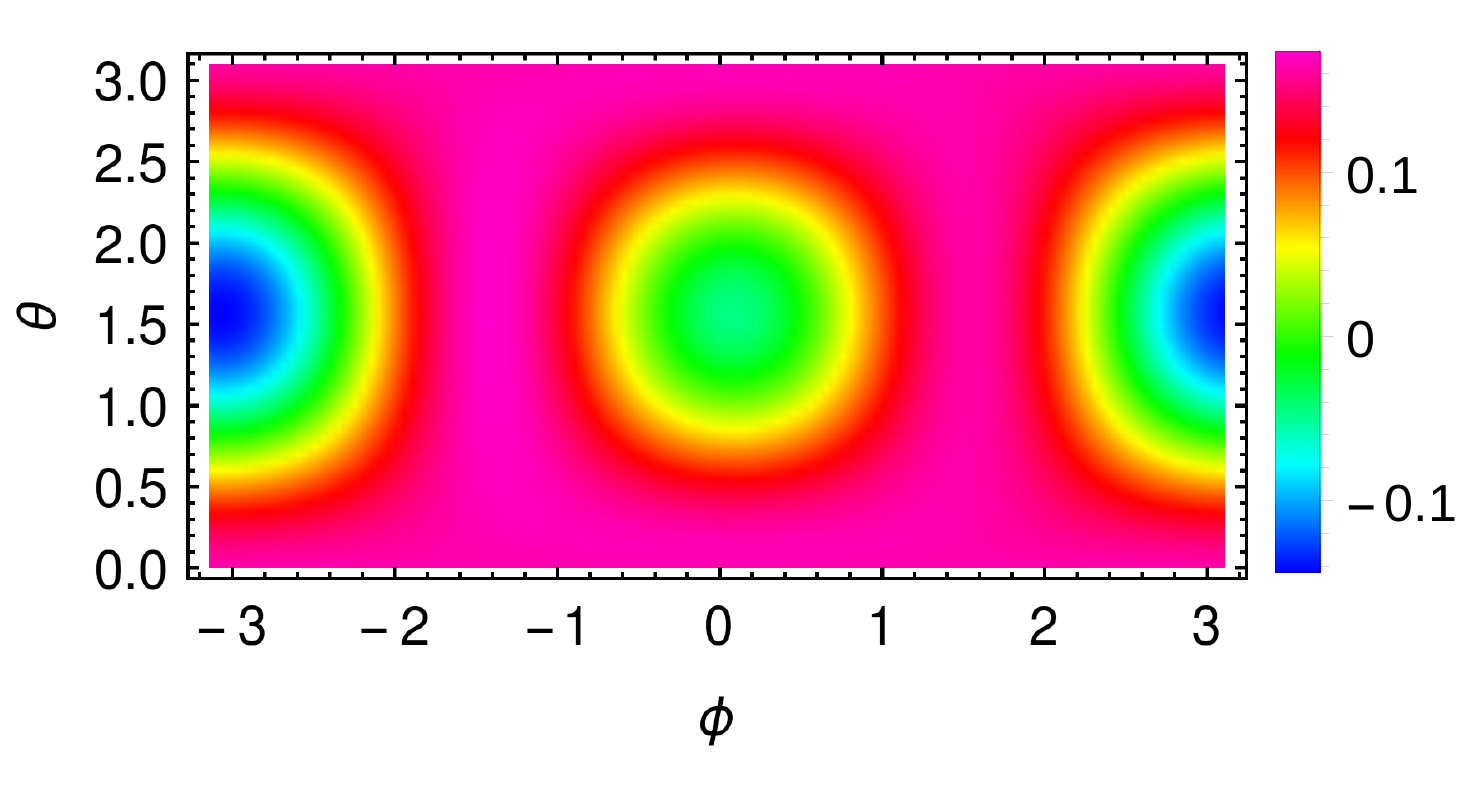}}
\subfloat[{\sf (b)}]{\includegraphics[scale=0.55]{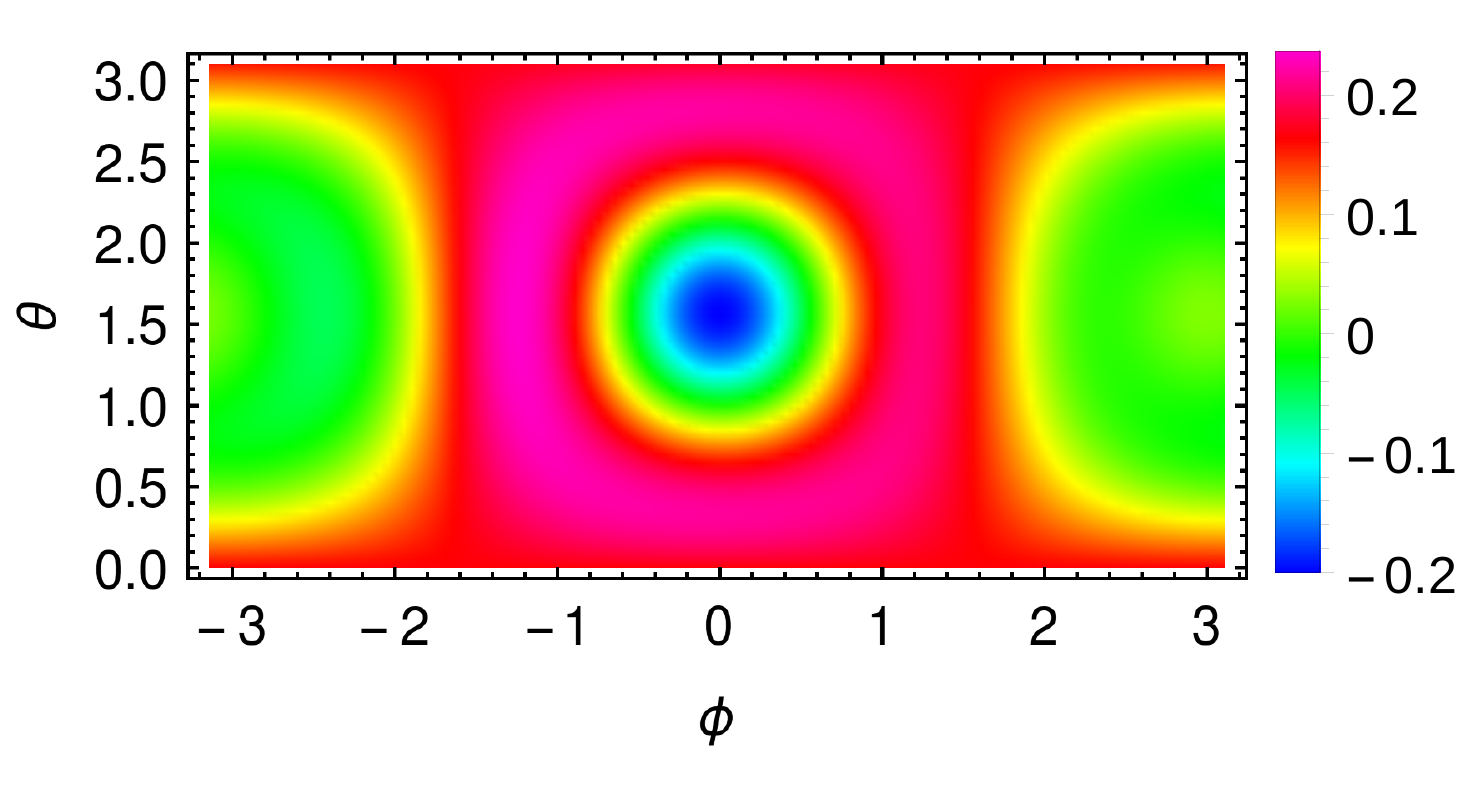}}
\caption{To study the delocalization in the phase space off the quantum revival times we study  the  spin quasiprobability $\mathrm{P}_{\mathcal{Q}}(\theta, \phi)$-representation for the factorized initial state (\ref{state-t0}) with the choice $\mathrm{c}=0$.  The diagrams {\sf (a, b)} refer to the cases of $s=1$ and $s=\frac{3}{2}$ for the times $\omega t = 2.022 \times 10^{6}$ and  $\omega t = 0.669 \times 10^{6}$ where the corresponding entropies are near their maximal values: $S(\rho^{(1)}_{\mathcal{Q}})=1.01065$ and $S(\rho^{(\frac{3}{2})}_{\mathcal{Q}})=1.22474$. The coupling strengths and parametric choices for the diagrams {\sf (a, b)} here are identical to those in Fig. \ref{entropy-1-tL}{\sf (a)} and Fig. \ref{entropy-3/2-tL}{\sf(a)}. Compared with the kitten like structure 
(Figs. \ref{kitten-1} and \ref{kitten-3/2}) the $\mathrm{P}_{\mathcal{Q}}(\theta, \phi)$-representations observed here have a broader spread.}
\label{P_func_max_ent}
	\end{center}
	\end{figure}

In the strong coupling regime the observed total and fractional revivals  in 
Fig. \ref{fig_corr} $(\mathsf{a}_{\mathsf{1}}, \mathsf{b}_{\mathsf{1}})$ are symbolized by nearly sinusoidal,
large amplitude and  short range oscillations in the time evolution of $g_{s}(t)$. These oscillations with time period 
$\sim \mathrm{O}(\omega^{-1})$ are visible in the insets of Fig. \ref{fig_corr} $(\mathsf{a}_{\mathsf{1}}, \mathsf{b}_{\mathsf{1}})$. Qualitatively this may be understood as follows. These revival times are rational multiples of $\mathrm{T}|_{\mathsf{quasiperiod}}$ given in (\ref{quasi-period}).  At the time of revival there is a drop in entropy and, consequently, there is an increase in the purity of the qudit quantum state while the composite bipartite system  remains almost disentangled. Relatively few modes of quantum fluctuations present at the instants of  revival interfere coherently and give rise to transient harmonic fluctuations observed at those times. Away from these periods the qudit inhabits  a highly mixed state 
and  the entropy returns to its near-maximal value. A comparatively larger number of fluctuation modes are produced and a randomization of their
phases leads them to largely annihilate each other.  This is evident (Fig. \ref{fig_corr}$(\mathsf{a}_{\mathsf{1}}, \mathsf{b}_{\mathsf{1}})$ ) in the collapse of the  fluctuations  of the function  $g_{s}(t)$ at instants apart from the  revival times. Another distinguishing feature of the short-lived revival times is that coherence of the quantum fluctuations localizes the phase space distributions in relatively small domains (Figs. \ref{kitten-1}, \ref{kitten-3/2}). Off the revival times, on the other hand, the lack of coherence of the fluctuations has a spreading effect on the phase space 
distributions and make them delocalized 
(Fig. \ref{P_func_max_ent}). Moreover, in support of the above description it is worth mentioning that similar oscillatory behavior is observed at the revival times  in the intensity for stimulated emission $\braket{S_{z}}$ plots with  period $\mathrm{T}|_{\mathsf{quasiperiod}}$. We, however, for the sake of brevity omit these plots. 

\par 

Another characteristic of the normalized correlation function $g_{s}(t)$ evident in Fig. \ref{fig_corr} $(\mathsf{a}_{\mathsf{1}}, \mathsf{b}_{\mathsf{1}})$ is that  successive antibunching and bunching of the emitted radiation  appear during the short range coherent oscillations generated at the moments of revival. For the correlation function in the range $g_{s}(t) < 1\;(g_{s}(t) > 1)$ 
antibunching (bunching) of the emission
process takes place. On the occasions of antibunching of the radiation the photoevents are said to be anticorrelated \textit{i.e.} occurrence of one makes the next one less likely. 
Oscillations  observed in the correlation function $g_{s}(t)$ at the instants of revival (insets of Fig. \ref{fig_corr} $(\mathsf{a}_{\mathsf{1}}, \mathsf{b}_{\mathsf{1}})$) demonstrate the consecutive display of antibunching and bunching effects. 
Collective interaction of the atoms with the photon field is known [\cite{ABNV1977}] to diminish the antibunching of radiation. In our context it is revealed in a comparison of the insets in Figs. \ref{fig_corr} $(\mathsf{a}_{\mathsf{1}})$ and 
$(\mathsf{b}_{\mathsf{1}})$. For the spin $s=\frac{3}{2}$ example the lowest values of $g_{s}(t)$ achieved during revivals are comparatively higher than those realized for spin $s=1$ case.  This points towards the reduction of antibunching effects with  increasing spin.

\par

As observed before  the ultrastrong coupling domain $\widetilde{\lambda} \sim \mathrm {O}(1)$ generates
a progressively large number of interaction modes  with  widely distributed  characteristic time scales.  As a consequence the phase correlations among the interacting modes endowed with  incommensurate frequencies  are completely lost. The randomization of the phases
eliminates  all quasiperodic patterns.  In particular, the time evolution of $g_{s}(t)$ in this domain exhibits 
(Fig. \ref{fig_corr} $(\mathsf{a}_{\mathsf{2}}, \mathsf{b}_{\mathsf{2}})$) chaotic behavior without any quantum collapse and revival structure. In this fully randomized realm it is, however, observed (Figs. \ref{fig_corr} $(\mathsf{a}_{\mathsf{2}})$ and 
$(\mathsf{b}_{\mathsf{2}})$) that for the $s=1$ case, in contrast to the higher spin $s=\frac{3}{2}$ example, the  correlation function  exhibits $g_{s}(t) < 1$ behavior far more frequently.  Therefore, the antibunching effect on the emitted photons survive in the chaotic regime for the low spin qudits, and gradually disappear for larger spin quantum numbers where  cooperative effects among the atoms become increasingly dominant.


\clearpage
\section{Quantum spin state tomography} 
\label{spin-tomogram}
\setcounter{equation}{0}
In the previous sections we have described the evolution of the interacting spin-oscillator system via  the phase space 
quasiprobability distributions. Various tomographic schemes, however, develop representations of quantum states of systems in 
terms of the measurable normalized probability densities. Advancing a formulation of  tomography of the spin 
states the authors of Ref. [\cite{MM1997}] produced an invariant expression for the elements of the density matrix of an 
arbitrary spin $s$ via the measurable probability of the projection of the spin in any direction. The probability distribution function is a diagonal density matrix element of the spin state in an arbitrarily rotated frame described by the corresponding Euler angles denoted below as 
$(\mathfrak{a}, \mathfrak{b}, \mathfrak{g})$.

\par
To describe the tomography scheme for the discrete spin variables the authors of Ref. [\cite{MM1997}, \cite{DM1997}] employed 
the Wigner $\mathrm{D}$-matrices which are the matrix elements of the irreducible representations of the rotation group: 
\bea
\mathrm{D}_{m'm}^{j}(\mathfrak{a}, \mathfrak{b}, \mathfrak{g}) &=& \exp(i \mathfrak{a} m')\, \exp(i \mathfrak{g} m)\,
\mathrm{d}_{m'm}^{j}(\mathfrak{b}),\nn \\
\mathrm{d}_{m'm}^{j}(\mathfrak{b}) &=&\sqrt{\tfrac{(j+m')! (j-m')!}{(j+m)! (j-m)!}} \left( \cos  \tfrac{\mathfrak{b}}{2}\right)^{m'+m} \left( \sin  \tfrac{\mathfrak{b}}{2}\right)^{m'-m} 
\; P_{j-m'}^{m'-m,m'+m}(\cos \mathfrak{b} ).
\label{Wigner-D}
\eea
Parametrized by  Euler angles the diagonal entries of the qudit density matrix provide [\cite{MM1997}, \cite{DM1997}] a positive definite 
probability distribution of the allowed spin components in a direction specified by corresponding rotations: 
\beq
\widetilde{\omega}^{(s)}\,(m;\mathfrak{a},\mathfrak{b},\mathfrak{g})\equiv\,\sum_{m',\,m'' = - s}^{s}\,\mathrm{D}_{m\,m'}^s\,(\mathfrak{a},\mathfrak{b},\,\mathfrak{g})
\;\left(\rho_{Q}\right)_{m'\,m''}^{(s)}\;\mathrm{D}_{m\,m''}^s\,(\mathfrak{a},\mathfrak{b}\,,\mathfrak{g})^*. 
\label{tomogram-def}
\eeq
 As the discrete probability distribution (\ref{tomogram-def}) is constructed via utilizing the representation (\ref{Wigner-D}) of the rotation group, its dependence on the Euler angle $\mathfrak{a}$ disappears while the normalization relation reads: 
 \bea
 \widetilde{\omega}^{(s)}\,(m;\mathfrak{a},\mathfrak{b},\mathfrak{g})\equiv \omega^{(s)}\,(m;\mathfrak{b}, \mathfrak{g}) ,
 \;\;\sum_{m=-s}^{s}\omega^{(s)}\,(m;\mathfrak{b}, \mathfrak{g})=1.
 \label{tomogram-norm}
 \eea
In the example studied here the above construction  (\ref{tomogram-def}) may be directly implemented as the evolution of the qudit density matrix is determined under the adiabatic approximation scheme. For instance, 
the qudit density matrix (\ref{density-1}) for $s=1$ case immediately provides the corresponding tomogram that may be considered as the reconstruction of the state via the positive definite probability distribution: 
\bea
\omega^{(1)}(m;\mathfrak{b},\mathfrak{g}) &=& \!\!\! \tfrac{1}{2} 
\left( \tfrac{\sin \mathfrak{b}}{2} \right)^{2m} (1-m)! (1+m)! 
\sum_{n,\widetilde{n}=0}^{\infty} 
\left\{ \left( \cot \tfrac{\mathfrak{b}}{2} \;\; P_{1-m}^{m-1,m+1} 
\left( \cos \mathfrak{b} \right) \right)^{2} \right.\times \nn \\
& \times & \!\!\!  \mathcal{B}_{+,n}^{(1)}(t) \mathcal{B}_{+,\widetilde{n}}^{(1)}(t)^{*} \, \delta_{n \widetilde{n}} + 
\left( \tan \tfrac{\mathfrak{b}}{2} \;\; P_{1-m}^{(m+1,m-1)} 
\left( \cos \mathfrak{b} \right) \right)^{2} \mathcal{B}_{-,n}^{(1)}(t) \mathcal{B}_{-,\widetilde{n}}^{(1)}(t)^{*} \delta_{n\widetilde{n}} \nn \\
& + & \!\!\! 2 \left( P_{1-m}^{m,m} 
\left( \cos \mathfrak{b} \right) \right)^{2} \mathcal{B}_{0,n}^{(1)}(t) \mathcal{B}_{0,\widetilde{n}}^{(1)}(t)^{*}
\, \delta_{n \widetilde{n}} + 2 \sqrt{2}  P_{1-m}^{m,m} \left( \cos \mathfrak{b} \right) \Big( \tan \tfrac{\mathfrak{b}}{2} \times \nn \\
& \times & \!\!\! P_{1-m}^{m+1,m-1} \left( \cos \mathfrak{b} \right) 
\mathrm{Re}\left( \exp\left( i \mathfrak{g} \right)
\mathcal{B}_{0,n}^{(1)}(t) \mathcal{B}_{-,\widetilde{n}}^{(1)}(t)^{*}\, \mathcal{G}_{\widetilde{n} n}(- \widetilde{\lambda})
\right) + \cot \tfrac{\mathfrak{b}}{2} \times \nn \\
& \times & \!\!\!  P_{1-m}^{m-1,m+1} \left( \cos \mathfrak{b} \right) 
\mathrm{Re}\left( \exp\left( i \mathfrak{g} \right)
\mathcal{B}_{+,n}^{(1)}(t) \mathcal{B}_{0,\widetilde{n}}^{(1)}(t)^{*}\, \mathcal{G}_{\widetilde{n} n}(- \widetilde{\lambda})
\right) \Big) \nn \\
& + & \!\!\! 2 \; P_{1-m}^{m-1,m+1} \left( \cos \mathfrak{b} \right)  P_{1-m}^{m+1,m-1} 
\left( \cos \mathfrak{b} \right)  \mathrm{Re} \Big( \exp\left(2 i \mathfrak{g} \right)  \times \nn \\
&\times & \left.\!\!\! \mathcal{B}_{+,n}^{(1)}(t) \mathcal{B}_{-,\widetilde{n}}^{(1)}(t)^{*}\, \mathcal{G}_{\widetilde{n} n}(- 2 \widetilde{\lambda}) \Big)
\right\}.
\label{Tomogram-s=1}
\eea
Continuing further we also utilize our approximate evaluation of the $s=\tfrac{3}{2}$ qudit density matrix 
given in (\ref{DenMtrx-3/2-t}) to procure the corresponding tomogram that expresses the evolution of the state 
in terms of the probability distribution in arbitrarily rotated frames characterized by the  Euler angles:
\bea
\omega^{(\frac{3}{2})}(m;\mathfrak{b},\mathfrak{g}) \!\!\! &=& \!\!\! \tfrac{1}{6} \left( \tfrac{\sin \mathfrak{b}}{2}\right)^{2\,m} \;\,\!\!\!
\left( \tfrac{3}{2} - m\right)! \left( \tfrac{3}{2} + m\right)! \!\!\! \sum_{n,\widetilde{n}=0}^{\infty}\left\{  \delta_{n\widetilde{n}} \Big( \cot^{3}  \tfrac{\mathfrak{b}}{2} \left( P_{\frac{3}{2}-m}^{m-\frac{3}{2},m+\frac{3}{2}}(\cos \mathfrak{b}) \right)^{2}
\mathrm{B}_{n,\widetilde{n}}^{(2,2)}(t)\right. \nn \\
&+& \!\!\! 3 \cot \tfrac{\mathfrak{b}}{2} \left( P_{\frac{3}{2}-m}^{m-\frac{1}{2},m+\frac{1}{2}}(\cos \mathfrak{b}) \right)^{2} \mathrm{B}_{n,\widetilde{n}}^{(1,1)}(t) +
3 \tan \tfrac{\mathfrak{b}}{2} \left( P_{\frac{3}{2}-m}^{m+\frac{1}{2},m-\frac{1}{2}}(\cos \mathfrak{b}) \right)^{2} \mathrm{B}_{n,\widetilde{n}}^{(-1,-1)}(t) \nn \\
&+& \!\!\! \tan^{3}  \tfrac{\mathfrak{b}}{2} \left( P_{\frac{3}{2}-m}^{m+\tfrac{3}{2},m-\frac{3}{2}}(\cos \mathfrak{b}) \right)^{2}
\mathrm{B}_{n,\widetilde{n}}^{(-2,-2)}(t) \Big) + 2 \sqrt{3} \; P_{\frac{3}{2}-m}^{m-\frac{3}{2},m+\frac{3}{2}}(\cos \mathfrak{b}) \Big( \cot^{2}  \tfrac{\mathfrak{b}}{2} \times \nn \\
&\times&  \!\!\! P_{\frac{3}{2}-m}^{m-\frac{1}{2},m+\frac{1}{2}}(\cos \mathfrak{b}) \; \mathrm{Re} \left( \exp(i \mathfrak{g})\, \mathrm{B}_{n,\widetilde{n}}^{(2,1)}(t)\,
\mathcal{G}_{\widetilde{n}n}(- \widetilde{\lambda}) \right) + 
\cot \tfrac{\mathfrak{b}}{2} \; P_{\frac{3}{2}-m}^{m+\frac{1}{2},m-\frac{1}{2}}(\cos \mathfrak{b}) \times \nn \\
&\times&  \!\!\!
\mathrm{Re} \left( \exp(2 i \mathfrak{g})\, \mathrm{B}_{n,\widetilde{n}}^{(2,-1)}(t)\,
\mathcal{G}_{\widetilde{n}n}(-2 \widetilde{\lambda}) \right) \Big) +  2 \sqrt{3} \;
P_{\frac{3}{2}-m}^{m+\frac{3}{2},m-\frac{3}{2}}(\cos \mathfrak{b}) \Big( \tan^{2}  \tfrac{\mathfrak{b}}{2} 
\times \nn \\
&\times& \!\!\!
P_{\frac{3}{2}-m}^{m+\frac{1}{2},m-\frac{1}{2}}(\cos \mathfrak{b}) \; \mathrm{Re} \left( \exp(i \mathfrak{g})\, \mathrm{B}_{n,\widetilde{n}}^{(-1,-2)}(t)\,
\mathcal{G}_{\widetilde{n}n}(- \widetilde{\lambda}) \right) +  \tan \tfrac{\mathfrak{b}}{2} \;
P_{\frac{3}{2}-m}^{m-\frac{1}{2},m+\frac{1}{2}}(\cos \mathfrak{b}) \times \nn \\
&\times& \!\!\!
\mathrm{Re} \left( \exp(2 i \mathfrak{g})\, \mathrm{B}_{n,\widetilde{n}}^{(1,-2)}(t)\,
\mathcal{G}_{\widetilde{n}n}(-2 \widetilde{\lambda}) \right) \Big) + 2 \;
P_{\frac{3}{2}-m}^{m+\frac{3}{2},m-\frac{3}{2}}(\cos \mathfrak{b}) \;
P_{\frac{3}{2}-m}^{m-\frac{3}{2},m+\frac{3}{2}}(\cos \mathfrak{b}) \times \nn \\
& \times & \!\!\! 
\mathrm{Re} \left( \exp(3 i \mathfrak{g})\, \mathrm{B}_{n,\widetilde{n}}^{(2,-2)}(t)\,
\mathcal{G}_{\widetilde{n}n}(-3 \widetilde{\lambda}) \right)
+  6 \; P_{\frac{3}{2}-m}^{m+\frac{1}{2},m-\frac{1}{2}}(\cos \mathfrak{b}) \;
P_{\frac{3}{2}-m}^{m-\frac{1}{2},m+\frac{1}{2}}(\cos \mathfrak{b}) \times \nn \\
& \times & \!\!\! \left.
\mathrm{Re} \left( \exp( i \mathfrak{g})\, \mathrm{B}_{n,\widetilde{n}}^{(1,-1)}(t)\,
\mathcal{G}_{\widetilde{n}n}(- \widetilde{\lambda}) \right) \right\}.
\label{Tomogram-s=3/2}
\eea

\par
  
Towards expressing the phase space quasiprobability densities via the true tomographic probability distribution explicitly determined here one may proceed as follows. Applying the orthogonality relations of the Wigner $3j$-coefficients [\cite{A1981}] the authors of 
Ref. [\cite{MM1997}] inverted the defining property (\ref{tomogram-def}) to express the qudit density matrix elements in the angular momentum basis: 
  \bea
  (-1)^{m''}\left({\rho_{Q}}\right)_{m'\,m''}^{(s)}&=&\sum_{\sigma=0}^{2\,s}\;\sum_{\widetilde{m}=-\sigma}^{\sigma}\;(2\,\sigma+1)^2\;
  \sum_{m=-s}^{s}\;(-1)^m\,\left(\begin{array}{clcr}s & {\;\;s} & \sigma\\m & -m & 0  \end{array}\right)\;
  \left(\begin{array}{clcr}s & {\;\;s} & \sigma\\m' & -m'' & \widetilde{m}  \end{array}\right) \times\nn \\
&& \times\,\int \omega^{(s)}\,(m;\mathfrak{b}, \mathfrak{g})\;
\mathrm{D}_{0\,\widetilde{m}}^\sigma\;(\mathfrak{a},\mathfrak{b},\mathfrak{g})\;
  \tfrac{\mathrm{d}\mathcal{W}}{8\,\pi^2},
\label{DenMa-Tom}  
  \eea
where the measure of the angular variables is given by $\int \mathrm{d}\mathcal{W}=\int_{0}^{2\pi}\mathrm{d} \mathfrak{a}\;\int_{0}^{\pi}\;\sin{\mathfrak{b}}\;\mathrm{d}\mathfrak{b}\;\int_{0}^{2\pi}\;\mathrm{d}\mathfrak{g}= 8 \pi^{2}$. Extending this 
approach we use the tomographic composition (\ref{DenMa-Tom}) of the qudit state to express its density matrix in the spherical tensor basis (\ref{Den-MaQ-SphTn}) as follows: 
  \beq
  \left({\varrho^{Q}}\right)_{k\,q}^{(s)}=(2\,k+1)^{\frac{3}{2}}\;
  \sum_{m}\;(-1)^{s-m+q}\,\left(\begin{array}{clcr}s & {\;\;s} & k\\m & -m & 0  \end{array}\right)\;
  \int \;\omega^{(s)}\,(m;\mathfrak{b}, \mathfrak{g})\;\mathrm{D}_{0\;-q}^{k}(\mathfrak{a},\mathfrak{b},\mathfrak{g})\;
  \tfrac{\mathrm{d}\mathcal{W}}{8\,\pi^2}.
  \label{tomogram-DenMa-kq}
  \eeq
The above integral on the Euler angular variables admits a consistency check between our expressions of the tomograms evaluated in (\ref{Tomogram-s=1}) and (\ref{Tomogram-s=3/2}) for the cases $s = 1,\tfrac{3}{2}$ respectively on one hand, and the corresponding expressions of the qudit density matrix in the spherical tensor basis produced in (\ref{DenMat_kq_s1}) and (\ref{DenMat_kq_s3/2}) on the other.  The phase space quasiprobabilities such as the  qudit $\mathrm{P}_{_{\mathcal{Q}}}$-representation, Wigner 
$\mathrm{W}_{_{\mathcal{Q}}}$-distribution, and the $\mathrm{Q}_{_{\mathcal{Q}}}$-function, given in equations (\ref{P-spin}), (\ref{W-spin}) and (\ref{Q-spin}) respectively, may now be explicitly formulated 
using the positive definite probability distribution $\omega^{(s)}(m;\mathfrak{b}, \mathfrak{g})$ associated with spin projections in arbitrarily rotated frames.

\par

\begin{figure}
\begin{center}
\captionsetup[subfigure]{labelformat=empty}
\hspace*{-2cm}
\subfloat[{\sf (a$_{1}$)}]{\includegraphics[scale=0.45]{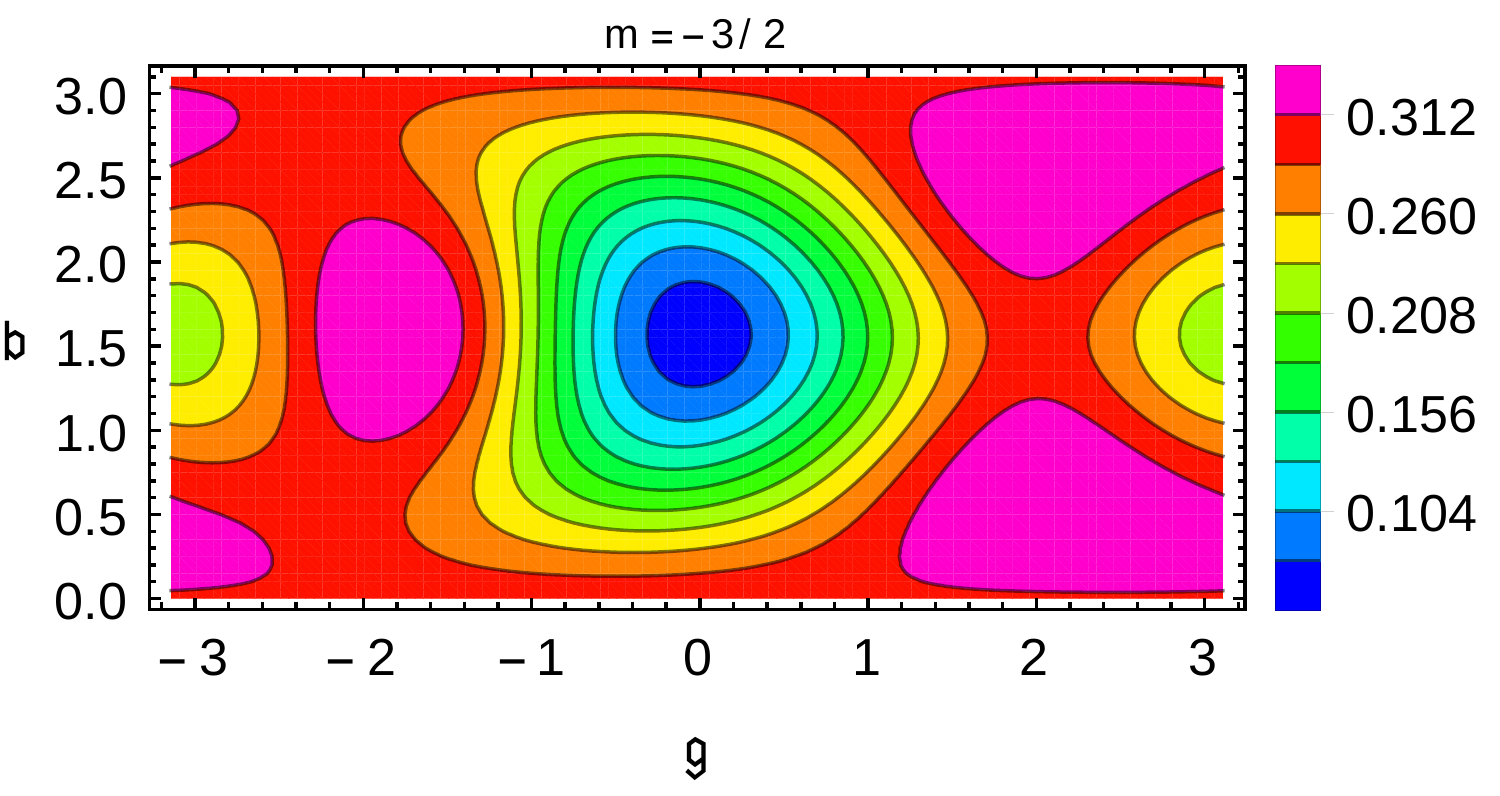}}
\subfloat[{\sf (a$_{2}$)}]{\includegraphics[scale=0.45]{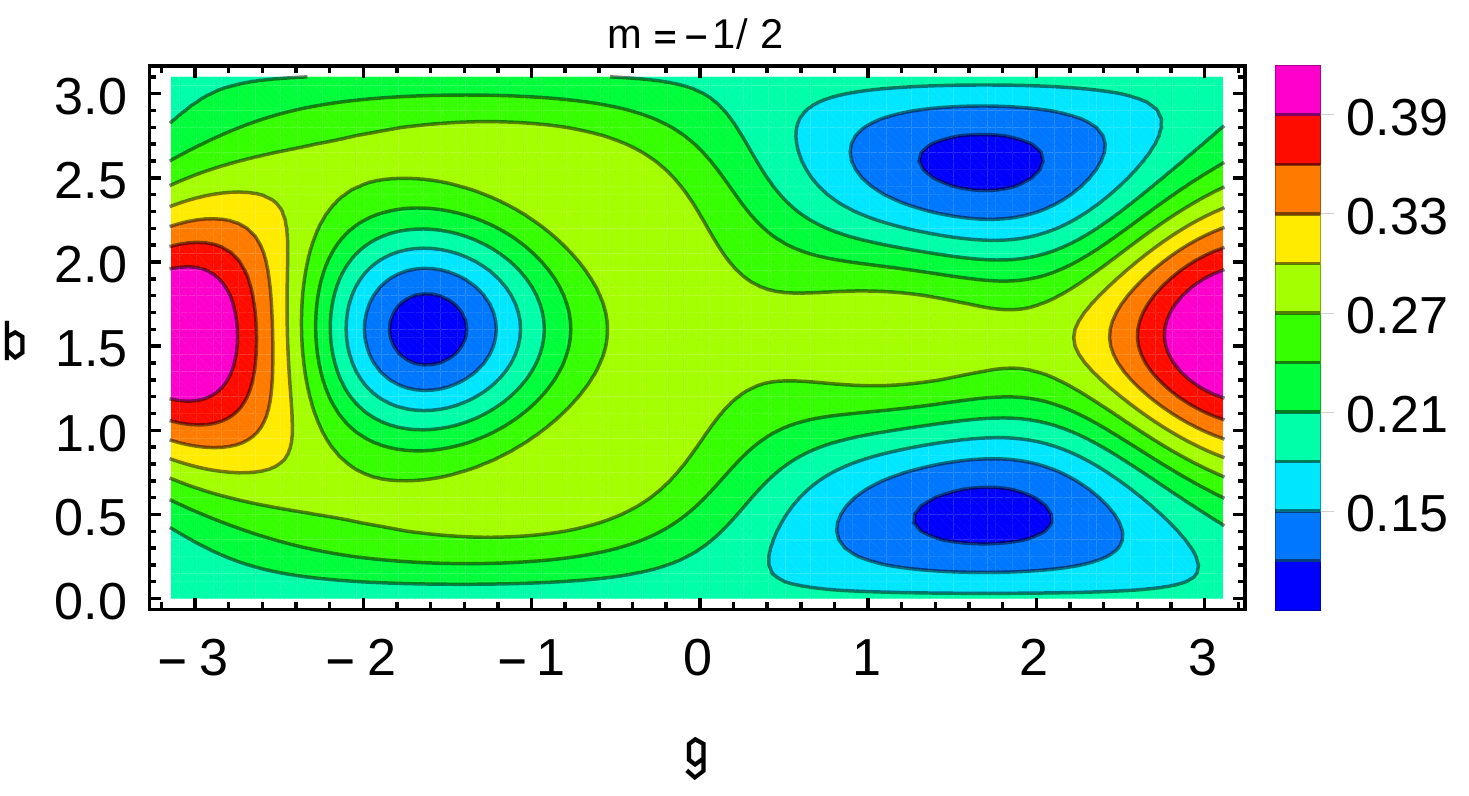}}\\
\hspace*{-2cm}
\subfloat[{\sf (a$_{3}$)}]{\includegraphics[scale=0.45]{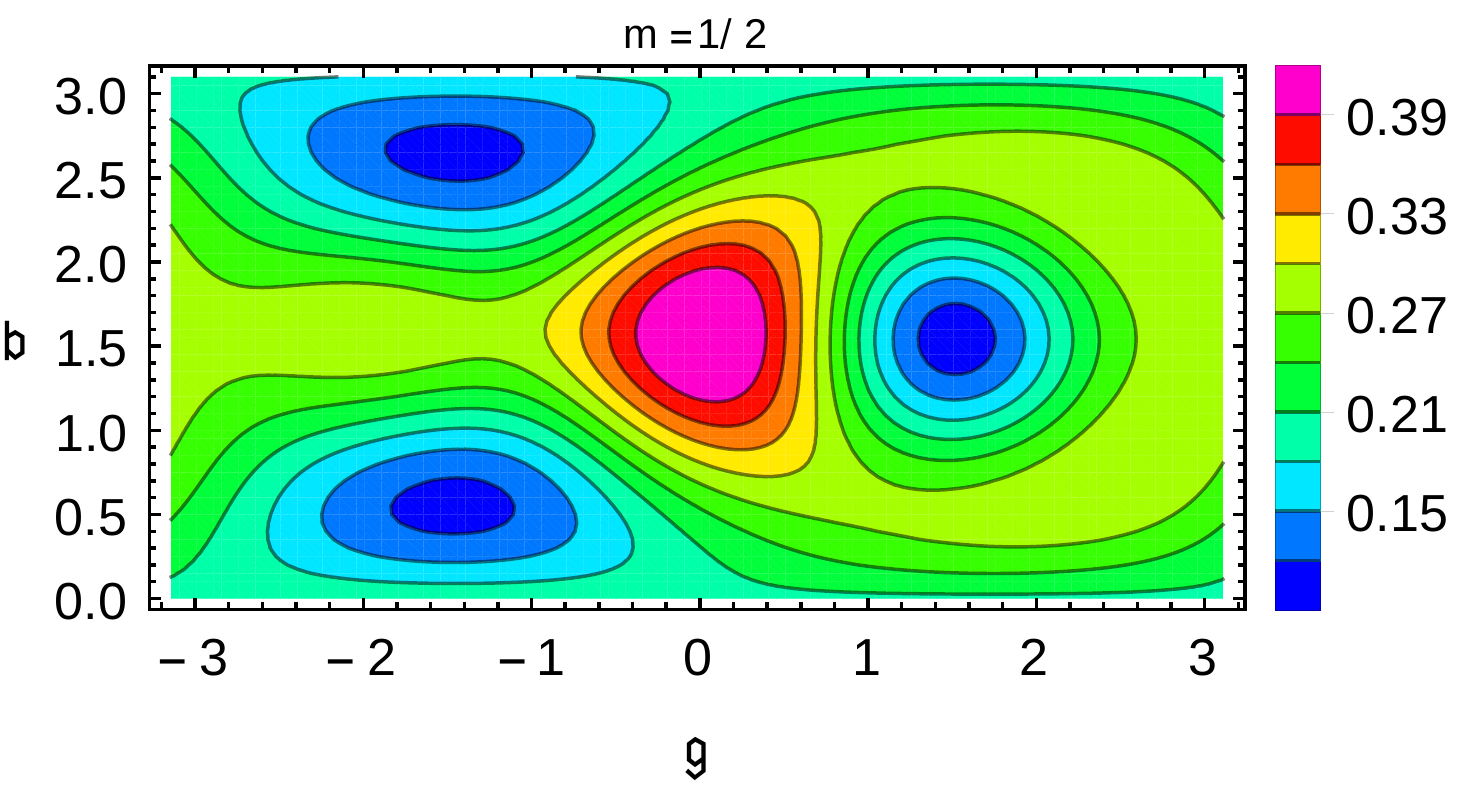}}
\subfloat[{\sf (a$_{4}$)}]{\includegraphics[scale=0.45]{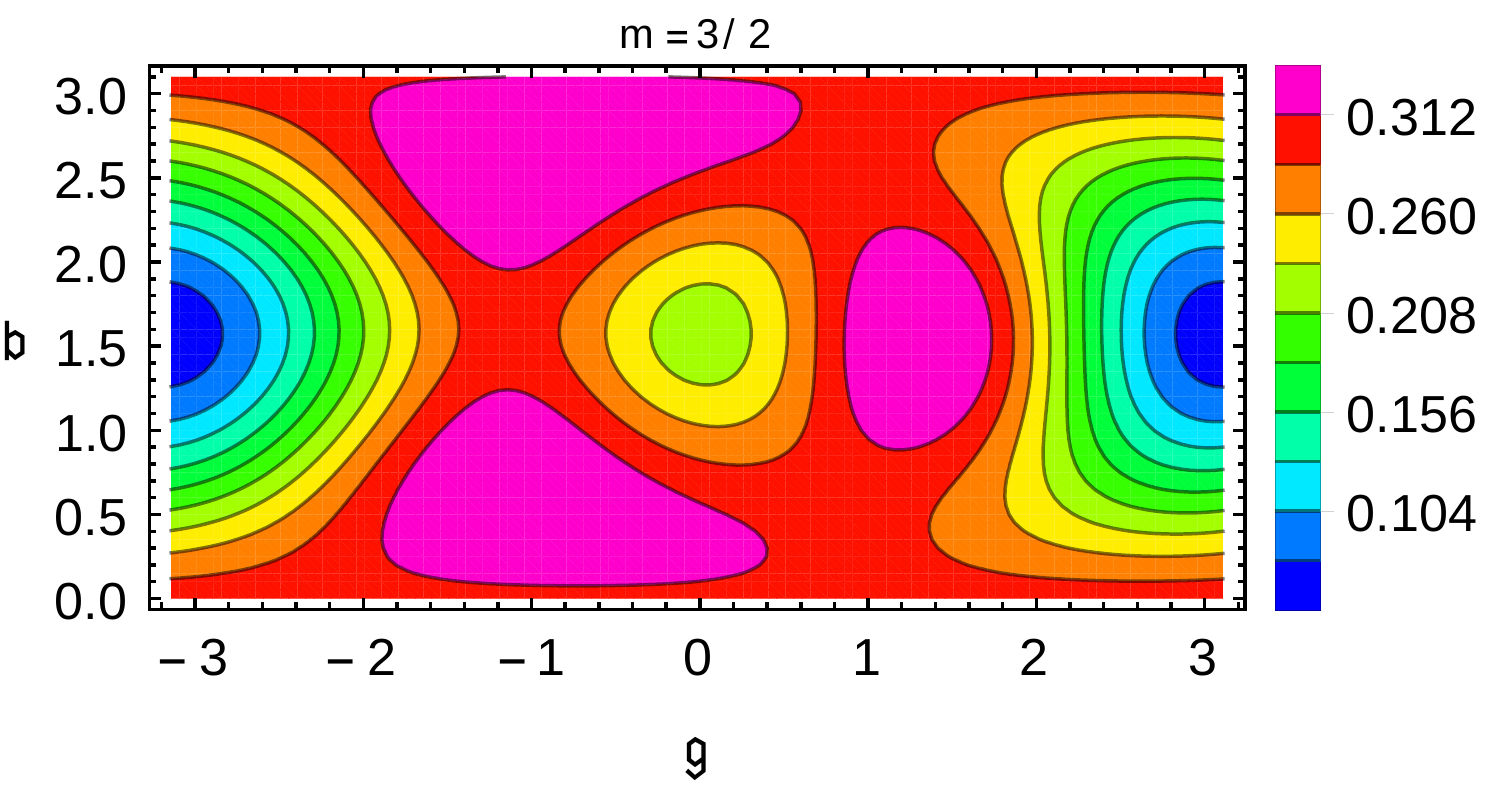}}\\
\hspace*{-2cm}
\subfloat[{\sf (b$_{1}$)}]{\includegraphics[scale=0.45]{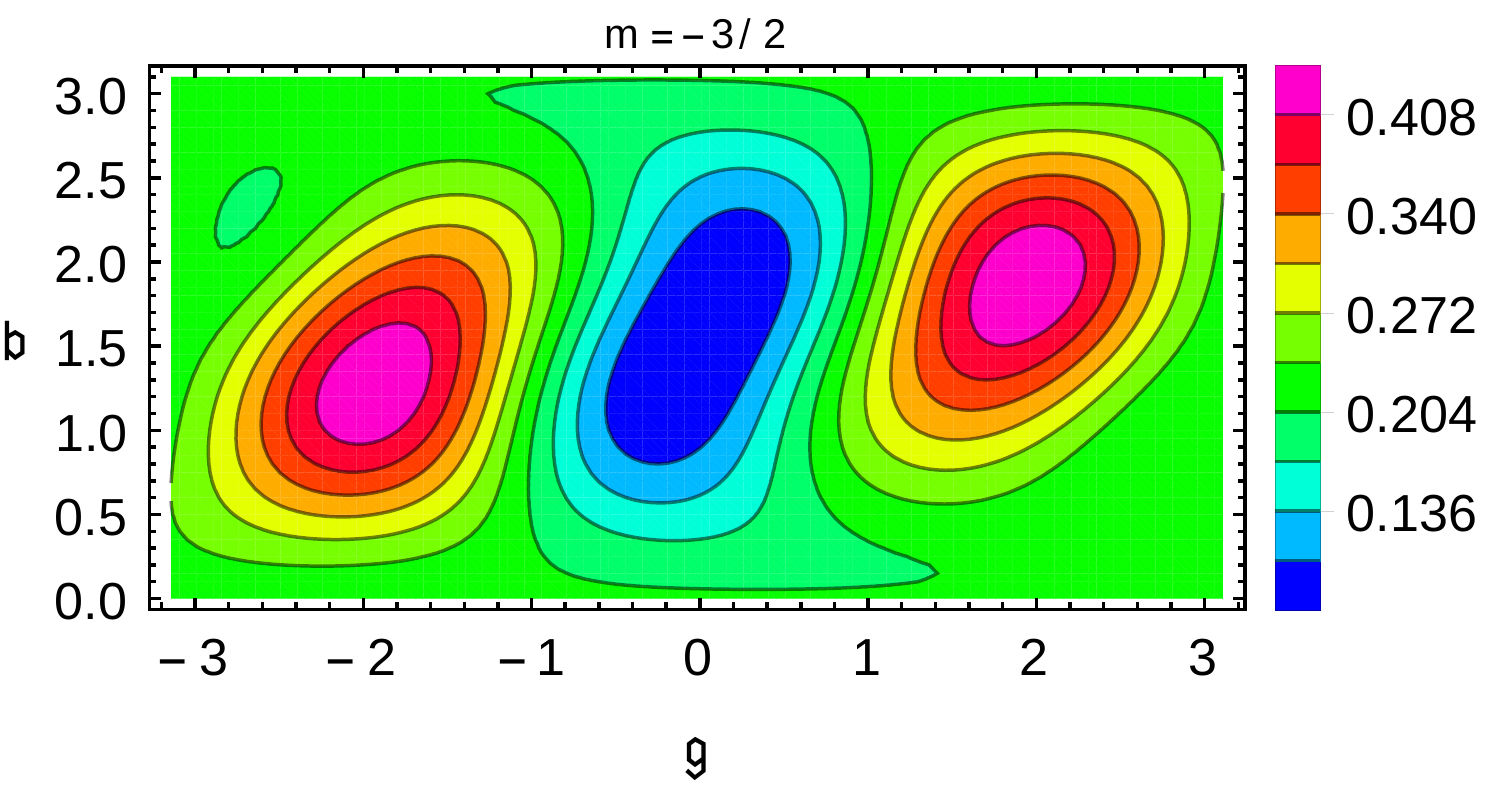}}
\subfloat[{\sf (b$_{2}$)}]{\includegraphics[scale=0.45]{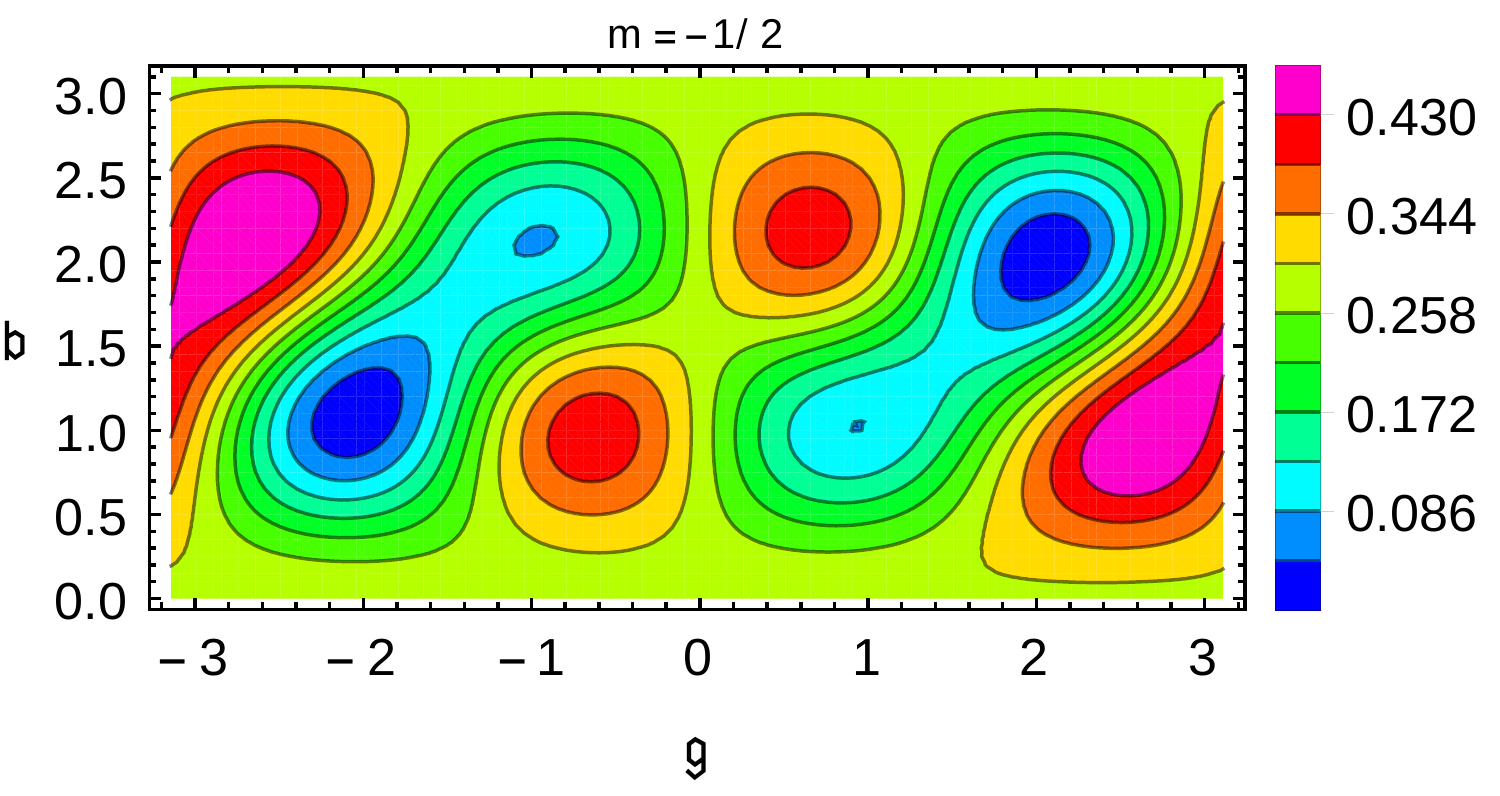}}\\
\hspace*{-2cm}
\subfloat[{\sf (b$_{3}$)}]{\includegraphics[scale=0.45]{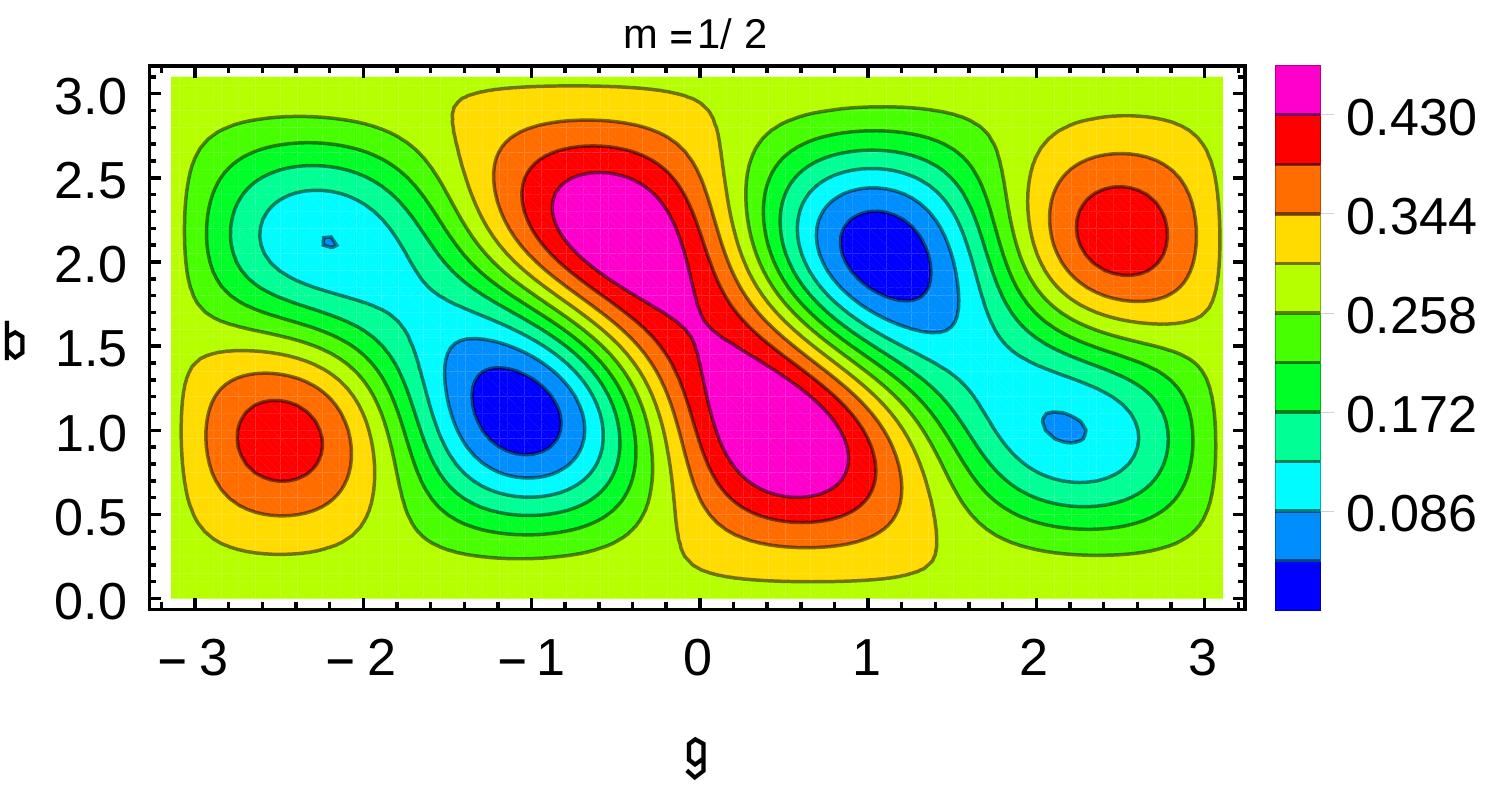}}
\subfloat[{\sf (b$_{4}$)}]{\includegraphics[scale=0.45]{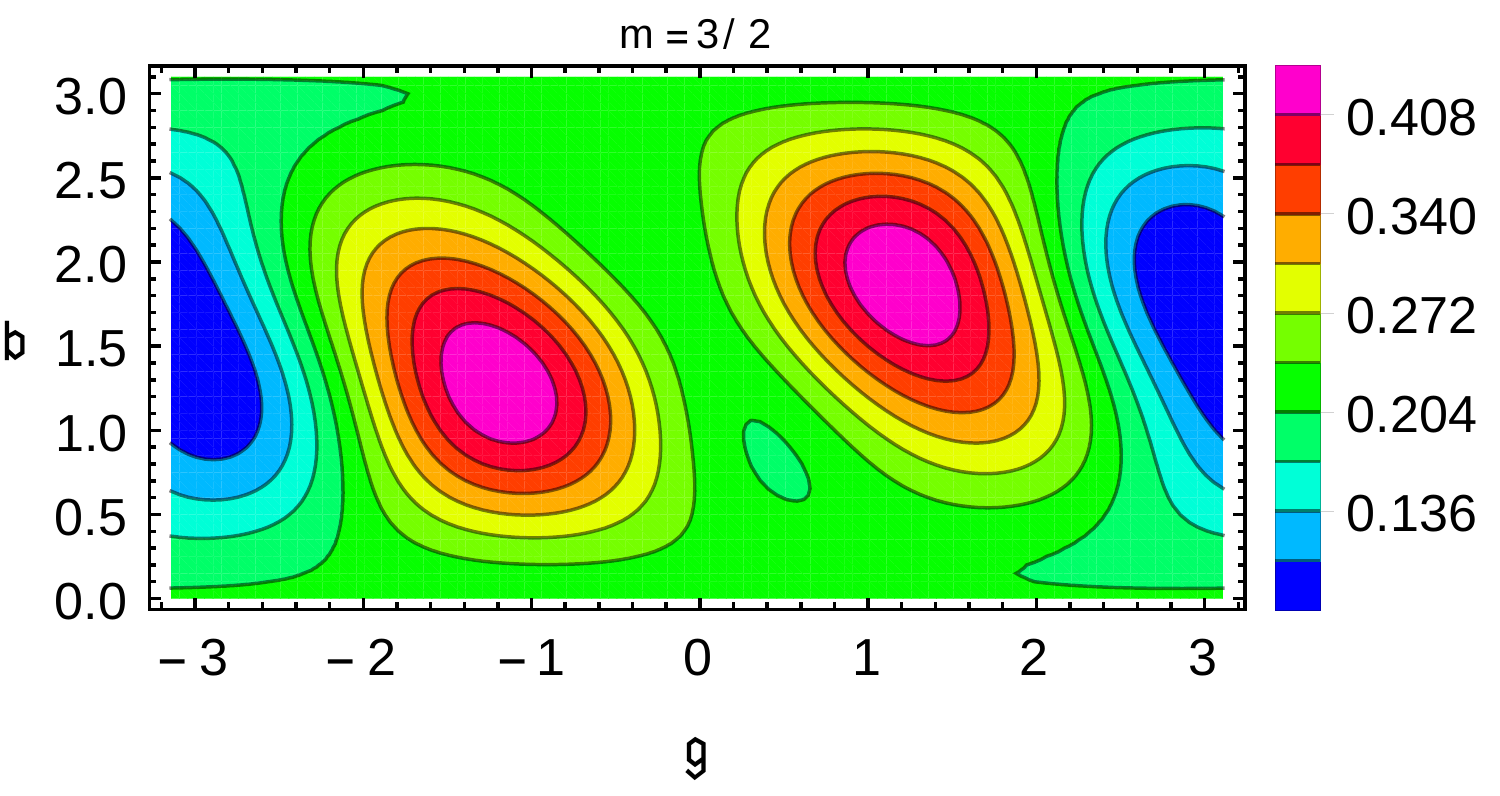}}
\caption{The tomogram $\omega^{(\frac{3}{2})}(m;\mathfrak{b},\mathfrak{g})$ for the $s=\tfrac{3}{2}$ case is considered for two examples. 
For the strong  coupling strength 
$\widetilde{\lambda}=0.002$ and a factorized initial state (\ref{state-t0}) maintained with $\mathrm{c}=0$, the parametric values are chosen as follows: $\Delta =0.15,\,\mathfrak{z}=0.1051,\,\alpha =3, r=0.2$. 
{\sf (a)}: The displayed $3$-kitten state is realized at time $t=1.397223 \times 10^7$, when the diagrams 
{\sf (a$_{1}$, a$_{2}$, a$_{3}$, a$_{4}$)} exhibit positive definite probability distributions for the spin projections $m = -\tfrac{3}{2},\ldots, \tfrac{3}{2}$, respectively. {\sf(b)}: On the other hand the $4$-kitten state develops at the time $t=5.242093 \times 10^6$. It is described via the illustrations  {\sf (b$_{1}$,\,b$_{2}$, b$_{3}$, b$_{4}$)} signifying the probabilities for the allowed values of the projection quantum number quoted therein. Complete separation of the lobes for the  $4$-kitten state
does not occur (say in {\sf (b$_{2}$, b$_{3}$)}). This suggests that the pertinent density matrix receives significant contributions from states other than the $4$-kitten state.}
\label{tomogram-3/2}
\end{center}
\end{figure}

Towards demonstrating the tomographic representations of the qudit states  we chose the $s=\tfrac{3}{2}$ example displayed in the Fig. \ref{tomogram-3/2}. For the selection of parameters given therein, we study the entropy $S(\rho_{\mathcal{Q}})$ of the state given in (\ref{entropy}) in the strong coupling regime ($\widetilde{\lambda}= 0.002)$. The entropy $S(\rho_{\mathcal{Q}})$  exhibits the 
quasiperiodicity discussed in Sec. \ref{spin-kitten}.
At the locally minimum configurations  of the entropy, where the spin kitten states have been observed to emerge, we consider the construction of the tomograms in Fig.  \ref{tomogram-3/2}{\sf (a, b)}. The diagrams {\sf (a$_{1}$,..., a$_{4}$)} in the
said figure specify the probability distribution for the projections of the spin variable corresponding to the qudit $3$-kitten state, whereas the diagrams {\sf (b$_{1}$,..., b$_{4}$)} similarly illustrate the tomographic composition of the spin $4$-kitten state. A comparison between the above two sets of diagrams suggests the following. While the $3$-kitten state is formed (Fig.  \ref{tomogram-3/2}{\sf (a)}) via the coherent superposition of largely the extremal spin component states $m = \pm \tfrac{3}{2}$, the $4$-kitten state (Fig.  \ref{tomogram-3/2}{\sf (b)}) owes its origin to a more complex superposition of  all the spin component states. A complete separation of lobes is not manifest for  the $4$-kitten state in the $s=\tfrac{3}{2}$ case as the density matrix  receives contributions from states other than the pure $4$-kitten state.
\newpage
\section{Generation of spin squeezed states}
\label{SpinSqueeze}
\setcounter{equation}{0}
In order to study the emergence of the spin squeezed states during the time evolution of the qudit-oscillator system we follow the description given by the authors of Ref. [\cite{KU1993}]. The spin is regarded [\cite{KU1993}] to be squeezed if  the variance of one spin component, perpendicular to the mean spin vector determined by the density matrix, assumes less value than the variance  for  a spin coherent state. The mean spin direction and its normal vectors are specified [\cite{SWYZ2006}, \cite{MWSN2011}] via the following triplet: 
\beq
\vec{n}_{1} \equiv \left( \sin \vartheta \cos \varphi , \sin \vartheta \sin \varphi , \cos \vartheta \right), \; 
\vec{n}_{2} \equiv \left( -\sin \varphi , \cos \varphi , 0 \right),\;
\vec{n}_{3} \equiv \left( -\cos \vartheta \cos \varphi , -\cos \vartheta \sin \varphi , \sin \vartheta \right),
\label{triplet}
\eeq
whose polar and azimuthal angles are characterized by the spin expectation values: 
\beq
\vartheta = \cos^{-1} \tfrac{\braket{S_{\mathsf{z}}}}{|\braket{\vec{S}}|}, \; 
\varphi = \begin{cases}
\cos^{-1} \tfrac{\braket{S_{\mathsf{x}}}}{|\braket{\vec{S}}| \sin \vartheta} \;\quad \qquad \mbox{if} 
\braket{S_{\mathsf{y}}} > 0, \\ 
2 \pi - \cos^{-1} \tfrac{\braket{S_{\mathsf{x}}}}{|\braket{\vec{S}}| \sin \vartheta } \quad \mbox{if} 
\braket{S_{\mathsf{y}}} \leq 0.
\end{cases}
\label{spin-angle}
\eeq
In (\ref{spin-angle}) we have used the notation $|\braket{\vec{S}}|\! = \!
\sqrt{\braket{S_{\mathsf{x}}}^{2}+\braket{S_{\mathsf{y}}}^{2}+\braket{S_{\mathsf{z}}}^{2}}$. An arbitrary vector normal to  the mean spin direction reads $\vec{n}_{_\perp} = \vec{n}_{2} \cos \chi + \vec{n}_{3} \sin \chi$, and the corresponding spin component is given by  $S_{_\perp} \equiv \vec{S} \cdot \vec{n}_{_\perp}$. The
defining property (\ref{spin-angle}) imposes the constraint  $\braket{S_{_\perp}} =0$, and, therefore, the dispersion of the normal spin component reads $\left( \Delta S_{_\perp}  \right)^{2} = \braket{S_{_\perp}^{2}}$.
Using the notation $S_{\vec{n}_{k}} = \vec{S} \cdot \vec{n}_{k}, k \in \{ 2,3 \}$ the minimum variance of the normal spin component is now given by [\cite{SWYZ2006}, \cite{MWSN2011}]
\beq
\min \left( \Delta S_{_\perp}  \right)^{2} = 
\frac{1}{2} \left( \braket{S^{2}_{\vec{n}_{2}}} + \braket{S^{2}_{\vec{n}_{3}}}  \right) 
- \frac{1}{2} 
\left[ \left( \braket{S^{2}_{\vec{n}_{2}}} - \braket{S^{2}_{\vec{n}_{3}}} \right)^{2} 
+ \braket{\left( S_{\vec{n}_{2}} S_{\vec{n}_{3}} + S_{\vec{n}_{3}} S_{\vec{n}_{2}} \right)}^{2} \right]^{\frac{1}{2}},
\label{min-var}
\eeq
The spin squeezing measure provided in Ref. [\cite{KU1993}] is the ratio of the above minimum dispersion with
the corresponding variance in a spin coherent state: $\xi^{2}_{s} = \tfrac{2 \min \left( \Delta S_{_\perp}  
\right)^{2}}{s}$. The spin squeezing is realized [\cite{KU1993}] when the quantum correlation reduces the fluctuations in one spin component normal to the mean spin direction less than its coherent state limit.

\par

\begin{figure}
	\begin{center}
		\captionsetup[subfigure]{labelformat=empty}
		\hspace*{0cm}
		\subfloat[{\sf (a$_{1}$)}]{\includegraphics[scale=0.37]{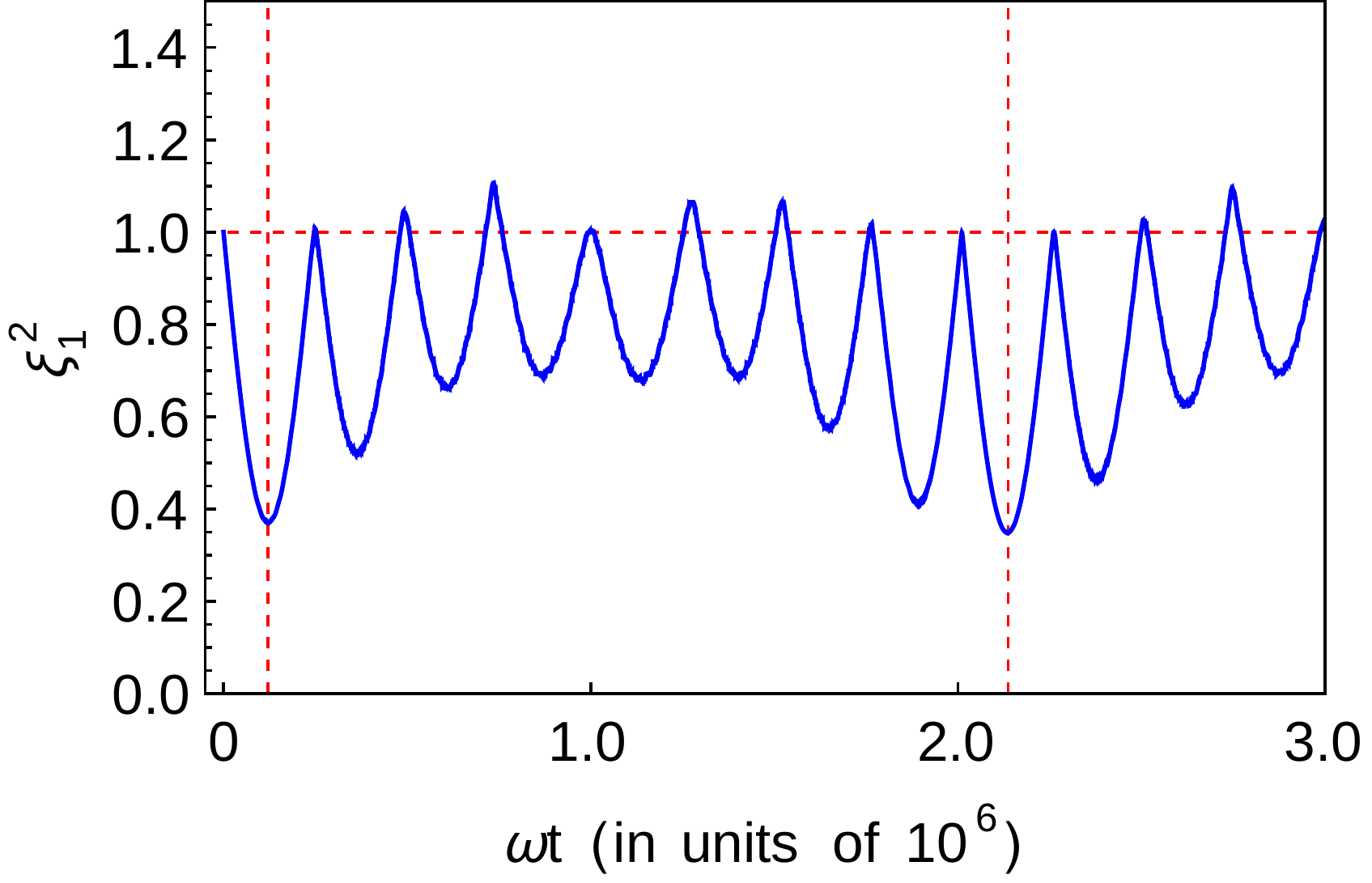}}
		\subfloat[{\sf (a$_{2}$)}]{\includegraphics[scale=0.25]{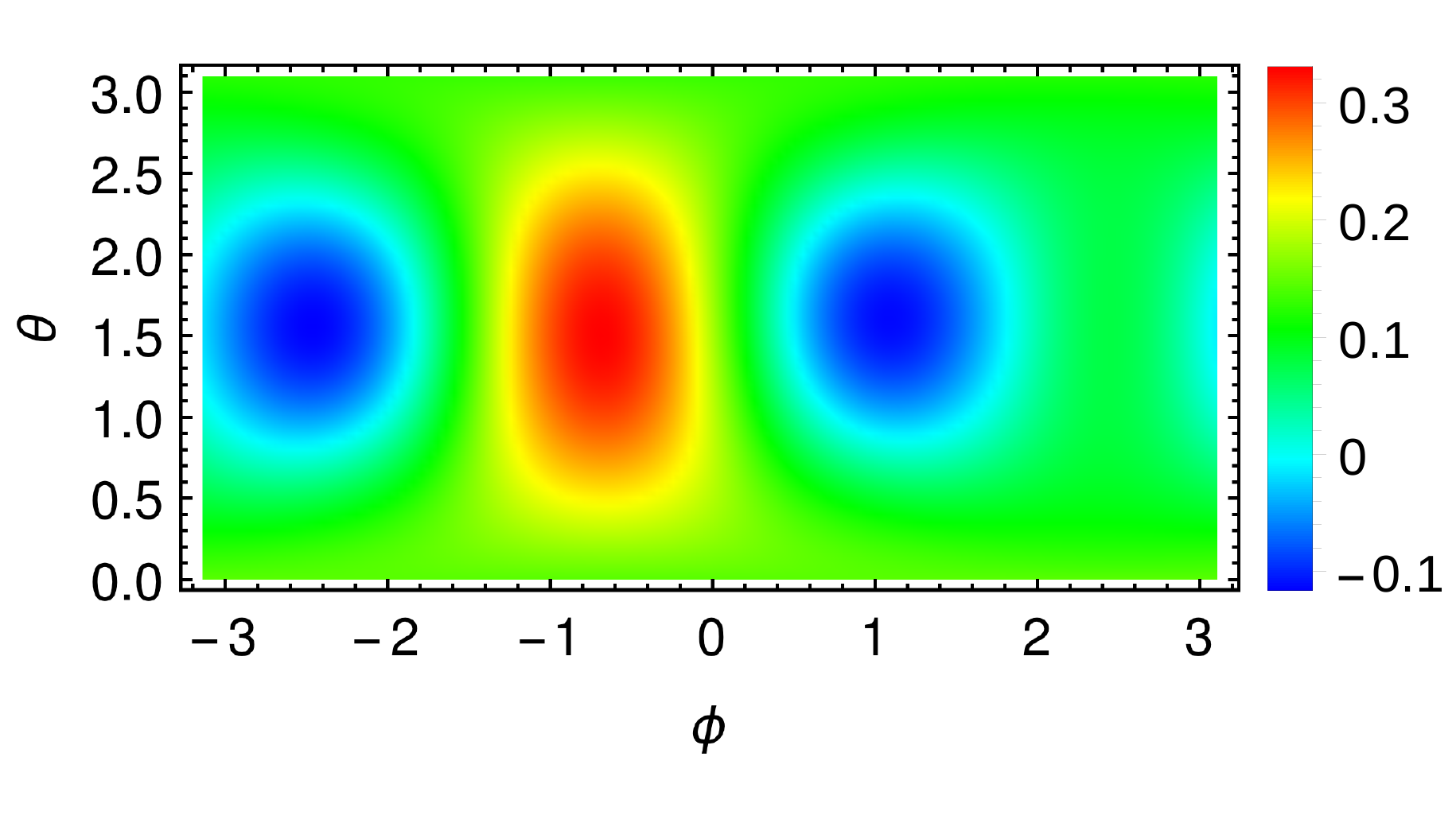}}
		\subfloat[{\sf (a$_{3}$)}]{\includegraphics[scale=0.25]{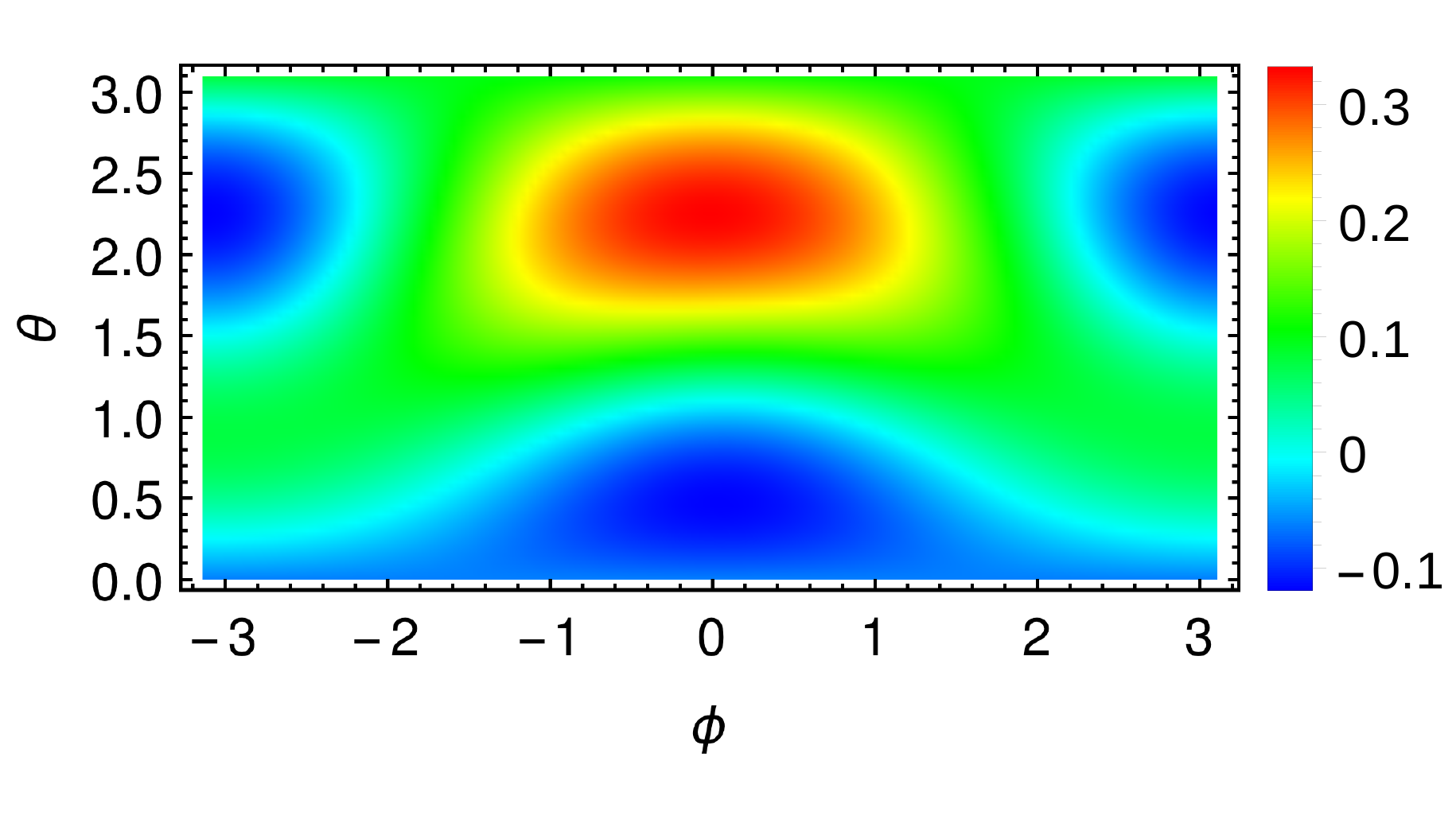}}\\
		\hspace*{0cm}
		\subfloat[{\sf (b$_{1}$)}]{\includegraphics[scale=0.40]{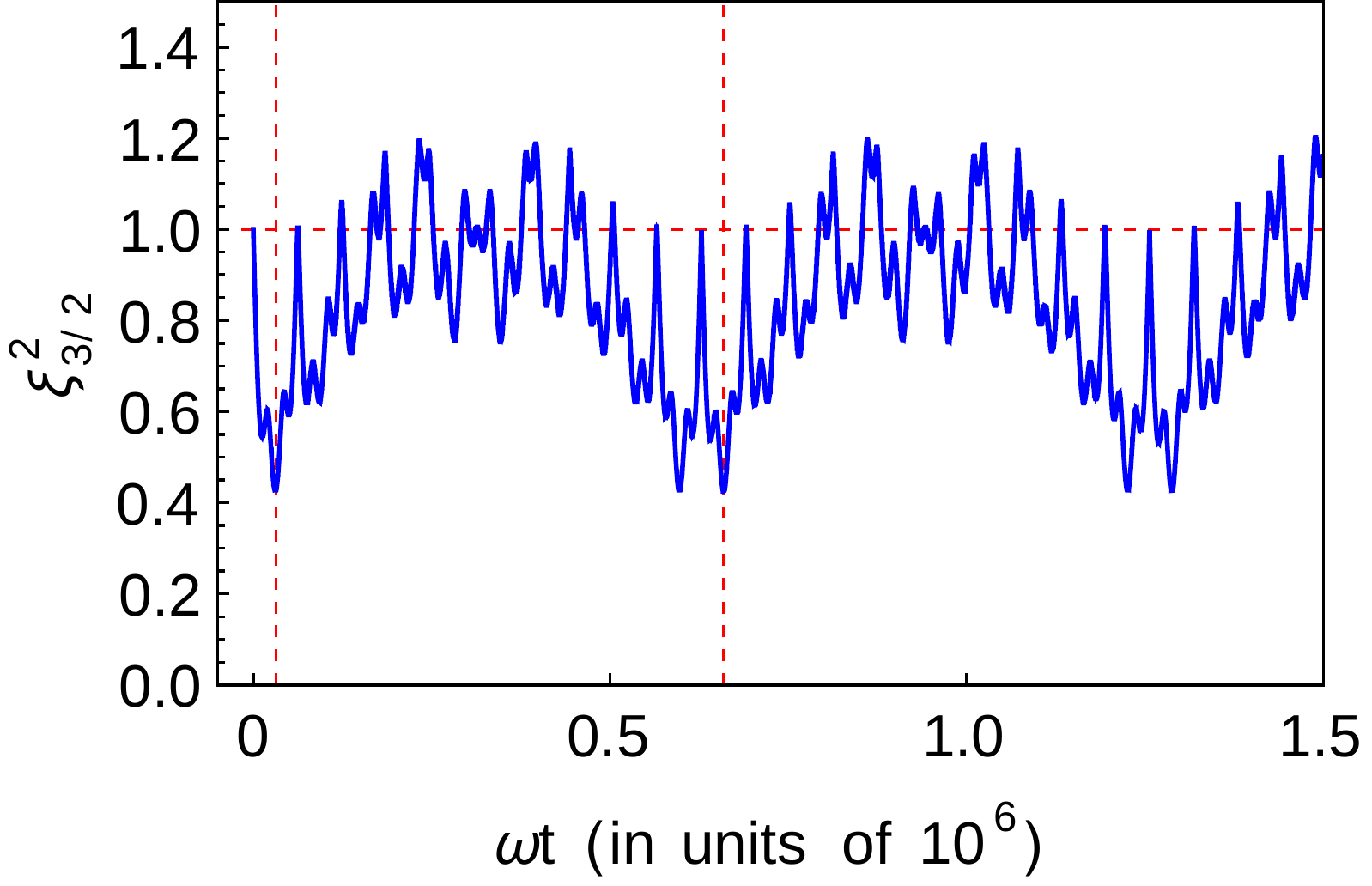}}
		\subfloat[{\sf (b$_{2}$)}]{\includegraphics[scale=0.32]{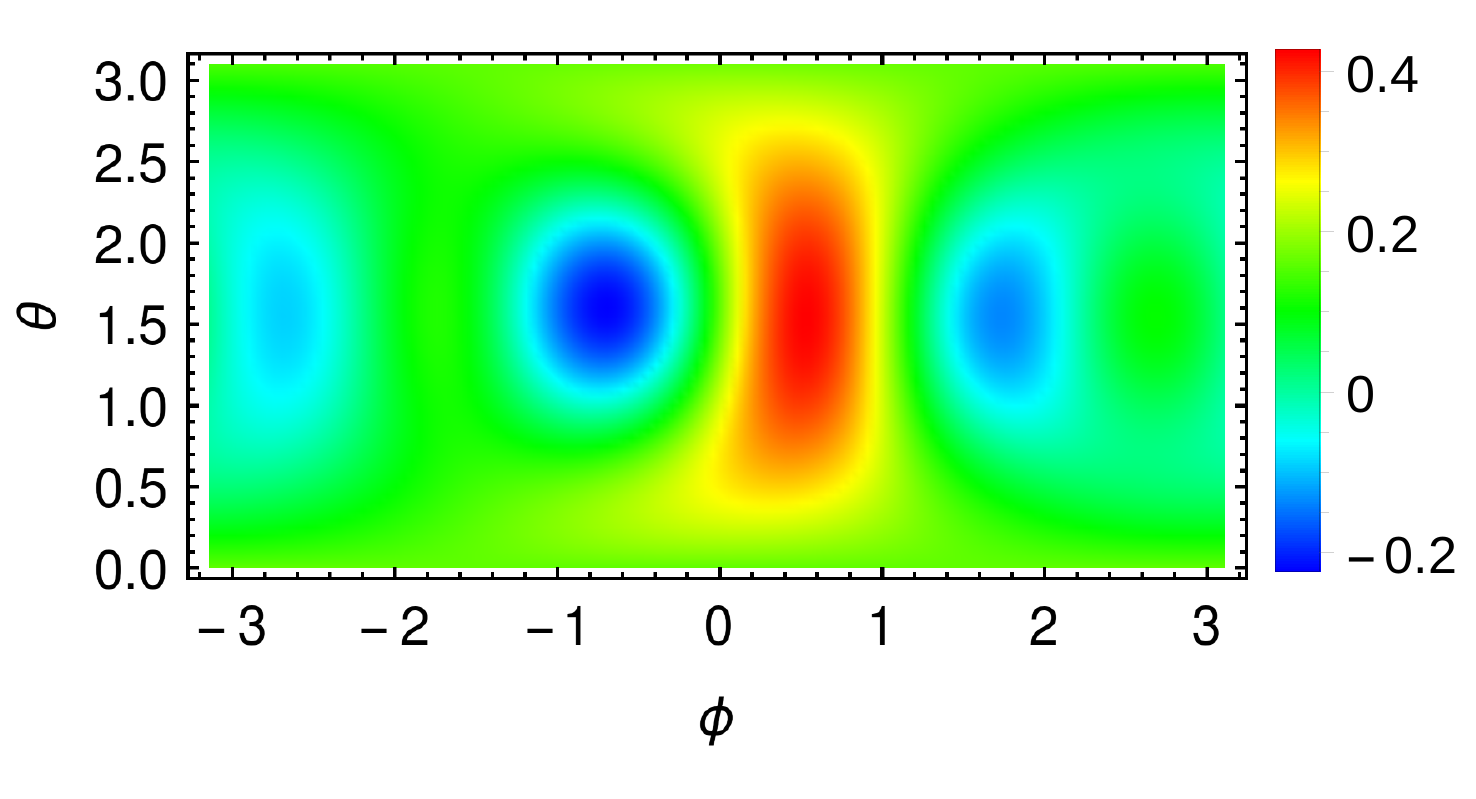}}
		\subfloat[{\sf (b$_{3}$)}]{\includegraphics[scale=0.32]{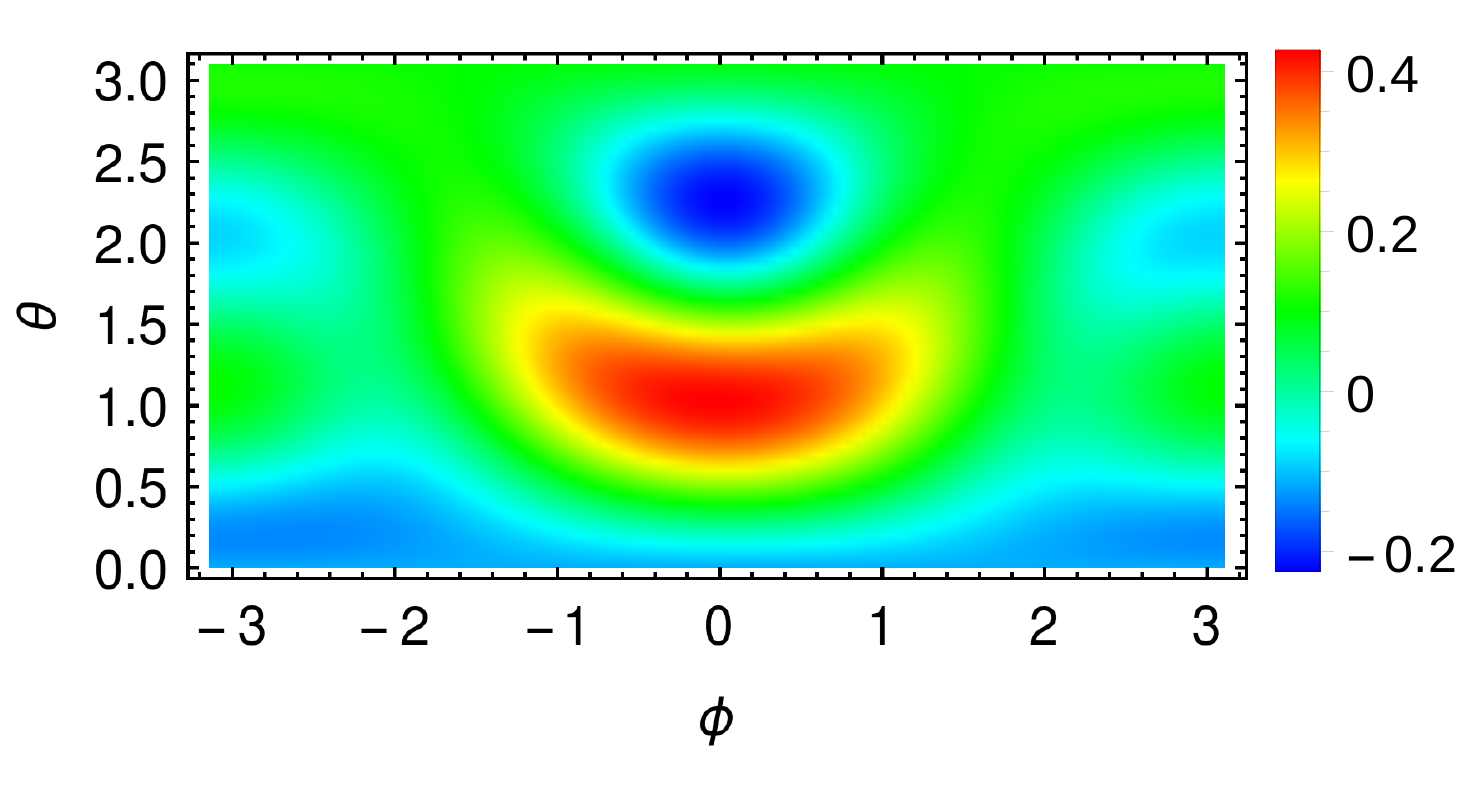}}\\
		\caption{The qudit Wigner $\mathrm{W}_{\mathcal{Q}}$-distributions are plotted for the initial state (\ref{state-t0}) with 
			$\mathrm{c}=0$, which represent a spin coherent state. {\sf(a)}: First row refers to spin $s=1$ case with the parametric choices $\Delta =0.12, 
			\widetilde{\lambda}=0.005, \mathfrak{z}=0.3249, \alpha =0.5, r=0$. Diagram $\mathsf{(a_{1})}$ depicts the long
			time evolution of the squeezing parameter, while $\mathsf{(a_{2} ,a_{3})}$ refer, consecutively, 
			to $\mathrm{W}_{\mathcal{Q}}$-distributions at times $t=120717,  \,t=2137004$. The corresponding squeezing coefficients 
			equal $\xi^{2}_{1} = 0.3712$ and  $\xi^{2}_{1}= 0.3492$. 
			{\sf(b)}: The  $\mathrm{W}_{\mathcal{Q}}$-distributions displayed in the second row study the $s=\tfrac{3}{2}$ example. Other parameters read $\Delta =0.1,\widetilde{\lambda}=0.01,
			\mathfrak{z}=0.3249,\alpha =0.5, r=0$.  The long
			time evolution of the squeezing parameter is portrayed in $\mathsf{(b_{1})}$, whereas $\mathsf{(b_{2}, b_{3})}$ represent 
			$\mathrm{W}_{\mathcal{Q}}$-distributions at respective times $t=31685, \, t=659013$ with the corresponding squeezing coefficients given by $\xi^{2}_{\frac{3}{2}} = 0.4242$ and $\xi^{2}_{\frac{3}{2}}= 0.4223$.} 
		\label{squeeze-longtime}
	\end{center}
\end{figure}

In the presence of the spin squeezing a quasiprobability density, say, the Wigner  $\mathrm{W}_{\mathcal{Q}}$-distribution
assumes an elliptical shape in contrast to an isotropic form that is evident for a spin coherent state. Quantum uncertainties are deformed by effective nonlinear interactions that twists the fluctuations as observed in Fig. 
\ref{squeeze-longtime}. Nonlinear interactions triggering the spin squeezing effect are produced in the low energy limit of the effective Hamiltonian for the bipartite process (\ref{H-Sp}) considered here. For instance, adopting the technic developed in [\cite{JJ2007}] we may obtain the lowest order of nonlinear interactions in the present model, which, in particular, includes a term 
$\sim \omega {\widetilde{\lambda}}^{2}S_{\mathsf{z}}^{2}$ that activates one axis twisting of the quasiprobability densities. This effective Hamiltonian has also been achieved [\cite{LL2013}] using another technic. The mean spin direction and the optimal squeezing direction vary with time. We also note that in conjunction with the spin squeezing various degrees of eddy like structures are present the  
$\mathrm{W}_{\mathcal{Q}}$-distributions. It is observed that with the dominance of the said swirl in the phase space distributions, limitations arise in the the minimum attainable uncertainty. This behavior, when present, causes relatively higher values of the squeezing parameter $\xi^{2}_{s}$. For the $s=1$ and $s=\frac{3}{2}$ example  we study  the evolutionary behavior of the squeezing parameter 
$\xi^{2}_{s}$  in Fig. \ref{squeeze-longtime}. With the choice of the parameter $\mathrm{c}=0$ the factorized initial state 
(\ref{state-t0}) does not experience any squeezing at $t=0$ (Fig. \ref{squeeze-longtime}). Owing to the nonlinear terms in the effective Hamiltonian squeezing develops dynamically for the evolving state.  
\newpage
\vspace*{-10cm}

\newpage
\section{Conclusion}
\label{concude}
Applying an adiabatic approximation method we study a hybrid qudit-oscillator  interacting system in the strong as well as the ultrastrong interaction regimes. Starting with a pure state of the bipartite system, we observe its evolution via the phase space dynamics. The quasiprobability distributions in the hybrid factorized phase space are constructed. The qudit and the oscillator phase space densities are procured via a dimensional reduction process achieved by integrating the phase space variables related to one sector.
In the strong coupling domain the system displays a quasiperiodic behavior when it returns close to its initial configuration. Starting with factorizable initial state we observe that at the local minimum values of the entropy atomic Schr\"{o}dinger kitten states form at times given by rational fractions of the period. This may be
evidenced via the spin phase space distributions, say the $\mathrm{P}_{\mathcal Q}(\theta,\phi)$-representation. These kitten states   embody coherent quantum superposition  and therefore reveal nonclassicality.  An alternate spin tomographic description expresses the evolution of the system via a positive definite probability distribution reflecting the diagonal elements of the spin density matrix in an arbitrarily rotated frame. Since our bipartite system resides in a pure state, the pertinent subsystems have equal entropy. In particular, the local minimum configurations of the entropy are experienced by both the subsystems concurrently. Therefore the transitory emergence of the spin kitten states and their oscillator counterparts accompany each other. This may be relevant
in the experimental observation of the spin kitten states. Another feature of nonclassicality observed is that 
due to presence of the quadratic terms of the spin generators in the effective Hamiltonian, the initial spin coherent state dynamically evolves to  squeezed spin states  
recurrently when the system undergoes quantum fluctuations. For the ultrastrong coupling strength the quasiperiodicity of the evolution disappears and the entropy shows stabilization in the presence of a randomized fluctuation. Moreover, both in the strong and ultrastrong coupling regimes antibunching of the emitted photons is observed particularly for the low spin ($s=1$) case.
  We also note that it is important to estimate the extent of nonclassicality of the quantum states in the ultrastrong coupling regime (Figs. \ref{entropy-1-tL}{\sf(b)} and \ref{entropy-3/2-tL}{\sf (b)}) where an equilibrium-like behavior sets in. This will be pursued elsewhere. 
\section*{Acknowledgments}
We are indebted for generous computational help from the Department of Central Instrumentation and Service Laboratory, University of Madras. One of us (MB) acknowledges the support from the University of Madras for granting a University Research Fellowship. Another author (RC) wishes to thank the Department of Nuclear Physics, University of Madras for kind hospitality.


\begin{thebibliography}{99}	
\bibitem{JC1963} E.T. James, F.W. Cummings, Proc. IEEE {\bf 51} 89 (1963).
 
\bibitem{ABS2002} A.D. Armour, M.P. Blencowe, K.C. Schwab, Phys. Rev. Lett. {\bf 88}, 148301 (2002). 

\bibitem{LaHaye2009} M.D. LaHaye, J. Suh, P.M. Echternach, K.C. Schwab, M.L. Roukes, Nature Lett. {\bf 459}, 960 (2009).

\bibitem{Niemczyk2010} T. Niemczyk, F. Deppe, H. Huebl, E.P. Menzel, F. Hocke, M.J. Schwarz, J.J. Garcia-Ripoli,
D. Zueco, T. H\"{u}mer, E. Solano, A. Marx, R. Gross, Nat. Phys. {\bf 6}, 772 (2010).

\bibitem{Forn-Diaz2010} P. Forn-D\'{i}az, J. Lisenfeld, D. Marcos, J. J. Garc\'{i}a-Ripoll,
E. Solano, C.J.P. M. Harmans, J. E. Mooij, Phys. Rev. Lett. {\bf 105}, 237001 (2010).

\bibitem{Anappara2009} A.A. Anappara, S. De Liberato, A. Tredicucci, C. Ciuti, G. Biasiol, L. Sorba, F. Beltram, Phys. Rev. B {\bf 79}, 201301(R) (2009). 

\bibitem{Todorov2010}  Y. Todorov, A. M. Andrews, R. Colombelli, S. De Liberato, C. Ciuti, P. Klang, G. Strasser, and C. Sirtori, Phys. Rev. Lett. {\bf 105}, 196402 (2010).

\bibitem{YN2005} J.Q. You, F. Nori, Phys. Today {\bf 58}, 42 (2005).

\bibitem{YN2011} J.Q. You, F. Nori, Nature {\bf 474}, 589 (2011).

\bibitem{NJBN2012} P.D. Nation, J.R. Johansson, M.P. Blencowe, F. Nori, Rev. Mod. Phys. {\bf 84}, 1 (2012). 

\bibitem{BAN2011} I. Buluta, S. Ashhab, F. Nori, Rep. Progr. Phys. {\bf 74}, 104401 (2011). 

\bibitem{BN2009} I. Buluta, F. Nori, Science {\bf 326}, 108 (2009).

\bibitem{GAN2014} I.M. Georgescu, S. Ashhab, F. Nori, Rev. Mod. Phys. {\bf 86}, 153 (2014). 

\bibitem{LMK2016} N. Lambert, Y. Matsuzaki, K. Kakuyanagi, N. Ishida, S. Saito, F. Nori, Phys. Rev. B {\bf 94}, 224510 (2016).

\bibitem{KFMM2015} E.O. Kiktenko, A.K. Fedorov, O.V. Man’ko, V.I. Man’ko, Phys. Rev. A {\bf 91}, 042312 (2015).

\bibitem{Lapkiewicz2011} R. Lapkiewicz, P. Li, C. Sch\"{a}ff, N.K. Langford, S. Ramelow, M. Wie\'{s}niak,  
A. Zeilinger, Nature {\bf 474}, 490 (2011).

\bibitem{ZAYN2014} Z.I. Ziang, S. Ashhab, J.Q. You, F. Nori, Rev. Mod. Phys. {\bf 86}, 153 (2014).

\bibitem {R1971} J.M. Radcliffe, J. Phys. A {\bf 4}, 313 (1971).

\bibitem {ACGT1972} F.T. Arecchi, E. Courtens, R. Gilmore, H. Thomas, Phys. Rev. A {\bf 6}, 2211 (1972).

\bibitem{AT1976} G.S. Agarwal, S.S. Trivedi, Opt. Comm. {\bf 18}, 417 (1976).

\bibitem{B2000} D. Braun, P.A. Braun, F. Haake, Opt. Comm. {\bf 179}, 411 (2000).

\bibitem{CNL2011} O. Casta$\tilde{\mbox{n}}$os, E.N. Achar, R.L. Pe$\tilde{\mbox{n}}$a, J.G. Hirsch, 
Phys. Rev. A {\bf 84}, 013819 (2011).

\bibitem{ABNV1977} G.S. Agarwal, A.C. Brown, L.M. Narducci, G. Vetri, Phys. Rev. A {\bf 15}, 1613 (1977).

\bibitem{D1980} P.D. Drummond, S.S. Hassan, Phys. Rev. A {\bf 22}, 662 (1980).

\bibitem{HR2006} S. Haroche, J.-M. Raimond,  \textit{Exploring the Quantum: Atoms, Cavities, and
Photons}, Oxford Univ. Press, Oxford (2006).

\bibitem{APS1997} G.S. Agarwal, R.R. Puri, R.P. Singh, Phys. Rev. A {\bf 56}, 2249 (1997).

\bibitem{GG1997} C.C. Gerry, R. Grobe, Phys. Rev. A {\bf 56}, 2390 (1997).

\bibitem{Leibfried2005} D. Leibfried, E. Knill, S. Seidelin, J. Britton, R.B. Blakestad, J. Chiaverini, D.B. Hume, W.M. Itano, J.D.
Jost, C. Langer, R. Ozeri, R. Reichle, D.J. Wineland, Nature {\bf 438}, 639 (2005).

\bibitem{McConnell2013} R. McConnell, H. Zhang, S. \'{C}uk, J. Hu, M.H. Schleier-Smith, V. Vuleti\'{c}, Phys. Rev. A {\bf 88}, 063802 (2013).

\bibitem{Chalopin2018} T. Chalopin, C. Bouazza, A. Evrard, V. Makhalov, D. Dreon, J. Dalibard, L. A. Sidorenkov, S. Nascimbene,
Nature Comm. {\bf 9}, 1 (2018).

\bibitem{Song2019} C. Song, K. Xu, H. Li, Y.R. Zhang, X. Zhang, W. Liu, Q. Guo, Z. Wang, W. Ren, J. Hao, H. Feng, H.
Fan, D. Zheng, D.W. Wang, H. Wang, S.Y. Zhu, Science {\bf 365}, 574 (2019)

\bibitem {KU1991} M. Kitagawa, M. Ueda, Phys. Rev. Lett. {\bf 67}, 1852 (1991).

\bibitem{WBIMH1992} D.J. Wineland, J.J. Bollinger, W.M. Itano, F.L. Moore, D.J. Heinzen, Phys. Rev. A {\bf 46}, 
R6797 (1992).

\bibitem {KU1993} M. Kitagawa, M. Ueda, Phys. Rev. A {\bf 47}, 5138 (1993).

\bibitem{Miranowicz2002} A. Miranowicz, S.K. \"{O}zdemir, Yu-xi Liu, M. Koashi, N. Imoto, Y. Hirayama,  Phys. Rev. A {\bf 65}, 062321 (2002).

\bibitem{WWYJ2002} T.-L. Wang, L.-N. Wu, W. Yang, G.-R. Jin, N. Lambert, F. Nori, New. J. Phys. {\bf 16}, 063039 (2014).

\bibitem {SWYZ2006} L. Song, X. Wang, D. Yan, Z. Zong, J. Phys. B {\bf 39}, 559 (2006).

\bibitem{CLMZ1998} J.I. Cirac, M. Lewenstein, K. M{\o}lmer, P. Zoller, Phys. Rev. A {\bf 57}, 1208  (1998).

\bibitem{JFIB2018} Y. Jing, M. Fadel, V. Ivannikov, T. Byrnes, \textit{Split spin-squeezed Bose-Einstein condensates},arXiv: 1808.10679 [quant-ph] (2018).

\bibitem{TAN2008} D. Tsomokos, S. Ashhab, F. Nori, New J. Phys. {\bf 10}, 113020 (2008).

\bibitem{ZYL2014} Y. Zhang, L. Yu, J.-Q. Liang, G. Chen, S. Jia, F. Nori, Sci. Rep. {\bf4}, 4083 (2014).

\bibitem{YNSZ2019} V. Macr{\`i}, F. Nori, S. Savasta, D. Zueco, \textit{Optimal spin squeezing in cavity-QED-based 
systems}, arXiv: 1902.10377 [quant-ph] 2019. 

\bibitem{PMM2005} V. Petersen, L.B. Madsen, K. M{\o}lmer, Phys. Rev. A {\bf 71}, 012312 (2005).

\bibitem{HSSP1999} J. Hald, J.L. S{\o}rensen, C. Schori, E.S. Polzik, Phys. Rev. Lett. {\bf 83}, 1319 (1999). 

\bibitem{MWSN2011} J. Ma, X. Wang, C.P. Sun, F. Nori, Phys. Rep. {\bf 509}, 89 (2011).  

\bibitem {IGMS2005} E.K. Irish, J. Gea-Banacloche, J. Martin, K.C. Schwab, Phys. Rev. B {\bf 72}, 195410 (2005).

\bibitem{AN2010} S. Ashhab, F. Nori, Phys. Rev. A {\bf 81}, 042311 (2010).

\bibitem{A1981} G.S. Agarwal, Phys. Rev. A {\bf 24}, 2889 (1981).
        
\bibitem{Gerry2005} C. Gerry, P. Knight, \textit{Introductory Quantum Optics}, Cambridge Univ. Press, Cambridge (2005).

\bibitem{MM1997} V.I. Man'ko, O.V. Man'ko, JETP, {\bf 85}, 430 (1997).

\bibitem{DM1997} V.V. Dodonov, V.I. Man'ko, Phys. Lett. A {\bf 229}, 335 (1997).
        
\bibitem{LL2013} K.M.C. Lee, C.K. Law, Phys. Rev. A {\bf 88}, 015802 (2013).

\bibitem{Talman1968} J.D. Talman, \textit{Special Functions: A Group Theoretic Approach}, Benjamin, New York (1968).

\bibitem{AAR1999} G.E. Andrews, R. Askey, R. Roy, \textit{Special Functions}, Cambridge Univ. Press, Cambridge (1999).

\bibitem{MK1993} H. Moya-Cessa and P.L. Knight, Phys. Rev. A {\bf 48}, 2479 (1993).

\bibitem{DAS1994} J.P. Dowling, G.S. Agarwal, W.P. Schleich, Phys. Rev. A {\bf 49}, 4101 (1994).

\bibitem{M1939} J. Meixner, Math. Z. {\bf 44}, 531 (1939). 

\bibitem{AL1970} H. Araki, E.H. Lieb, Comm. Math. Phys. {\bf 18}, 160 (1970).

\bibitem{DMMW2000} V.V. Dodonov, O.V. Man’ko, V.I. Man’ko, A. W\"{u}nsche, J. Mod. Opt. {\bf 47}, 633 (2000).

\bibitem{RS2007} E. Romera, F. de los Santos, Phys. Rev. Lett. {\bf 99}, 263601 (2007).

\bibitem{AP1989} I.S. Averbukh, N.F. Perelman, Phys. Lett. A {\bf 139}, 449 (1989).

\bibitem{W1984} K. W\'{o}dkiewicz, Opt. Comm. {\bf 51}, 198 (1984).

\bibitem{JJ2007} D.F.V. James, J. Jerke, Can. J. Phys. {\bf 85}, 625 (2007).

\end{thebibliography}
\end{document}